\documentstyle [12pt,eqsecnum,aps,amsfonts] {revtex}
\input epsf
\newcommand{\beq}{\begin{equation}}
\newcommand{\eeq}{\end{equation}}
\newcommand{\beqs}{\begin{eqnarray}}
\newcommand{\eeqs}{\end{eqnarray}}

\begin{document}
\tighten
\draft
\preprint{YITP-SB-99-31, BNL-HET-99/35}

\baselineskip 6.0mm

\title{Exact Potts Model Partition Functions on Ladder Graphs}

\author{Robert Shrock$^{(a)}$\thanks{(a): permanent address; 
email: robert.shrock@sunysb.edu}}

\address{(a) \ C. N. Yang Institute for Theoretical Physics \\
State University of New York \\
Stony Brook, N. Y. 11794-3840}

\address{(b) \ Physics Department \\
Brookhaven National Laboratory \\
Upton, NY  11973}

\maketitle

\vspace{5mm}

\begin{abstract}

We present exact calculations of the partition function $Z$ of the $q$-state
Potts model and its generalization to real $q$, the random cluster model, for
arbitrary temperature on $n$-vertex ladder graphs with free, cyclic, and
M\"obius longitudinal boundary conditions. These partition functions are
equivalent to Tutte/Whitney polynomials for these graphs.  The free energy is
calculated exactly for the infinite-length limit of these ladder graphs and the
thermodynamics is discussed.  By comparison with strip graphs of other widths,
we analyze how the singularities at the zero-temperature critical point of the
ferromagnet on infinite-length, finite-width strips depend on the width.  We
point out and study the following noncommutativity at certain special values
$q_s$: $\lim_{n \to \infty} \lim_{q \to q_s} Z^{1/n} \ne \lim_{q \to q_s}
\lim_{n \to \infty} Z^{1/n}$.  It is shown that the Potts/random cluster
antiferromagnet on both the infinite-length line and ladder graphs with cyclic
or M\"obius boundary conditions exhibits a phase transition at finite
temperature if $0 < q < 2$, but with unphysical properties, including negative
specific heat and non-existence, in the low-temperature phase, of an $n \to
\infty$ limit for thermodynamic functions that is independent of boundary
conditions.  Considering the full generalization to arbitrary complex $q$ and
temperature, we determine the singular locus ${\cal B}$ in the corresponding
${\mathbb C}^2$ space, arising as the accumulation set of partition function
zeros as $n \to \infty$.  In particular, we study the connection with the $T=0$
limit of the Potts antiferromagnet where ${\cal B}$ reduces to the accumulation
set of chromatic zeros.  Certain properties of the complex-temperature phase
diagrams are shown to exhibit close connections with those of the model on the
square lattice, showing that exact solutions on infinite-length strips provide
a way of gaining insight into these complex-temperature phase diagrams.

\end{abstract}


\vspace{16mm}

\pagestyle{empty}
\newpage

\pagestyle{plain}
\pagenumbering{arabic}
\renewcommand{\thefootnote}{\arabic{footnote}}
\setcounter{footnote}{0}

\section{Introduction}

The $q$-state Potts model has served as a valuable model for the study of phase
transitions and critical phenomena \cite{potts,wurev}.  On a lattice, or, more
generally, on a graph $G$, at temperature $T$, this model is defined by the
partition function 
\beq
Z(G,q,v) = \sum_{ \{ \sigma_n \} } e^{-\beta {\cal H}}
\label{zfun}
\eeq
with the (zero-field) Hamiltonian
\beq
{\cal H} = -J \sum_{\langle i j \rangle} \delta_{\sigma_i \sigma_j}
\label{ham}
\eeq
where $\sigma_i=1,...,q$ are the spin variables on each vertex $i \in G$;
$\beta = (k_BT)^{-1}$; and $\langle i j \rangle$ denotes pairs of adjacent
vertices.  The graph $G=G(V,E)$ is defined by its vertex set $V$ and its edge 
set $E$; we denote the number of vertices of $G$ as $n=n(G)=|V|$ and the
number of edges of $G$ as $e(G)=|E|$.  We use the notation
\beq
K = \beta J
\label{kdef}
\eeq
\beq
a = u^{-1} = e^K
\label{a}
\eeq
and 
\beq
v = a-1
\label{v}
\eeq
so that the physical ranges are (i) $a \ge 1$, i.e., $v \ge 0$ corresponding to
$\infty \ge T \ge 0$ for the Potts ferromagnet, and (ii) $0 \le a \le 1$,
i.e., $-1 \le v \le 0$, corresponding to $0 \le T \le \infty$ for the Potts 
antiferromagnet. An equivalent expression for $Z$ is
\beq
Z(G,q,v) = \sum_{\{\sigma_i \}} \prod_{\langle i j \rangle} 
(1+v\delta_{\sigma_i,\sigma_j}) \ . 
\label{zv}
\eeq
One defines the (reduced) free energy per site $f=-\beta F$,
where $F$ is the actual free energy, via 
\beq
f(\{G\},q,v) = \lim_{n \to \infty} \ln [ Z(G,q,v)^{1/n}]  \ . 
\label{ef}
\eeq

Let $G^\prime=(V,E^\prime)$ be a spanning subgraph of $G$, i.e. a subgraph 
having the same vertex set $V$ and an edge set $E^\prime \subseteq E$. Then 
$Z(G,q,v)$ can be written as the sum \cite{birk}-\cite{kf} 
\beq
Z(G,q,v)=\sum_{G^\prime \subseteq G} q^{k(G^\prime)}v^{e(G^\prime)}
\label{cluster}
\eeq 
where $k(G^\prime)$ denotes the number of connected components of $G^\prime$. 
The formula (\ref{cluster}) enables one to generalize
$q$ from ${\mathbb Z}_+$ to ${\mathbb R}_+$ (keeping $v$ in its physical
range). This generalization is the random cluster model \cite{kf}. The formula
(\ref{cluster}) shows that $Z(G,q,v)$ is a polynomial in $q$ and $v$ 
(equivalently, $a$) with maximum degrees 
$\max\{deg_q(Z(G,q,v))\}=n(G)$ and $\max\{deg_v(Z(G,q,v))\}=e(G)$.  The 
minumum degrees are $\min\{deg_q(Z(G,q,v))\}=k(G)$, which is
equal to 1 for the graphs of interest here (since they are connected), and 
$\min\{deg_v(Z(G,q,v))\}=0$, so 
\beq
Z(G,q,v) = \sum_{r=k(G)}^{n(G)}\sum_{s=0}^{e(G)}z_{rs} q^r v^s
\label{zpol}
\eeq
with $z_{rs} \ge 0$. 

The Potts model partition function on a graph $G$ is essentially equivalent to
the Tutte polynomial \cite{tutte1}-\cite{tutte5} and Whitney rank polynomial
\cite{whit}, \cite{wurev}, \cite{bbook}-\cite{boll} for this graph, as
discussed in the appendix.  As a consequence, there are many interesting
connections between properties of this partition function and various
graph-theoretic quantities. 

The Potts model has never been solved exactly for arbitrary temperature on
lattices of dimensionality $d \ge 2$ except for special $d=2$, $q=2$ case in
which it is equivalent to the solvable 2D Ising model \cite{onsager} (with the
redefinition $J_{Potts} = 2J_{Ising}$).  Knowledge about the Potts model
includes exact calculations of the critical exponents and critical value of the
free energy for the 2D ferromagnet for the range $q \le 4$ where it has a
second-order transition; conformal algebra properties for the same range of
$q$; the latent heat at the transition point for $q \ge 5$; and certain
formulas for the critical point \cite{wurev,cft}. There is thus motivation for
studies that can give further insight into the Potts model.  Among these are
exact results for the partition function and free energy that one can obtain
for infinite-length, finite-width strips with various boundary conditions.  We
shall present such results in this paper.

\vspace{10mm}

One of the interesting features of the Potts model is that the antiferromagnet
(AF) exhibits nonzero ground state entropy (without frustration) for
sufficiently large $q$ on a given lattice or graph $G$, and serves as a
valuable model for the study of this phenomenon.  The phenomenon of nonzero
ground state entropy, $S_0 > 0$, is an exception to the third law of
thermodynamics \cite{lieb,al}.  This is equivalent to a ground state
degeneracy per site (vertex), $W > 1$, since $S_0 = k_B \ln W$.  The
zero-temperature partition function of the above-mentioned $q$-state Potts
antiferromagnet (PAF) on $G$ satisfies
\beq 
Z(G,q,T=0)_{PAF} \equiv Z(G,q,v=-1)=P(G,q)
\label{zp}
\eeq
where $P(G,q)$ is the chromatic polynomial (in $q$) expressing the number of 
ways of coloring the vertices of the graph $G$ with $q$ colors such that no 
two adjacent vertices have the same color \cite{birk,rrev,rtrev}.  The minimum 
(integral) number of colors necessary for this coloring is the chromatic 
number of $G$, denoted $\chi(G)$.  Thus 
\beq
W(\{G\},q)= \lim_{n \to \infty}P(G,q)^{1/n}
\label{w}
\eeq
where we use the symbol $\{G\}$ to denote $\lim_{n \to \infty}G$ for a given
family of graphs. 

Since $Z(G,q,v)$ is a polynomial in $q$ and $v$, or equivalently, $a$, one can
generalize $q$ from ${\mathbb Z}_+$ not just to ${\mathbb R}_+$ but to
${\mathbb C}$ and $a$ from its physical ferromagnetic and antiferromagnetic
ranges $1 \le a \le \infty$ and $0 \le a \le 1$ to $a \in {\mathbb C}$.  A
subset of the zeros of $Z$ in the two-complex dimensional space ${\mathbb C}^2$
defined by the pair of variables $(q,a)$ can form an accumulation set in the $n
\to \infty$ limit, denoted ${\cal B}$, which is the continuous locus of points
where the free energy is nonanalytic. As will be discussed below, this locus is
determined as the solution to a certain $\{G\}$-dependent equation.  For a
given value of $a$, one can consider this locus in the $q$ plane, and we denote
it as ${\cal B}_q(\{G\},a)$.  In the special case $a=0$ ($v=-1$) where the
partition function is equal to the chromatic polynomial, the zeros in $q$ are
the chromatic zeros, and ${\cal B}_q(\{G\},a=0)$ is their continuous
accumulation set in the $n \to \infty$ limit \cite{bds}- \cite{ss}; we have
determined these accumulation sets exactly for various families of graphs in a
series of papers. Other properties of chromatic zeros such as zero-free regions
for general graphs, are of mathematical interest (see, e.g.,
\cite{bl,read91,zerofree}), although we shall not focus on them here.  For a
given value of $q$, we shall study the continuous accumulation set of the zeros
of $Z(G,q,v)$ in the $a$ plane; this will be denoted ${\cal B}_a(\{G\},q)$.  It
will often be convenient to consider the equivalent locus in the $u=1/a$ plane,
namely ${\cal B}_a(\{G\},q)$.  We shall sometimes write ${\cal B}_q(\{G\},a)$
simply as ${\cal B}_q$ when $\{G\}$ and $q$ are clear from the context, and
similarly with ${\cal B}_{a}$ and ${\cal B}_{a}$.  A subtlety in the 
definition of this locus will be discussed in the next section.

One gains a unified understanding of the separate loci ${\cal B}_q(\{G\})$ for
fixed $a$ and ${\cal B}_a(\{G\})$ for fixed $q$ by relating these as different
slices of the locus ${\cal B}$ in the ${\mathbb C}^2$ space defined by $(q,a)$.
This is similar to the insight that one gained in studies of accumulation sets
of zeros of the partition functions of the Ising model in the ${\mathbb C}^2$
space defined by $(a,\mu)$, where $\mu=e^{-2\beta H}$ with $H$ being the
external field \cite{ipz,ih}, which generalized the Yang-Lee zeros
(zeros in $\mu$ for fixed physical $a$) \cite{ly} and Fisher zeros (zeros in
$a$ for fixed $H$, often $H=0$) \cite{fisher}.

In our earlier works on ${\cal B}_q(\{G\})$ for $a=0$, we had denoted the
maximal region in the complex $q$ plane to which one can analytically continue
the function $W(\{G\},q)$ from physical values where there is nonzero ground
state entropy as $R_1$.  The maximal value of $q$ where ${\cal B}$ intersects
the (positive) real axis was labelled $q_c(\{G\})$.  Thus, region $R_1$
includes the positive real axis for $q > q_c(\{G\})$.  Correspondingly, in our
works on complex-temperature properties of spin models, we had labelled the
complex-temperature extension (CTE) of the physical paramagnetic phase as 
(CTE)PM, which will simply be denoted PM here, the extension being understood,
and similarly with ferromagnetic (FM) and antiferromagnetic (AFM); other
complex-temperature phases, having no overlap with any physical phase, were
denoted $O_j$ (for ``other''), with $j$ indexing the particular phase
\cite{chisq}.  Here we shall continue to use this notation for
the respective slices of ${\cal B}$ in the $q$ and $a$ or $u$ planes.

\vspace{6mm}

In this paper we shall present exact calculations of the Potts/random cluster
partition function for strips of the square lattice with arbitrary length $L_x$
and width $L_y=2$, i.e ladder graphs, having free, periodic (= cyclic), and
M\"obius longitudinal ($x$-direction) boundary conditions.  These families of
graphs are denoted, respectively, as $S_{m}$ (for open \underline strip), $L_m$
(for ladder), and $ML_m$ (for M\"obius ladder), where $L_x=m+1$ for $S_m$
(following our labelling convention in \cite{strip}) and $L_x=m$ for $L_m$ and
$ML_m$.  One has $n(S_m)=2(m+2)$ and $n(L_m)=n(ML_m)=2m$.  It will also be
instructive to use the well-known solutions for the partition function on the
tree and circuit graphs to illustrate some points.  We shall discuss several
items and investigate several questions about the Potts/random cluster model:

\begin{enumerate} 

\item 

We analyze the thermodynamic behavior of the Potts model (for $q \ge 2$) on the
infinite-length, $L_y=2$ strip and compare it with the known behavior on the
line.  In particular, we study the zero-temperature critical point of the Potts
ferromagnet and discuss how the critical singularities (which are essential
singularities in temperature) depend on $q$ and the width of the strip.  For
reference, one may recall that as one part of his original paper, Onsager
used his closed-form solution to the partition function of the Ising model to
study it for $L_y \times \infty$ strips of the square lattice \cite{onsager}.  
The difference, of course, is that for the Potts model with $q \ne 2$, one 
does not have a general closed-form solution for $Z$ on a $L_y \times L_x$ 
grid with arbitrary $L_y$ and $L_x$; indeed if one did, one would have solved
the model on the square lattice. 

\item 

We shall show that the Potts/random cluster model with $0 < q < 2$ on the $n
\to \infty$ limit of the circuit, ladder, and M\"obius ladder graphs exhibits a
finite-temperature phase transition but with unphysical properties in the
low-temperature phase, including negative specific heat, negative partition
function, and non-existence of an $n \to \infty$ limit that is independent of 
boundary conditions. 

\item

We shall point out and illustrate a certain noncommutativity in the definition
of the free energy for the Potts/random cluster model. 

\item 

For a given family of graphs $G$ and its $n \to \infty$ limit, $\{G\}$, we
shall explore the nature of the nonanalyticities of the free energy in $q$ and
the temperature variable $u$.  Some questions related to this are: what is the 
locus ${\cal B}_q$ for various values of $u$ and the locus ${\cal B}_u$ for
various values of $q$ (and the loci $({\cal B}_u)_{qn}$ and $({\cal
B}_u)_{nq)}$, for the special values $q=q_s$ where these differ)?  The strip
graphs that we consider here are useful for this study since they are wide
enough to exhibit a number of important features but narrow enough so that the
terms, denoted $\lambda_j$, whose powers contribute to the partition function
for a given family of graphs (see eq. (\ref{zgsum})) can be calculated as
explicit algebraic functions.  As our previous studies of asymptotic limits of
chromatic polynomials on various infinite-length, finite-width strips have
shown \cite{strip,wcy,pm,bcc}, as the width of the strip increases, one
encounters algebraic equations defining the $\lambda_j$'s that increase in
degree so that these $\lambda_j$ can involve cube roots, fourth roots, and, for
equations higher than fourth degree, one cannot obtain exact analytic
expressions for them, making it more cumbersome to calculate the
locus ${\cal B}$, which is the solution to the degeneracy in magnitude of
different dominant $\lambda_j$'s.

\item

Starting from our previous determination of ${\cal B}_q(\{G\})$ for the
zero-temperature limit of the Potts antiferromagnet, we explore how this locus
changes as on increases $T$ in the range $0 < T \le \infty$. In cases where
this locus separates the $q$ plane into different regions for $T=0$, does it
continue to do so?  How does the point $q_c(\{G\})$ vary with temperature?
Since we are now dealing with a singular locus in ${\mathbb C}^2$, we can
investigate how the various features of the slice of ${\cal B}(\{G\})$ in the
$q$ plane relate to features of the slice ${\cal B}(\{G\})$ in the plane of the
temperature variable, $u$ (or the equivalent variable $a$). 

\item 

We shall show that certain features of the complex-temperature phase diagrams
of the infinite-length, finite-width strip graphs considered here can give 
insight into analogous features of the Potts model on the square lattice.  

\item

Just as was true for our earlier studies of chromatic polynomials and their
asymptotic limits, it is of interest to explore the effects of different
boundary conditions on the singular locus ${\cal B}(\{G\})$, and we do this. 

\item 

For mathematically inclined readers, we shall give the Tutte polynomials that
are equivalent to the Potts model partition functions that we have calculated
for the cyclic and M\"obius strip graphs and extract special values that are of
specific graph-theoretic interest.

\end{enumerate}

  Conference reports on some of the material in this paper were given in
\cite{bcc,tw}.  In collaboration with H. Kluepfel, we have also carried out
calculations of Potts/random cluster partition functions for general $T$ and
$q$ on finite patches of several 2D lattices \cite{ks} (see also \cite{kc}); 
our work here complements these calculations on finite patches in that we 
obtain exact results for strip graphs of arbitrarily great length, 
and the nonanalyticities in the limit $n \to \infty$.

\section{Some General Considerations}

\subsection{Basic Properties of $Z$}

For our later analysis, it will be useful to record some basic properties of
the Potts model partition function.  Assuming $q > 0$, one observes that for
the Potts ferromagnet, since $v > 0$, each term in the sum (\ref{cluster}) is
positive, and consequently, $Z(G,q,v)$ does not have any zeros on the positive
real $q$ axis for the physical temperature range.  On higher-dimensional
lattices where the ferromagnet has a finite-temperature phase transition, zeros
will coalesce and pinch the real positive $q$ axis to form a region boundary,
but this does not happen in the 1D case and the infinite-length, finite-width
strips, which are quasi-1D systems.

One may ask what factors $Z$ has in general.  From eq. (\ref{cluster}) it 
follows that for an arbitrary graph $G$, 
\beq
Z(G,q=0,v)=0
\label{zq0}
\eeq
and since $Z(G,q,v)$ is a polynomial, this implies that $Z(G,q,v)$ always has 
an overall factor of $q$.  For the families of graphs studied here, this is, in
general, the only overall factor that $Z(G,q,v)$ has.  For the special case
$v=-1$, the resultant chromatic polynomial $Z(G,q,v=-1)=P(G,q)$ has the 
additional factors $\prod_{s=1}^{\chi(G)-1}(q-s)$. 
Another general result is 
that 
\beq
Z(G,q=1,v)=\sum_{G^\prime \subseteq G} v^{e(G^\prime)} = a^{e(G)} \ . 
\label{zq1}
\eeq
For temperature $T=\infty$, i.e., $v=0$, we have 
\beq
Z(G,q,v=0)=q^{n(G)} \ . 
\label{za1}
\eeq
For the Ising case $q=2$, if $G$ is bipartite,
\beq
Z(G_{bip.},q=2,a)=a^{2e(G_b)}Z(G_{bip.},q=2,1/a) 
\label{zbipq2}
\eeq
where we have written $Z$ here as a function of $a$.  This is the well-known
equivalence of the Ising ferromagnet (F) and antiferromagnet (AF) on a
bipartite lattice, when one makes the replacement $J \to -J$, and hence $K \to
-K$, $a \to 1/a$.  As a consequence, the zeros of $Z$ in $a$ for $q=2$ are
invariant under the inversion mapping $a \to 1/a$.  Among the families of
graphs considered here, the following are bipartite (equivalently, have
chromatic number $\chi=2$): tree graph $T_m$ for any $m$; circuit graph $C_m$
and cyclic ladder graph $L_m$ with $L_y=2$ for even $m$; and M\"obius ladder
graph $ML_m$ for odd $m$.  In contrast, the cyclic ladder graph $L_m$ for odd
$m$ and the M\"obius ladder graph $ML_m$ for even $m$ have $\chi=3$.  

For a strip of the
square lattice with width $L_y$ and cyclic or M\"obius longitudinal boundary
conditions, the average coordination number (degree in the graph-theoretic
terminology) $\Delta_{ave.} = 2 \lim_{n(G) \to \infty} e(G)/n(G)$ is 
\beq
\Delta_{ave.} = 4-\frac{2}{L_y}  \ . 
\label{delta}
\eeq
For the corresponding strip of width $L_y$
and free boundary conditions, this formula for $\Delta$ also holds in the $L_x
\to \infty$ limit.  Another
consequence of the symmetry (\ref{zbipq2}) is that for the Ising case the 
internal energy $U$ and specific heat $C$ satisfy the relations 
\beq 
U(G_{bip},q=2,J)_{F} = U(G_{bip.},q, J \to -J)_{AF} - \frac{\Delta_{ave.}J}{2}
\label{ising_urel}
\eeq
and 
\beq
C(G_{bip},q=2,J)_{F} = C(G_{bip.},q, J \to -J)_{AF}
\label{ising_crel}
\eeq
(where we have taken the $n \to \infty$ limit, in which these results are
independent of the boundary conditions for physical values of temperature).  

Another basic property, evident from eq. (\ref{cluster}), is that (i) the zeros
of $Z(G,q,v)$ in $q$ for real $v$ and hence also the continuous accumulation
set ${\cal B}_q$ are invariant under the complex conjugation $q \to q^*$; (ii)
the zeros of $Z(G,q,v)$ in $v$ or equivalently $a$ for real $q$ and hence also
the continuous accumulation set ${\cal B}_a$ are invariant under the complex
conjugation $a \to a^*$.

\subsection{Noncommutativity in the Random Cluster Model} 

Just as we showed the 
importance of noncommutative limits in our earlier work on chromatic
polynomials (eq. (1.9) in Ref. \cite{w}), so also we encounter an analogous 
noncommutativity here for the general partition function (\ref{cluster}) of 
the random cluster model: at certain special points $q_s$ (typically 
$q_s=0,1...,\chi(G)$) one has 
\beq
\lim_{n \to \infty} \lim_{q \to q_s} Z(G,q,v)^{1/n} \ne
\lim_{q \to q_s} \lim_{n \to \infty} Z(G,q,v)^{1/n} \ . 
\label{fnoncomm}
\eeq
Clearly, no such issue of noncommutativity arises if one restricts to 
positive integer $q$ values and uses the Potts model definition (\ref{zfun}),
(\ref{ham}).  However, for the general random cluster model, whenever 
$Z(G,q,v)$ has a factor $(q-q_s)^{\mu_s}$ with finite multiplicity 
$\mu_s$ (where typically $\mu_s=1$ here), one encounters this 
noncommutativity.  It can also occur when 
such a factor appears as a coefficient of one of the terms 
$(\lambda_{G,j})^m$ contributing to $Z(G,q,v)$ (see eq. (\ref{zgsum}) below).
The reason for this noncommutativity is analogous to that which we discussed 
earlier in the special case ($a=0$) of the chromatic polynomial \cite{w}; it 
is a consequence of the basic result
\beq
\lim_{n \to \infty} (q-q_s)^{\mu_s/n} = \cases{1 & if $q \ne q_s$ \cr
                                     0 & if $q=q_s$ \cr} \ . 
\label{basiclim}
\eeq
We shall illustrate this with our exact results to be discussed below. 
Because of
this noncommutativity, the formal definition (\ref{ef}) is, in general,
insufficient to define the free energy $f$ at these special points $q_s$; it is
necessary to specify the order of the limits that one uses in eq. 
(\ref{fnoncomm}).  We denote the two
definitions using different orders of limits as $f_{qn}$ and $f_{nq}$:
\beq
f_{nq}(\{G\},q,v) = \lim_{n \to \infty} \lim_{q \to q_s} n^{-1} \ln Z(G,q,v)
\label{fnq}
\eeq
\beq
f_{qn}(\{G\},q,v) = \lim_{q \to q_s} \lim_{n \to \infty} n^{-1} \ln Z(G,q,v) \
. 
\label{fqn}
\eeq
For the zero-temperature Potts/random cluster antiferromagnet case $a=0$ 
($v=-1$), the same ordering ambiguity affects the formal equation (\ref{w}). 
In Ref. \cite{w} and our subsequent works on chromatic polynomials and the
above-mentioned zero-temperature antiferromagnetic limit, it was convenient 
to use 
the ordering $W(\{G\},q_s) = \lim_{q \to q_s} \lim_{n \to \infty} P(G,q)^{1/n}$
since this avoids certain discontinuities in $W$ that would be present with the
opposite order of limits.   In the present work on the full
temperature-dependent random cluster model partition function, we shall 
consider both orders of limits and comment on the differences where 
appropriate.  Of course in discussions of the usual $q$-state Potts model (with
positive integer $q$), one automatically uses the definition in eq. 
(\ref{zfun}) with (\ref{ham}) and no issue of orders of limits arises, as it
does in the random cluster model with real $q$. 

As a consequence of the noncommutativity (\ref{fnoncomm}), it follows that for
the special set of points $q=q_s$ one must distinguish between (i) $({\cal
B}_a(\{G\},q_s))_{nq}$, the continuous accumulation set of the zeros of
$Z(G,q,v)$ obtained by first setting $q=q_s$ and then taking $n \to \infty$,
and (ii) $({\cal B}_a(\{G\},q_s))_{qn}$, the continuous accumulation set of the
zeros of $Z(G,q,v)$ obtained by first taking $n \to \infty$, and then taking $q
\to q_s$.  For these special points, 
\beq 
({\cal B}_a(\{G\},q_s))_{nq} \ne ({\cal B}_a(\{G\},q_s))_{qn} \ . 
\label{bnoncomm}
\eeq

 From eq. (\ref{zq0}), it follows that for any $G$, 
\beq
\exp(f_{nq})=0 \quad {\rm for} \quad q=0
\label{fnqq0}
\eeq
and thus 
\beq
({\cal B}_a)_{nq} = \emptyset \quad {\rm for} \quad q=0 \ . 
\label{bnq0}
\eeq
However, for many families of graphs, including the circuit graph $C_n$, and
cyclic and M\"obius strips $L_m$ and $ML_m$, if we take $n \to \infty$ first
and then $q \to 0$, we find that $({\cal B}_a)_{qn}$ is nontrivial. 
For these families of graphs, with this order 
of limits, although the free energy is nonanalytic at $q=0$, it is 
continuous, and we find that, in general, $\exp(f_{qn}) \ne 0$ at $q=0$. 
Similarly, from (\ref{zq1}) we have, for any $G$, 
\beq
({\cal B}_a)_{nq} = \emptyset \quad {\rm for} \quad q=1
\label{bnq1}
\eeq
since all of the zeros of $Z$ occur at the single discrete point $a=0$ (and in
the case of a graph $G$ with no edges, $Z=1$ with no zeros).  However, as the
simple case of the circuit graph below will show, $({\cal B}_a)_{qn}$ is, in
general, nontrivial. 

We shall also find that $({\cal B}_a)_{nq} \ne ({\cal B}_a)_{qn}$ for $q=2$ for
the infinite-length, $L_y=2$ width strip graphs $\{L\}$ and $\{ML\}$.  
As stated, this noncommutativity can, in general, occur at integer values of
$q$ up to and including $q=\chi(G)$.  However, although $\chi=3$ for $C_m$ and
$L_m$ with odd $m$ and for $ML_m$ with even $m$, there is no noncommutativity
at $q=3$ in these cases.  This can be seen as a consequence of the fact that 
one can take the limit $m \to \infty$ on even values of $m$.

In the $q=2$ Ising case, as a consequence of the relation (\ref{zbipq2}), the
locus $({\cal B}_a)_{nq}$ is invariant under the inversion map $a \to 1/a$ for
the $n \to \infty$ limit of a sequence of bipartite graphs of type $G$:
\beq
({\cal B}_a)_{nq}(\{G_{bip.}\}) \quad {\rm is \ invariant \ under} \quad a 
\to \frac{1}{a} \quad {\rm if} \quad q=2 
\label{bq2sym}
\eeq
where $\{G_{bip.}\}$ means that for the family of graphs of type $G$, one can
take the limit $n \to \infty$ with a sequence of bipartite members of the
family $G$.  (For example, for the circuit graphs $C_n$, one can do this by
taking $n \to \infty$ on even values, and so forth for other families.)  
As our explicit results for the strips $\{L\}$ and $\{ML\}$ below will show, 
the locus obtained with the opposite order of limits, 
$({\cal B}_a)_{nq}(\{G_{bip.}\})$, does not, in general, have this inversion 
symmetry, even if $q=2$. 

Concerning the cases where the continuous locus ${\cal B}$ may be the nullset
$\emptyset$, some examples from chromatic polynomials may be useful.  For the
complete graph on $n$ vertices $K_n$ (defined as the graph in which each point
is connected to every other point with edges), if $a=0$, then
$Z(K_n,q,v=-1)=P(K_n,q)=\prod_{s=0}^{n-1}(q-s)$, so that as $n \to \infty$,
there is no continous accumulation set of the chromatic zeros, so ${\cal
B}_q=\emptyset$ .  Another example is provided by the strip $S_m$ (see 
eq. (\ref{qbp}) below).

\subsection{Definition of Free Energy for Complex $q$ and $K$}

Another matter concerns the definition of $f$ away from physical values of
$q$ and $K$, where $Z(G,q,a)$ can be negative or complex.  In these ranges of 
$q$ and $K$, there is no canonical choice of which $1/n$'th root, i.e., which
value of $r$, to pick in eq. (\ref{fnq}) or (\ref{fqn}): 
\beq
Z(G,q,a)^{1/n} = \{ |Z(G,q,a)|^{1/n}e^{(\phi+2\pi i r)/n} \} \ , 
\quad r=0, \ 1,...,n-1 \ . 
\label{zphase}
\eeq 
where $\phi=\arg(Z)$.  Thus, we start by considering the free energy $f$
defined for sufficiently large (physical) $T$ and integer $q \ge 2$ and define
the maximal region in the $q$ plane for fixed $a$ or in the $a$ plane for fixed
$q$ that can be reached by analytic continuation of this function.  As noted,
this is labelled the region $R_1$ in the $q$ plane and the PM phase (and its
complex-temperature extension) in the $u$ plane. In these regions, the
canonical phase choice in (\ref{zphase}) is clearly that given by $r=0$.  This
would also be true in physical low-temperature broken-symmetry phases such as
ferromagnetic (FM) or antiferromagnetic (AFM) phases and their
complex-temperature extensions, as discussed in \cite{chisq}; however, such
phases do not occur in the 1D and quasi-1D strip graphs considered here.
However, in general, in complex-temperature O phases in the $u$ plane and $R_j$
regions with $j \ne 1$ in the $q$ plane, only the quantity $|e^f|=\lim_{n \to
\infty}|Z(G,q,a)|^{1/n}$ can be determined unambiguously.

\subsection{General Form of $Z$ for Recursively Defined Graphs}

We find that a general form for the Potts model partition function for the
strip graphs considered here, or more generally, for recursively defined
families of graphs comprised of $m$ repeated subunits (e.g. the columns of
squares of height $L_y$ vertices that are repeated $L_x$ times to form an $L_x
\times L_y$ strip of a regular lattice with some specified boundary
conditions), is 
\beq 
Z(G,q,v) =
\sum_{j=1}^{N_\lambda} c_{G,j} (\lambda_{G,j}(q,v))^m
\label{zgsum}
\eeq
where $N_\lambda$ depends on $G$.  The formula
(\ref{zgsum}) can be understood from the fact that for the cyclic case, for 
$q \in {\mathbb Z}_+$, $Z(G,q,v)$ can be expressed as the trace of a transfer 
matrix ${\cal T}$: 
\beq
Z(G,q,v) = Tr({\cal T}^m) \ . 
\label{ztrace}
\eeq
Having obtained $Z(G,q,v)$ in this manner, one can then generalize $q$ from 
${\mathbb Z}_+$ to ${\mathbb R}_+$. 
For a strip of a regular lattice, this transfer matrix has dimensions ${\cal N}
\times {\cal N}$, where ${\cal N}$ denotes the number of possible spin
configurations along a transverse slice of the strip; for example, for the
square strips of interest here (or for triangular strips, written in the form
of a square strip with additional diagonal edges), ${\cal N}=q^{L_y}$.  
Clearly, ${\cal T}$ is a symmetric matrix each
of whose elements is either 1 or a (positive) power of $a=v+1$.  For 
physical temperature, for which $a \ge 0$, ${\cal T}$ is real and hence can 
be diagonalized by an orthogonal transformation $O$, yielding ${\cal N}$ 
eigenvalues (some of which may coincide).  Generically, the multiplicity of a
given eigenvalue $\lambda_{G,j}$, which yields the coefficient $c_j$ for these
cyclic graphs, depends on $q$ but is independent of $v$.  The result
(\ref{zgsum}) applies for both free and periodic transverse boundary
conditions, given that one uses periodic (cyclic) longitudinal boundary
conditions.  Similar arguments based on the transfer matrix yield this
result if one uses free transverse and M\"obius longitudinal, and periodic
transverse and M\"obius longitudinal (i.e. Klein bottle) boundary conditions. 

For strips with open longitudinal
boundary conditions, $Z(G,q,v)$ is not a trace, but instead, is given by 
\beq
Z(G,q,v) = \sum_{{\tilde \sigma}_i,{\tilde \sigma}_f} 
\langle \tilde \sigma_i | {\cal T}^{L_x} | \tilde \sigma_f \rangle 
\label{zopen}
\eeq 
where ${\tilde \sigma}_i$ and ${\tilde \sigma}_f$ denote the states of the
$L_y$ spins on the initial (i) and final (f) transverse edges of the strip. As
our explicit solution (given below) for the open ladder graph $S_m$ shows, 
here $c_j$ can depend on both $q$ and $v$.

For recursively defined families of graphs $G$, the result (\ref{zgsum}) is a 
generalization to the case of the Potts/random cluster model partition 
function $Z(G,q,v)$ of the Beraha-Kahane-Weiss result \cite{bkw} that the 
chromatic polynomial $P(G,q)=Z(G,q,v=-1)$ can be written in the form 
\beq
P(G,q) = \sum_{j=1}^{N_{\lambda,P}} c_{P,G,j} (\lambda_{P,G,j})^m \ . 
\label{pgsum}
\eeq
Since $P(G,q)$ is a special case of $Z(G,q)$, it follows that 
\beq
N_{\lambda,P} \le N_{\lambda} \ . 
\label{nlamrel}
\eeq
For example, in two well-known cases, (i) tree graph: 
$N_{\lambda}=N_{\lambda,P}=1$, (ii) circuit graph: 
$N_{\lambda}=N_{\lambda,P}=2$. We find (iii) for the open $L_y=2$ ladder 
$S_m$, $N_{\lambda}=2$, while $N_{\lambda,P}=1; (iv) for the $cyclic and 
M\"obius strips, $N_{\lambda}=6$ while \cite{bds} $N_{\lambda,P}=4$. 
For the subset of the $\lambda_{G,j}$'s that remain for $P(G,q)$, the
coefficient functions $c_{P,G,j}=c_{G,j}(v=-1)$; as remarked above, for the
cyclic case and our M\"obius strip, $c_{G,j}$ are independent of $v$. 

As $m \to \infty$, for a given point $(q,v)$ in the ${\mathbb C}^2$ space of 
variables, one $\lambda_j$ will dominate this sum; we denote this as the
``leading term'' $\lambda_\ell$, where $\ell$ stands for leading.  As one 
moves to another point $(q^\prime,v^\prime)$, it may happen that there is a 
change in the dominant $\lambda$, from $\lambda_\ell$ to, say, 
$\lambda_\ell^\prime$. 
Consequently, there is a nonanalytic change in the free energy $f$ as it
switches from being determined by the first dominant $\lambda_\ell$ to 
being determined by $\lambda_\ell^\prime$.  Thus, the equation for the 
continuous nonanalytic locus ${\cal B}$ across which $f$ is nonanalytic, is 
(with the $\{G\}$ dependence indicated explicitly)
\beq
{\cal B}(\{G\}): \quad |\lambda_{G,\ell}|=|\lambda_{G,\ell}^\prime| \ . 
\label{degeneq}
\eeq  
Although $f$ is nonanalytic across ${\cal B}$, a consequence of
eq. (\ref{degeneq}) is that $|e^f|$ is continuous across this locus. 
This is the
generalization of the analogous phenomenon for the asymptotic limit of 
chromatic polynomials, or equivalently, the $T=0$ limit of the Potts
antiferromagnet \cite{bkw,readcarib,read91,w}, i.e. the property that 
$W(\{G\},q)$ is nonanalytic (although its magnitude is continuous) across 
${\cal B}_q(\{G\})$.  As noted above, the locus
${\cal B}$ forms as the continuous accumulation set of the zeros of $Z(G,q,v)$
in the 2-complex dimensional space $(q,v)$ as $n(G) \to \infty$.  This again
generalizes the earlier analysis for chromatic polynomials, where ${\cal B}_q$
formed as the continuous accumulation set of the zeros of the chromatic
polynomial in the single complex variable $q$.

It is straightforward to generalize the transfer matrix formalism and hence 
eqs. (\ref{ztrace}) and (\ref{zopen}) to the case of the Potts model in an
external magnetic field $H$, where the Hamiltonian is 
${\cal H} = -J\sum_{\langle i j \rangle } \delta_{\sigma_i \sigma_j} - 
H\sum_i \delta_{\sigma_i,\sigma_0}$ (with $\sigma_0$ chosen, say, as 1).
Hence, the full generalization for recursively defined families of graphs is,
with $\eta=e^{\beta H}$, 
\beq
Z(G,q,v,\eta) =\sum_{j=1}^{N_\lambda} c_{G,j} (\lambda_{G,j}(q,v,\eta))^m \ . 
\label{zghsum}
\eeq
One could proceed to study the singular locus ${\cal B}$ in the 
${\mathbb C}^3$ space of the variables $(q,v,\eta)$ which forms as the
continuous accumulation set of the zeros of $Z$, and the various slices in the
$q$, $a$, and $\eta$ planes.  Here we restrict ourselves to the zero-field 
case, $\eta=1$. 

Typically, for a given point $a \in {\mathbb C}$, there will be an infinite set
of points in the $q$ plane lying on ${\cal B}$, and for a given point $q \in
{\mathbb C}$, there will be an infinite set of points in the $a$ plane lying on
${\cal B}$.  Again, usually (we have commented on some exceptions above, where
accumulation sets of zeros are discrete), we find that ${\cal B}_q$ and ${\cal
B}_a$ form curves (and possible line segments) in the respective $q$ and $x=a$
or $u$ planes.  This follows from the property that ${\cal B}$ is the solution
of an algebraic equation expressing the degeneracy of the leading $\lambda$'s
contributing to $Z$.  (For higher dimensional lattices the equations defining 
${\cal B}$ can be transcendental instead of algebraic.) 
The loci ${\cal B}_q$ and ${\cal B}_x$, $x=a$ or $u$, may
be connected, or may consist of several disconnected components; our results
illustrate both types of behavior. 

The Potts ferromagnet has a zero-temperature phase transition in the $L_x \to
\infty$ limit of the strip graphs considered here, and this has the consequence
that for cyclic and M\"obius boundary conditions, ${\cal B}$ passes through the
$T=0$ point $u=0$.  It follows that ${\cal B}$ is noncompact in the $a$ plane.
Hence, it is usually more convenient to study the slice of ${\cal B}$ in the
$u=1/a$ plane rather than the $a$ plane.  In the Ising case $q=2$, ${\cal
B}_a={\cal B}_u$ and so both are noncompact.  For the ferromagnet, since $a \to
\infty$ as $T \to 0$ and $Z$ diverges like $a^{e(G)}$ in this limit, we shall
use the reduced partition function $Z_r$ defined by
\beq
Z_r(G,q,v)=a^{-e(G)}Z(G,q,v)=u^{e(G)}Z(G,q,v)
\label{zr}
\eeq
which has the finite limit $Z_r \to 1$ as $T \to 0$.  For a general strip 
graph $(G_s)_m$ of type $G_s$ and length $L_x=m$, we can write 
\beqs
Z_r((G_s)_m,q,a) & = & u^{e((G_s)_m)}\sum_{j=1}^{N_\lambda} c_{G_s,j} 
(\lambda_{G_s,j})^m \equiv \sum_{j=1}^{N_\lambda} c_{G_s,j}
(\lambda_{G_s,j,u})^m
\label{zu}
\eeqs
with 
\beq
\lambda_{G_s,j,u}=u^{e((G_s)_m)/m}\lambda_{G_s,j} \ . 
\label{lamu}
\eeq
For example, for the strips of the square 
lattice with periodic longitudinal boundary conditions and free transverse
boundary conditions, and of width $L_y$ vertices, we have
$e(sq(L_y)_m)=(2L_y-1)m$, so the prefactor in (\ref{lamu}) is $u^{2L_y-1}$.

\section{1D Case with Free Boundary Conditions}

We first briefly discuss two cases that illustrate
some important features in their simplest contexts.  We begin with Potts/random
cluster model on a line of $n$ vertices, or, more generally, a tree graph,
$T_n$. One has the well-known result
\beq
Z(T_n,q,v) = q(q+v)^{n-1} . 
\label{ztn}
\eeq
This case 
illustrates the general feature that the antiferromagnetic random cluster
model for real positive non-integral $q$ fails to satisfy the usual statistical
mechanical requirement that the partition function is positive, and hence does
not, in general admit a Gibbs measure \cite{ssbounds}.  Specifically, here we
have
\beq 
Z(T_n,q,v) < 0 \quad {\rm for} \ \ n \ \ {\rm even} \ \ {\rm and} \ \ 
q+v < 0 \ . 
\label{ztreeneg}
\eeq
These negative values of $Z(T_n,q,v)$ occur at physical finite temperature
if $0 < q < 1$ and $n$ is even, since the condition $q+a-1<0$ is equivalent to 
$T<T_{un}$, where 
\beq
T_{un}=\frac{J}{k_B\ln(1-q)}=\frac{|J|}{k_B\ln \Bigl ( \frac{1}{1-q} \Bigr )} 
\ , \quad 0 < q < 1 \ , \quad J < 0.
\label{tun}
\eeq
Although in this case one could restore positivity by requiring that $n$ be 
odd, we shall show that there are further pathologies associated with this 
temperature. 

The continuous locus ${\cal B}=\emptyset$ since the accumulation set of the
zeros of $Z$ in $q$ is the discrete point $q=-v$.  For $q \ne 0$, the 
limits in the 
definitions (\ref{fqn}) and (\ref{fnq}) commute, and the free energy is (with
$v=a-1$) 
\beq
f_{qn}=f_{nq} \equiv f = \ln(q+a-1) \ . 
\label{ft}
\eeq
The physical thermodynamic behavior for this case will be compared below with
that for the width $L_y=2$ strips.  For this purpose, we record the internal 
energy, 
\beq
U = -\frac{Ja}{q+a-1} 
\label{ut}
\eeq
and the specific heat, 
\beq
C=\frac{k_BK^2(q-1)a}{(q+a-1)^2} \ . 
\label{ct}
\eeq 
Note that the specific heat (\ref{ct}) is positive if $q>1$ but is negative 
and hence
unphysical for all temperatures if $q < 1$, in both the ferromagnetic and
antiferromagnetic cases.  Thus the pathological nature of the range $0 < q < 1$
is manifested in the negative specific heat even for temperatures above the
value $T_{un}$ in eq. (\ref{ztreeneg}) below which $Z$ can be negative.  Also,
note that in the antiferromagnetic case there are unphysical divergences of $U$
and $C$ at $T=T_{un}$.  

For $q=0$, the noncommutativity (\ref{fnoncomm}) occurs, and one has
$\exp(f_{nq})=0$ but $\exp(f_{qn}) = (a-1)$.  This simple case demonstrates
that the noncommutativity at a special point $q_s$ can occur even when this
point is not the the singular locus $({\cal B})_{qn}$ or $({\cal B})_{nq}$.
In passing we note that a study of the $q$-state Potts
model on the Bethe lattice (tree graph in which the interior vertices all have
the same coordination number) has been carried out in \cite{ww}.

\section{1D Case with Periodic Boundary Conditions}

\subsection{General}

The Potts/random cluster model on the circuit graph $C_n$, or equivalently, the
1D line with periodic boundary conditions, is probably the simplest case with a
nontrivial locus ${\cal B}$.  The Tutte polynomial for this graph is well known
\cite{bbook}, and the corresponding Potts model partition function is 
\beq 
Z(C_n,q,a) = (q+v)^n + (q-1)v^n \ .
\label{zcn}
\eeq 
As noted above, the Potts ferromagnet has a zero-temperature critical point,
and this is also true of the antiferromagnet in the $q=2$ case where these are
equivalent.  In the antiferromagnetic case, there is nonzero ground 
state entropy, $S=k_B\ln(q-1)$ if $q > 2$.  

For comparison with the $L_y=2$ results to be given below, we recall some of
the thermodynamic properties of the 1D Potts model.  Here we take $q \ge 2$
where there is no pathological behavior (see below) and restrict to physical
values of $J$ and $T$; the resulting thermodynamic functions are then
independent of whether one uses periodic or free boundary conditions and were
given above for $f$, $U$, and $C$ in eqs. (\ref{ft})-(\ref{ct}).  The internal
energy and specific heat have the high-temperature expansions 
\beq
U=-\frac{J}{q}\biggl [ 1+\frac{(q-1)}{q}K + O(K^2) \biggr ]
\label{uly1ht}
\eeq
\beq
C=\frac{k_B(q-1)K^2}{q^2}\biggl [ 1 + \frac{(q-2)}{q}K + O(K^2) \biggr ]
\label{chigh}
\eeq
Recall that the $T \to 0$ limit corresponds to $K \to \infty$, i.e. $u \to 0$,
for the ferromagnet ($J>0$) and to $K \to -\infty$, i.e., $a \to 0$, for the
antiferromagnet ($J<0$).  The low-temperature expansions for these two 
cases are different:
\beq
U = -J\biggl [ 1 - (q-1)e^{-K} + O(e^{-2K}) \biggr ]  \quad {\rm as} \quad 
K \to \infty 
\label{ulowfm}
\eeq
\beq
U = \frac{(-J)e^K}{(q-1)}\biggl [ 1 - \frac{1}{q-1}e^K + O(e^{2K}) 
\biggr ] \quad {\rm as} \quad K \to -\infty 
\label{ulowafm}
\eeq
\beq
C = k_B(q-1)K^2 e^{-K} \biggl [ 1 -2(q-1)e^{-K} + O(e^{-2K}) \biggr ]
\quad {\rm as} \quad K \to \infty
\label{clowfm}
\eeq
\beq
C =\frac{k_BK^2e^K}{(q-1)}\biggl [ 1 - \frac{2}{q-1}e^K + 
O(e^{2K}) \biggr ] \quad {\rm as} \quad K \to -\infty \ . 
\label{clowafm}
\eeq
Note that in the Ising case $q=2$, these expansions satisfy the symmetry 
relations (\ref{ising_urel}) and (\ref{ising_crel}).

\vspace{8mm}

\begin{figure}
\vspace{-4cm}
\centering
\leavevmode
\epsfxsize=4.0in
\begin{center}
\leavevmode
\epsffile{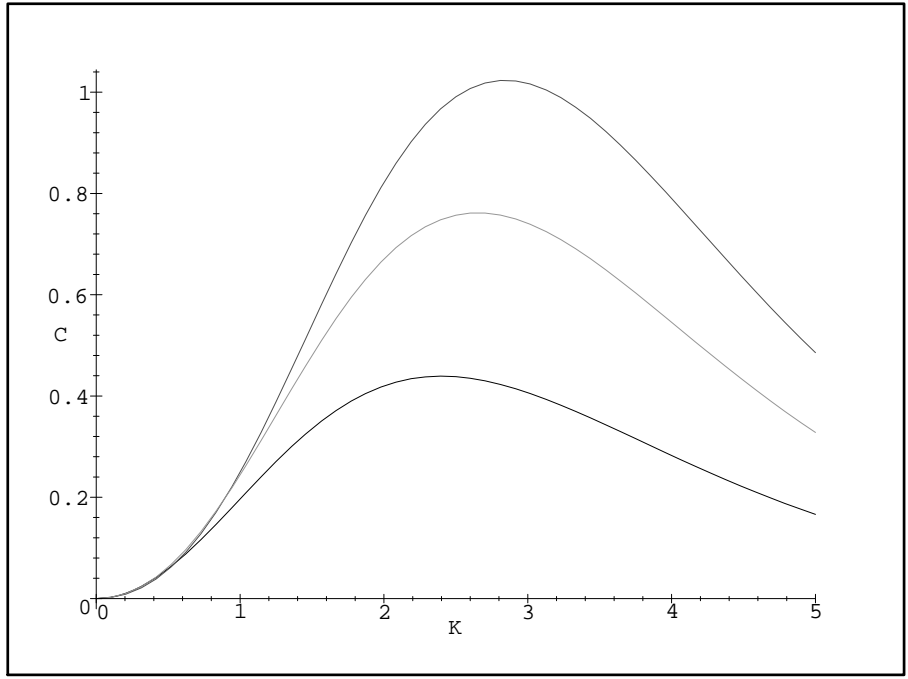}
\end{center}
\vspace{-2cm}
\caption{\footnotesize{Specific heat for the 1D Potts ferromagnet as a function
of $K=J/(k_BT)$. Going from bottom to top in order of the heights of the 
maxima, the curves are for $q=2,3,4$.}}
\label{cfmline}
\end{figure}

\vspace{8mm}

\begin{figure}
\vspace{-4cm}
\centering
\leavevmode
\epsfxsize=4.0in
\begin{center}
\leavevmode
\epsffile{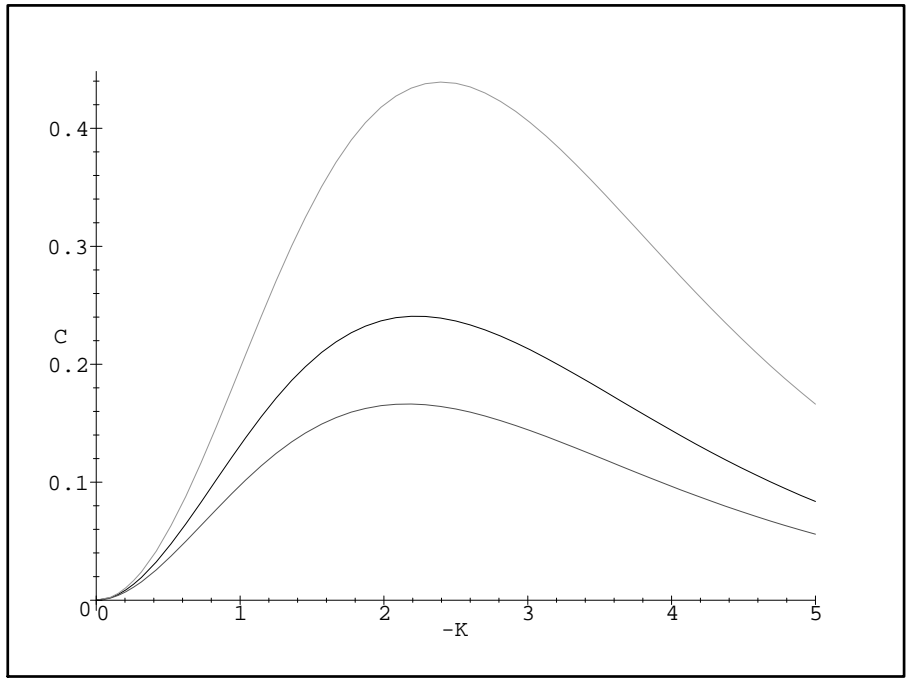}
\end{center}
\vspace{-2cm}
\caption{\footnotesize{Specific heat for the 1D Potts antiferromagnet as a
function of $-K = -J/(k_BT)$.  Going from top to bottom, the curves are for 
$q=2,3,4$.}}
\label{cafmline}
\end{figure}

\vspace{8mm}

  In Figs. \ref{cfmline}
and \ref{cafmline} we plot $C$ for the ($n \to \infty$ limit of the) 
ferromagnetic (F) and antiferromagnetic (AF) cases (with $k_B=1$). 
In the antiferromagnetic case, $C$ is a decreasing function of $q$ for all
$0 < T < \infty$.  In the ferromagnetic case, $C$ increases (decreases) with 
$q$ at low (high) temperatures and the curves for two different values $q=q_1$
and $q=q_2$ are equal at the temperature $T_{cr}=J/(k_BK_{cr})$, where 
\beq
K_{cr}=\frac{1}{2}\ln \Bigl [ (q_1-1)(q_2-1) \Bigr ] \ . 
\label{acr}
\eeq 
Thus, $K_{cr} \simeq 0.35, \ 0.55, \ 0.90$ for $q=2,3,4$. 
The specific heat has a maximum at a temperature $K_{cmax}$ 
given by the solution of the equation $(q-1)(K+2)+(2-K)e^{K}=0$.  
Some illustrative values for Figs. \ref{cfmline} and \ref{cafmline} are
(i) $K_{cmax} \simeq \pm 2.4$ (F,AF) for $q=2$; (ii) $K_{cmax} \simeq 2.7$ (F),
$K_{cmax} \simeq -2.2$ (AF) for $q=3$; (iii) $K_{cmax} \simeq 2.85$ (F), 
$K_{cmax} \simeq -2.15$ (AF) for $q=4$. For $q \ge 3$,
the value of $C$ at this maximum increases (decreases) with $q$ for the
ferromagnet (antiferromagnet).

\subsection{${\cal B}_q(\{C\})$ for fixed $a$}

Returning to the study of ${\cal B}_q$ in the complex $q$ plane as a function
of $a$, we note that the solution of the degeneracy equation $|q+a-1|=|a-1|$ 
determines the locus ${\cal B}_q(\{C\})$ to be the circle centered at $q=1-a$ 
of radius $|1-a|$: 
\beq
q=(1-a)(1+e^{i\theta}) \ , 0 \le \theta < 2\pi \ . 
\label{qcircb}
\eeq
For $a < 1$ and $a > 1$, this circle has support in the right-hand and
left-hand half planes $Re(q) \ge 0$ and $Re(q) \le 0$, respectively; for any
$a$, it always passes through the origin.  The locus ${\cal B}_q$ intersects 
the real $q$ axis at $q=0$ and at $q=q_c(\{C\})$, where
\beq
q_c(\{C\})=2(1-a) \ . 
\label{qcc}
\eeq 
In general, for finite $n$, the zeros of $Z$ do not lie exactly on ${\cal B}$.
We have shown, however, that in the $T=0$ limit of the Potts antiferromagnet,
i.e., for $a=0$, the zeros of $Z(C_n,q,a=0)=P(C_n,q)$ do lie exactly on the
locus ${\cal B}$, which, for this case is the circle $|q-1|=1$ \cite{w,wc}. As
$T$ increases from 0 to infinity for the Potts antiferromagnet, i.e. as $a$
increases from 0 to 1, the radius and center of the circle both decrease from 
1 to 0 so that it contracts to the origin at $a=1$.  As $a$ increases above 1
through real values, i.e. as $T$ decreases from $\infty$ to 0 for the
ferromagnetic case, the circle is located in the $Re(q) \le 0$ half-plane, with
its center moving leftward and its radius increasing as a function of $a$. In
this ferromagnetic case, the crossing point given by $q_c(\{C\})$ occurs on 
the negative real $q$ axis; ${\cal B}_q$ does not cross the positive real $q$
axis. 

\subsection{${\cal B}_u(\{C\})$ for fixed $q$}

We first consider values of $q \ne 0,1$, so that no noncommutativity occurs, 
and $({\cal B}_u)_{nq}=({\cal B}_u)_{qn} \equiv {\cal B}_u$. As discussed 
above, it is convenient to use the $u$ plane since ${\cal B}_u$ is compact in
this plane, except for the case $q=2$, whereas ${\cal B}_a$ is noncompact
because of the ferromagnetic zero-temperature critical point at $u=1/a=0$. 
For $q \ne 0,1,2$, ${\cal B}_u$ is the circle \cite{is1d} 
\beq
{\cal B}_u: \ \ u=\frac{1}{(q-2)}(-1+e^{i\omega}) \ , \quad 
0 \le \omega < 2\pi \ . 
\label{ucircle}
\eeq
The exterior of this circle is the (complex-temperature extension of the) PM 
phase, and its interior is an O phase.  

For the ferromagnet, the fact that the singular locus ${\cal B}_u$ passes
through the $T=0$ point $u=0$ for the 1D Potts model with periodic boundary
conditions, while for the same model with free boundary conditions, ${\cal
B}_u$ does not pass through $u=0$, means that the use of periodic boundary
conditions yields a singular locus that manifestly incorporates the
zero-temperature critical point, while this is not manifest in ${\cal B}_u$
when calculated using free boundary conditions.  As we shall show, this
continues to be true concerning the longitudinal boundary conditions when one
considers the Potts ferromagnet on the $L_y=2$ strip graphs.  This leads us to
one of the important conclusions of this work, namely that although
calculations of the free energy of the Potts model on infinite-length,
finite-width strips with periodic boundary conditions in the longitudinal
direction (the direction in which the strip length goes to infinity) are more
difficult than with free longitudinal boundary conditions, the extra work is
worth it since the resulting locus ${\cal B}$ incorporates this feature of
passing through $u=0$, corresponding to the zero-temperature critical point of
the ferromagnet, if one uses periodic longitudinal boundary conditions.  From
our studies in the different, although related, context of chromatic
polynomials \cite{pg,wcy,pm,tk,bcc}, we reached the analogous conclusion that
although the calculation of $P(G,q)$ for strip graphs of a regular lattice is
more complicated when one uses periodic longitudinal boundary conditions, the
resultant singular locus ${\cal B}_q$ has the advantage of incorporating more
of the features expected of the infinite-width limit, i.e. the full
two-dimensional lattice.  One such expectation, based on calculations of
chromatic polynomials and the resultant $W$ functions of eq. (\ref{w}) for
infinite-length, finite-width strips as the width increased, was that ${\cal
B}_q$ passes through $q=0$ and that this nonanalytic locus separates the region
including the interval $0 < q < q_c(\{G\})$ on the real axis from the outlying
region for sufficiently large $|q|$ \cite{strip,strip2}; this is also in
agreement with the calculation in \cite{baxter} for the triangular lattice.
However, for finite width strips, ${\cal B}_q$ consists of arcs \cite{strip}
(and possible closed regions, as in Fig. 4 of \cite{strip2}) which do not pass
through $q=0$ and do not have this enclosure property.

The circle (\ref{ucircle}) crosses the real axis at $u=0$ and at 
\beq
u_c(\{C\}) = \frac{1}{a_c(\{C\})} = -\frac{2}{q-2} 
\label{ucc}
\eeq
(cf. eq. (\ref{qcc})). 
The point $u_c(\{C\})$ occurs at complex temperature for $q > 2$ and physical 
temperature for $q < 2$. We shall comment further below on the case 
$0 < q < 2$.  In the $a$ plane, ${\cal B}_a$ is the vertical line 
\beq
{\cal B}_a: \ \ Re(a)=a_c = 1 - \frac{q}{2}, \quad -\infty \le Im(a) \le \infty
\ . 
\label{baline}
\eeq
The phase with $Re(a) > a_c$, to the right of this line, is the 
(complex-temperature extension of the) PM phase, while the phase to the left of
the line is the O phase. 
As $q \to 2$, the radius of the circle (\ref{ucircle}) goes to infinity, and at
$q=2$, ${\cal B}_u$ is identical to ${\cal B}_a$ by the symmetry relation
(\ref{bq2sym}), forming the full imaginary $u$ axis.  

We next consider the special values $q=0$ and 1 for which noncommutativity
occurs.  For $q=0$, $e^{f_{nq}}=0$ as in (\ref{fnqq0}) while in the PM phase
with $Re(a) > 1$, $e^{f_{qn}}=a-1$ and in the O phase with $Re(a) < 1$,
$|e^{f_{qn}}|=|a-1|$; the locus $({\cal B}_a)_{nq}=\emptyset$, while $({\cal
B}_a)_{qn}$ is given by (\ref{baline}) as the vertical line $Re(a)=1$.  For
$q=1$, the coefficient multiplying the second term in $Z(C_n,q,v)$ vanishes,
and $Z(G,q=1,v)=a^n$, a special case of (\ref{zq1}).  Here $e^{f_{nq}}=a$ while
in the PM phase defined by $Re(a) > a_c=1/2$, we have $e^{f_{qn}}=a$ and in the
O phase defined by $Re(a) < 1/2$, we have $|e^{f_{qn}}|=|a-1|$; $({\cal
B}_a)_{nq}=\emptyset$ since all of the zeros of $Z$ occur at the discrete point
$a=0$, while $({\cal B}_a)_{qn}$ is given by eq. (\ref{baline}) as the vertical
line $Re(a)=1/2$.

\subsection{Phase Transition for Antiferromagnetic Case with $0 < q < 2$}

For the range $0 < q < 2$, and the antiferromagnetic case $J < 0$, the 
nonanalyticity in the free energy at $a=a_c=(2-q)/2$ in (\ref{ucc}) occurs at 
the physical temperature 
\beq
T_p =  \frac{|J|}{k_B\ln \Big ( \frac{2}{2-q} \Bigr )}  \ , \quad 0 < q < 2 \ .
\label{tp}
\eeq
Therefore, the generalization of the Potts antiferromagnet to
real positive $q$ defined by the random cluster representation (\ref{cluster})
has a finite-temperature phase transition in the $n \to \infty$ limit of the
circuit graph, i.e. in 1D with periodic boundary conditions.  (For the special
value $q=q_s=1$, it is understood that one takes $n \to \infty$ first and then
$q \to 1$, i.e., one uses $f_{qn}$; with the other order, $q \to 1$ and then $n
\to \infty$, $f_{nq}$ is analytic and there is no phase transition.) 
The phase transition at $T=T_p$ is not a
counterexample to the usual theorem that a one-dimensional spin model with
short-ranged interactions does not have any phase transition at finite
temperature, because the existence of this transition is inextricably connected
with the failure of positivity for $Z$ and hence the absence of a Gibbs
measure, which are implicit requirements for the applicability of the
above-mentioned theorem. As $q$ decreases from 2 to 0, the phase transition 
temperature $T_p$ increases from 0 to infinity.  

In the high-temperature paramagnetic phase $T > T_p$, the free 
energy, internal energy, and specific heat are given by the same expressions as
for the $n \to \infty$ limit of the Potts/random cluster model on the tree
graph, eqs. (\ref{ft}), (\ref{ut}), and (\ref{ct}), respectively.  Hence, even
in the high-temperature phase, one has an unphysical negative specific heat if
$q < 1$. In the low-temperature O phase, strictly speaking, only $|Z|$ can be 
determined: $|\exp(f_{qn})|=|a-1|$, but with an appropriate choice of
multiplicative phase, we can choose 
\beq
f=\ln(1-a) \ , \quad T < T_p
\label{fqncno}
\eeq
and hence 
\beq
U=\frac{Ja}{1-a} \ , \quad T < T_p 
\label{uclow}
\eeq
\beq
C=-\frac{k_BK^2a}{(1-a)^2} \ , \quad T < T_p \ . 
\label{cclow}
\eeq
Thus, for all $q$ in the range $0 < q < 2$ where there
is a finite-temperature phase transition, the low-temperature phase has
pathological property that the specific heat is negative.  The phase transition
itself is first-order, with latent heat
\beq
  \lim_{T \to T_p^+}U - \lim_{T \to T_p^-}U = \frac{2|J|(2-q)}{q} \ . 
\label{latentc}
\eeq
A basic pathology of the low-temperature phase of this antiferromagnet, i.e.,
the phase where $|v|>|q+v|$, is that $Z$ can be negative.  For sufficiently
large $n$, this occurs for $1<q<2$ if $n$ is odd and for $0<q<1$ if $n$ 
is even.  

Thus, if one restricts to $q > 1$, this 1D antiferromagnetic random cluster
model satisfies, at least in the high-temperature phase, the requirement that
the specific heat is positive and, for the interval $1 < q < 2$, has a
(first-order) finite-temperature phase transition; however, even if one
restricts the approach to the $n \to \infty$ limit to even values of $n$, the
low-temperature phase is unphysical because of the negative specific heat. One
also observes that the results for the free energy and associated thermodynamic
functions are the same for the $n \to \infty$ limit of the tree graph and the
circuit graph, i.e. are independent of whether one uses free or periodic
boundary conditions, if $T > T_p$, but differ for $T < T_p$, so that the
existence of the low-temperature phase in the case of periodic boundary
conditions also means that the $n \to \infty$ limit does not exist owing to
different results obtained with different boundary conditions.  The
non-existence of a well-defined $n \to \infty$ limit for the random cluster
model with non-integral $q$ has been noted previously in \cite{ssbounds}.  For
positive integer $q$, the (zero-field) $q$-state Potts model is invariant under
the operations of the permutation group $S_q$; however, this symmetry group is
not defined for non-integral $q$.  In any case, the usual Peierls argument
shows that even if one could define some notion of a symmetry of $Z$ for
non-integral $q$, this symmetry could not be broken spontaneously in the phase
transition for this 1D system or, indeed, for the random cluster model on an
infinite-length, finite-width strip, to be discussed below.

A further generalization of this 1D random cluster model is to keep $q$ real
but let it be negative; in this case the model with the ferromagnetic sign of
the coupling, $J>0$, formally has a nonanalyticity in the free energy at a
positive finite value of the parameter $T$ given by $k_BT=J/\ln [(2-q)/2]$.
However, since this model does not, in general, have a positive $Z$, one 
cannot really refer to this parameter as a physical temperature and we shall
not discuss this case further.

\subsection{Other slices of ${\cal B}(\{C\})$}

So far we have considered the ``orthogonal'' slices of ${\cal B}$ obtained by
holding either $q$ or $q$ constant.  A different type of slice is obtained if
one has $q$ and $a$ satisfy some functional relation.  As an illustration of
this, we consider perhaps the simplest case, namely the linear relation $a+q =
c$, where $c \in {\mathbb C}$ is a constant.  If one treats the $a$ and $q$
variables as the ``horizontal'' and ``vertical'' axes (actually planes, in
terms of real variables), then the condition $a+q=c$ is an affine
translation of a diagonal slice of the complex locus ${\cal B}$. The
resultant ${\cal B}_a$ is the solution of the equation $|c-1|=|a-1|$, which is
a circle centered at $a=1$ with radius $|c-1|$.  The corresponding ${\cal B}_q$
is a circle centered at $q=c-1$ with radius $|c-1|$.  These circles in the $a$
and $q$ planes pass through $a=0$ and $q=0$, respectively.

\section{Square Strip with Free Longitudinal Boundary Conditions}

In this section we present the Potts model partition function 
$Z(S_m,q,v)$ for the $L_y=2$ strip of the square lattice $S_m$ with arbitrary 
length $L_x=m+1$ (i.e., containing $m+1$ squares) and free transverse and
longitudinal boundary conditions.
One convenient way to express the results is in terms of a generating 
function:
\beq
\Gamma(S,q,v,z) = \sum_{m=0}^\infty Z(S_m,q,v)z^m \ . 
\label{gammazfbc}
\eeq
We have calculated this generating function using the deletion-contraction 
theorem for the corresponding Tutte polynomial $T(S_m,x,y)$ and then 
expressing the result in terms of the variables $q$ and $v$.  We find 
\beq
\Gamma(S,q,v,z) = \frac{ {\cal N}(S,q,v,z)}{{\cal D}(S,q,v,z)}
\label{gammazcalc}
\eeq
where
\beq
{\cal N}(S,q,v,z)=A_{S,0}+A_{S,1}z
\label{numgamma}
\eeq
with
\beq
A_{S,0}=q(v^4+4v^3+6qv^2+4q^2v+q^3)
\label{as0}
\eeq
\beq
A_{S,1}=-q(v+1)(v+q)^3v^2
\label{as1}
\eeq
and
\beq
{\cal D}(S,q,v,z) = 1-(v^3+4v^2+3qv+q^2)z+(v+1)(v+q)^2v^2z^2 \ . 
\label{dengamma}
\eeq
(The generating function for the Tutte polynomial 
$T(S_m,x,y)$ is given in the appendix.) Writing 
\beq
{\cal D}(S,q,v,z) = \prod_{j=1}^2 (1-\lambda_{S,j}z)
\label{ds}
\eeq
we have
\beq
\lambda_{S,(1,2)} = \frac{1}{2}(T_{S12} \pm \sqrt{R_{S12}} \ )
\label{lams}
\eeq
where
\beq
T_{S12}=v^3+4v^2+3qv+q^2
\label{t56}
\eeq
and
\beq
R_{S12}=v^6+4v^5-2qv^4-2q^2v^3+12v^4+16qv^3+13q^2v^2+6q^3v+q^4 \ . 
\label{rs12}
\eeq

In \cite{hs} we presented a formula to obtain the chromatic polynomial for a
recursive family of graphs in the form (\ref{pgsum}) starting from the
generating function.  It will be useful to give here the generalization of
this formula for the full Potts partition function.  For a strip (recursive) 
graph with the labelling conventions used here, the generating function can 
be written as 
\beq
\Gamma(G,q,v,z) = \frac{{\cal N}(G,q,v,z)}{{\cal D}(G,q,v,z)}
\label{gammagen}
\eeq
with
\beq
{\cal N}(G,q,v,z) = \sum_{j=0}^{d_{\cal N}} A_{G,j}z^j
\label{n}
\eeq
and
\beqs
{\cal D}(G,q,v,z) & = & 1 + \sum_{j=1}^{d_{\cal D}} b_{G,j}z^j \cr\cr
& = & \prod_{j=1}^{d_{\cal D}}(1-\lambda_{G,j}z)
\label{d}
\eeqs
where 
\beq
d_{\cal N}(G) = deg_z({\cal N}(G))
\label{degn}
\eeq
\beq
d_{\cal D}(G) = deg_z({\cal D}(G))
\label{degd}
\eeq
Then the formula is 
\beq
Z(G_m,q,v) = \sum_{j=1}^{d_{\cal D}} \Biggl [ \sum_{s=0}^{d_{\cal N}}
A_{G,j} \lambda_j^{d_{\cal D}-s-1} \Biggr ]
\Biggl [ \prod_{1 \le i \le d_{\cal D}; i \ne j}
\frac{1}{(\lambda_{G,j}-\lambda_{G,i})} \Biggr ] \lambda_{G,j}^m
\label{chrompgsumlam}
\eeq
For our present open strip $S_m$, we have 
\beq
Z(S_m,q,v) = \frac{(A_{S,0}\lambda_{S,1} + A_{S,1})}
{(\lambda_{S,1}-\lambda_{S,2})}\lambda_{S,1}^m +
 \frac{(A_{S,0} \lambda_{S,2} + A_{S,1})}
{(\lambda_{S,2}-\lambda_{S,1})}\lambda_{S,2}^m
\label{pgsumkmax2}
\eeq 
(which is symmetric under $\lambda_{S,1} \leftrightarrow \lambda_{S,2}$).
This shows that the $c_{G,j}$ can depend on both $q$ and $v$ for open strip
graphs.  Although both the $\lambda_{S,j}$'s and the coefficient functions
involve the square root $\sqrt{R_{S12}}$ and are not polynomials in $q$ and 
$v$,
the theorem on symmetric functions of the roots of an algebraic equation
\cite{uspensky} guarantees that $Z(S_m,q,v)$ is a polynomial in $q$ and $v$ (as
it must be by (\ref{cluster}) since the coefficients of the powers of $z$ in
the equation (\ref{d}) defining these $\lambda_{S,j}$'s are polynomials in
these variables $q$ and $v$.  This is a generalization of our discussion in
\cite{pm} from the special case of chromatic polynomials to the general case of
the Potts/random cluster partition function.

As will be shown below, the singular locus ${\cal B}_u$ consists of arcs that 
do not separate the $u$ plane into different regions, so that the PM phase 
and its complex-temperature extension occupy all of this plane, except for 
these arcs.  For physical temperature and positive integer $q$, the (reduced) 
free energy of the Potts model in the limit $n \to \infty$ is given by 
\beq
f = \frac{1}{2}\ln \lambda_{S,1} \ . 
\label{fs}
\eeq
This is analytic for all finite
temperature, for both the ferromagnetic and antiferromagnetic sign of the
spin-spin coupling $J$.  The internal energy and specific heat can be 
calculated in a straightforward manner from the free energy (\ref{fs}); since
the resultant expressions are somewhat cumbersome, we do not list them here. 
We find that for $q < 2$, in both the ferromagnetic and antiferromagnetic case,
for sufficiently low temperature, the specific heat is negative, and hence 
the random cluster model is unphysical for $q < 2$ on this family of graphs.

Let us define 
\beq
D_k(q) = \frac{P(C_k,q)}{q(q-1)} = 
\sum_{s=0}^{k-2}(-1)^s {{k-1}\choose {s}} q^{k-2-s}
\label{dk}
\eeq
and $P(C_k,q)$ is the chromatic polynomial for the circuit
(cyclic) graph $C_k$ with $k$ vertices,
\beq
P(C_k,q) = (q-1)^k + (q-1)(-1)^k
\label{pck}
\eeq
so that $D_2=1$, $D_3=q-2$,
\beq
D_4=q^2-3q+3
\label{d4}
\eeq
and so forth for other $D_k$'s. 
In the $T=0$ Potts antiferromagnet limit $v=-1$, 
$\lambda_{S,1}=D_4$ and $\lambda_{S,2}=0$, so that eq. (\ref{gammazcalc})
reduces to the generating function for the chromatic polynomial 
for this open square strip (cf. eq. (2.16) in \cite{strip}) 
\beq
\Gamma(S,q,v=-1;z) = \frac{q(q-1)D_4}{1-D_4z}
\label{gammacp}
\eeq
where
Equivalently, the chromatic polynomial is 
\beq
P(S_m,q) = q(q-1)(D_4)^{m+1} \ . 
\label{psq}
\eeq
For the ferromagnetic case with general $q$, in the low-temperature limit 
$v \to \infty$,
\beq
\lambda_{S,1} = v^3+3v^2+(q+2)v+ O(1) \ , \quad 
\lambda_{S,2} = v^2+2(q-1)v+O(1) \quad {\rm as} \quad v \to \infty
\label{lamsvinf}
\eeq
so that $|\lambda_{S,1}|$ is never equal to $|\lambda_{S,2}|$ in this
limit, and hence ${\cal B}_u$ does not pass through the origin of the
$u$ plane for the $n \to \infty$ limit of the open square strip:
\beq
u=0 \not\in {\cal B}_u(\{S\}) . 
\label{unotinbs}
\eeq
In contrast, as will be shown below, ${\cal B}_u$ does pass through
$u=0$ for this strip with cyclic or M\"obius boundary conditions. 
For our later discussion, we record here the expressions for the
$\lambda_{S,j}$'s for the Ising case, $q=2$: 
\beq
\lambda_{S,(1,2)} = \frac{1}{2}(v+2)\Bigl [ v^2+2v+2 \pm 
(v^4+4v^2+8v+4)^{1/2} \Bigr ] \ . 
\label{lamsq12}
\eeq

\subsection{${\cal B}_q(\{S\})$ for fixed $a$} 

\begin{figure}
\vspace{-4cm}
\centering
\leavevmode
\epsfxsize=4.0in
\begin{center}
\leavevmode
\epsffile{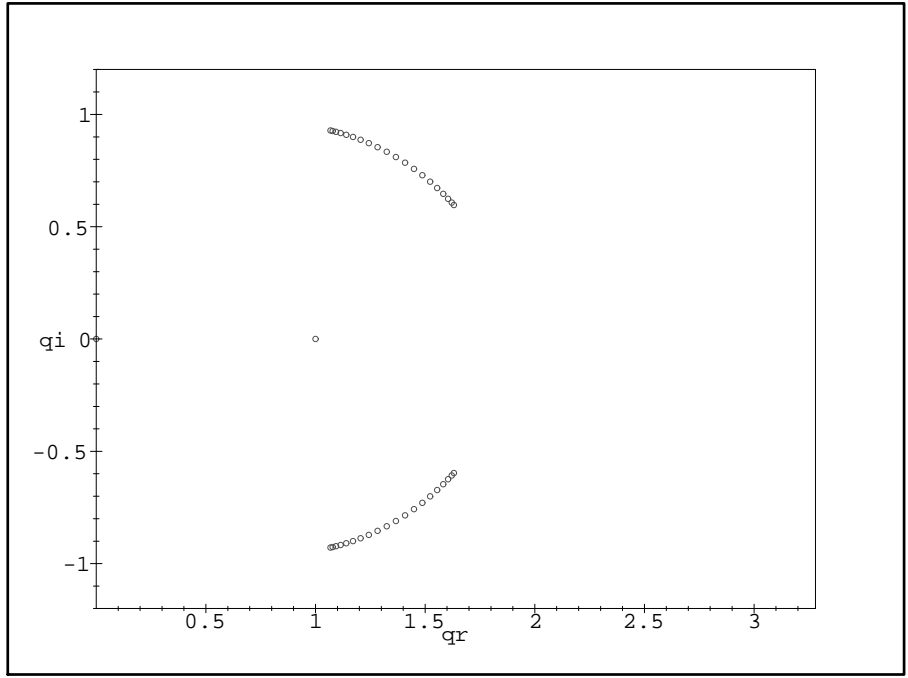}
\end{center}
\vspace{-2cm}
\caption{\footnotesize{Zeros of $Z(S_m,q,a)$ in the $q$ plane for
$a=0.1$.  For this and the other figures on zeros of $Z(S_m,q,a)$, we use 
$m=19$, i.e., $n=42$.  The axis labels are 
$qr \equiv Re(q)$ and $qi \equiv Im(q)$ here and in other $q$-plane plots.}}
\label{sqffq1}
\end{figure}

\begin{figure}
\vspace{-4cm}
\centering
\leavevmode
\epsfxsize=4.0in
\begin{center}
\leavevmode
\epsffile{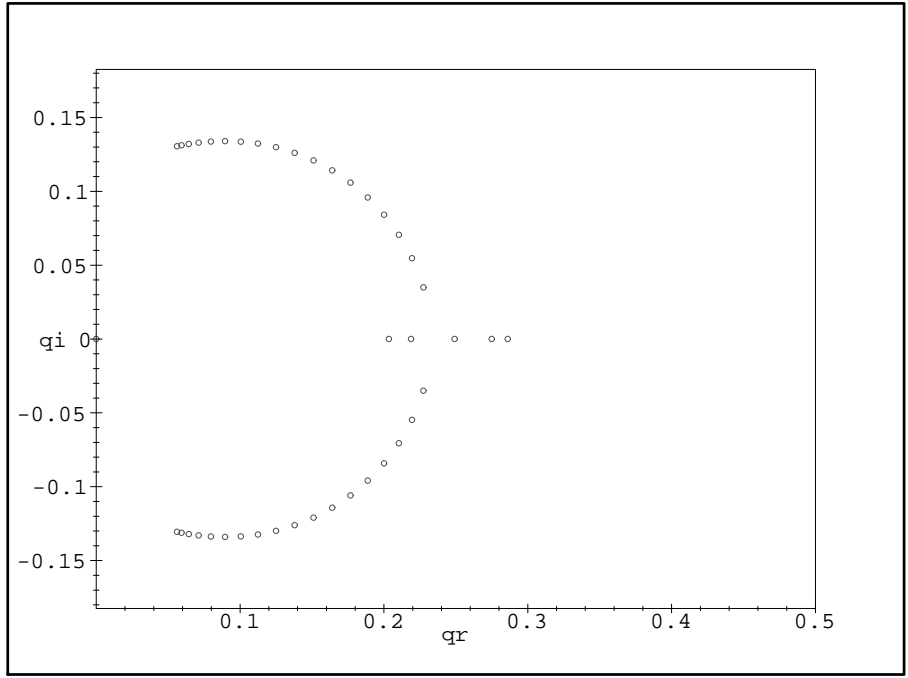}
\end{center}
\vspace{-2cm}
\caption{\footnotesize{Same as Fig. \ref{sqffq1} for $a=0.9$.}}
\label{sqffq2}
\end{figure}

\begin{figure}
\vspace{-4cm}
\centering
\leavevmode
\epsfxsize=4.0in
\begin{center}
\leavevmode
\epsffile{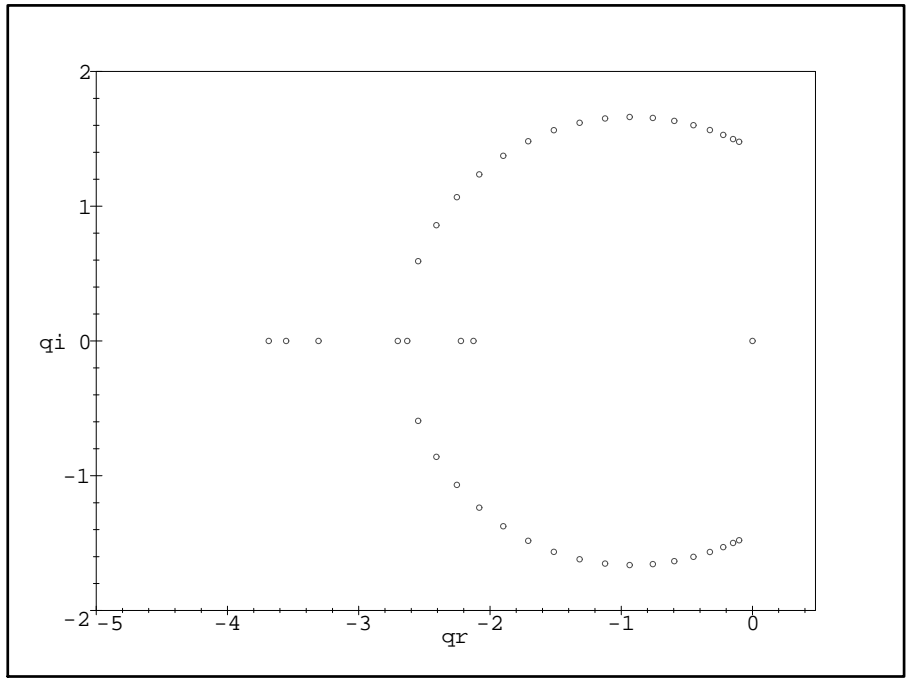}
\end{center}
\vspace{-2cm}
\caption{\footnotesize{Same as Fig. \ref{sqffq1} for $a=2$.}}
\label{sqffq3}
\end{figure}

We discuss here the continuous locus ${\cal B}_q(\{S\})$ in the $q$ plane for 
various values of $a$.  For the chromatic polynomial case $a=0$
($v=-1$), ${\cal B}=\emptyset$, since
$W(\{S\},q) = (D_4)^{1/2}$ has only the discrete branch point
singularities (zeros) at 
\beq
q_{bp}, \ q_{bp}^* = 1+e^{\pm i\pi/3} \ . 
\label{qbp}
\eeq
However, for $a \ne 0$, the situation is qualitatively different; 
${\cal B}_q$ is nontrivial.  As $a$ increases above 0, the locus ${\cal B}_q$ 
forms two complex-conjugate (c.c.) arcs, as shown, for $a=0.1$, in Fig. 
\ref{sqffq1}. For
small $a$, these arcs lie near to the circle $|q-1|=1$; as $a$
decreases, they shorten and as $a \to 0$, they degenerate to the c.c. 
points $q_{bp},q_{bp}^*$ in eq. (\ref{qbp}).  The 
endpoints of the arcs are the (finite) branch 
point singularities of $\lambda_{S,j}$, $j=1,2$, arising from the zeros of 
the square root in (\ref{lams}); for example, for $a=0.1$, these endpoints
occur at $q \simeq 1.0654 + 0.9293i$ and $q \simeq 1.6346 + 0.59275i$, together
with their complex conjugates.  From Fig. \ref{sqffq1}, one can see that the 
density of chromatic zeros is greatest at the endpoints and minimal at the
centers of the arcs.  As $a$ increases, these arcs extend downward toward the 
positive real $q$ axis.  As $a$ reaches the value $a=9/16$, the arcs touch the
real axis at $q=63/64=0.984375$, thereby joining to form a single 
self-conjugate arc with endpoints at $q, q^*=(21\pm 14\sqrt{6}i)/64
\simeq 0.3281 \pm 0.5358i$. As $a$ increases above the value 9/16, 
${\cal B}_q$ consists of the self-conjugate arc and a line segment on the real 
axis, which spreads out from the point $q=63/64$.  
This corresponds to the fact that for $a \ge 9/16$, the
expression in the square root in eq. (\ref{lams}) has real as well as complex
zeros.  An illustration is given for $a=0.9$ in Fig. \ref{sqffq2}. 
As $a \to 1$, the locus ${\cal B}_q$ shrinks in toward the origin.
For the ferromagnetic range $a > 1$, ${\cal B}_q$ is 
located in the $Re(q) \le 0$ half-plane and forms c.c. arcs together with a 
line segment on the negative real axis, as illustrated for $a=2$ in Fig. 
\ref{sqffq3}. 

\vspace{12mm}

\vspace{8mm}

\vspace{8mm}

One can also consider negative real values of $a$, which correspond to complex
temperature. As $a$ decreases from 0 through real values, 
${\cal B}_q$ forms arcs, as was the case when $a$ 
increased from 0; these arcs have endpoints at the branch point 
singularities of $\lambda_{S,j}$ and elongate as $a$ moves downward in the
range $-1 < a < 0$.  As $a$ reaches $-1$, these arcs touch the positive real 
$q$ axis at the point $q=2$ and join to form a single self-conjugate arc (with
endpoints at $4 \pm 4i$), but as $a$ decreases below $-1$, the arcs retract 
from the real axis to form two c.c. parts again.  It is straightforward to 
consider complex values of $a$ also, but we shall restrict ourselves to real 
$a$ here.

\subsection{${\cal B}_u(\{S\})$ for fixed $q$}

\begin{figure}
\vspace{-4cm}
\centering
\leavevmode
\epsfxsize=4.0in
\begin{center}
\leavevmode
\epsffile{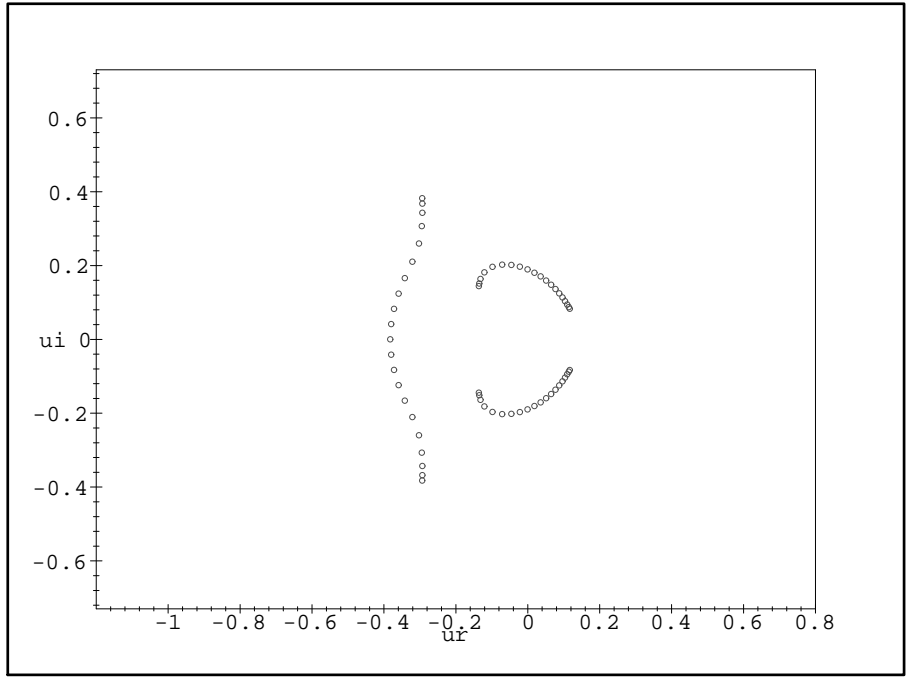}
\end{center}
\vspace{-2cm}
\caption{\footnotesize{Zeros of $Z(S_m,q,a)$ in the $u$ plane for $q=10$ and 
$m=19$ ($n=42$).  The axis labels are 
$ur \equiv Re(u)$ and $ui \equiv Im(u)$ here and in other $u$-plane plots.}}
\label{sqffuq10}
\end{figure}

For our analysis of ${\cal B}_u(\{S\})$ we start with large $q$.  Here 
${\cal B}_u$ consists of a self-conjugate arc that crosses the real
$u$ axis, together with a complex-conjugate pair of arcs that are concave
toward the real $u$ axis.  As $q \to \infty$, these arcs all shrink and move in
toward the origin of the $u$ plane.  This limit thus commutes with the result
of taking $q \to \infty$ first before taking $n \to \infty$; in this case, 
\beq
\lambda_{S,1} = q^2+3vq+O(1) \ , \quad \lambda_{S,2} = v^2(1+v) + O \Bigl (
\frac{1}{q} \Bigr ) \quad {\rm as} \quad q \to \infty
\label{lams1qinf}
\eeq
so that the degeneracy equation $|\lambda_{S,1}|=|\lambda_{S,2}|$ has no
solution for $q \to \infty$.  In Fig. \ref{sqffuq10} we show the
complex-temperature zeros of $Z$ for a typical value, $q=10$, calculated for 
$L_x=m+1=20$, i.e., $n=42$.  With this large a value of $n$, these
zeros occur close to the asymptotic locus ${\cal B}_u$ and give an 
adequate indication of its location.  The self-conjugate arc crosses
the real $u$ axis at $u\simeq -0.3954$ where the quantity
$T_{S12}$ in eq. (\ref{t56}) vanishes, so that $|\lambda_{S,1}|=
|\lambda_{S,2}|$.  The endpoints of this arc occur at two of the zeros of the
square root in eq. (\ref{lams}), viz., $u \simeq -0.2937 \pm 0.3870i$.  The two
c.c. arcs have their endpoints at the four other zeros of this square root, at
$u \simeq -0.1361 \pm 0.14245i$ and $u \simeq 0.1178 \pm 0.8130i$.  

\begin{figure}
\vspace{-4cm}
\centering
\leavevmode
\epsfxsize=4.0in
\begin{center}
\leavevmode
\epsffile{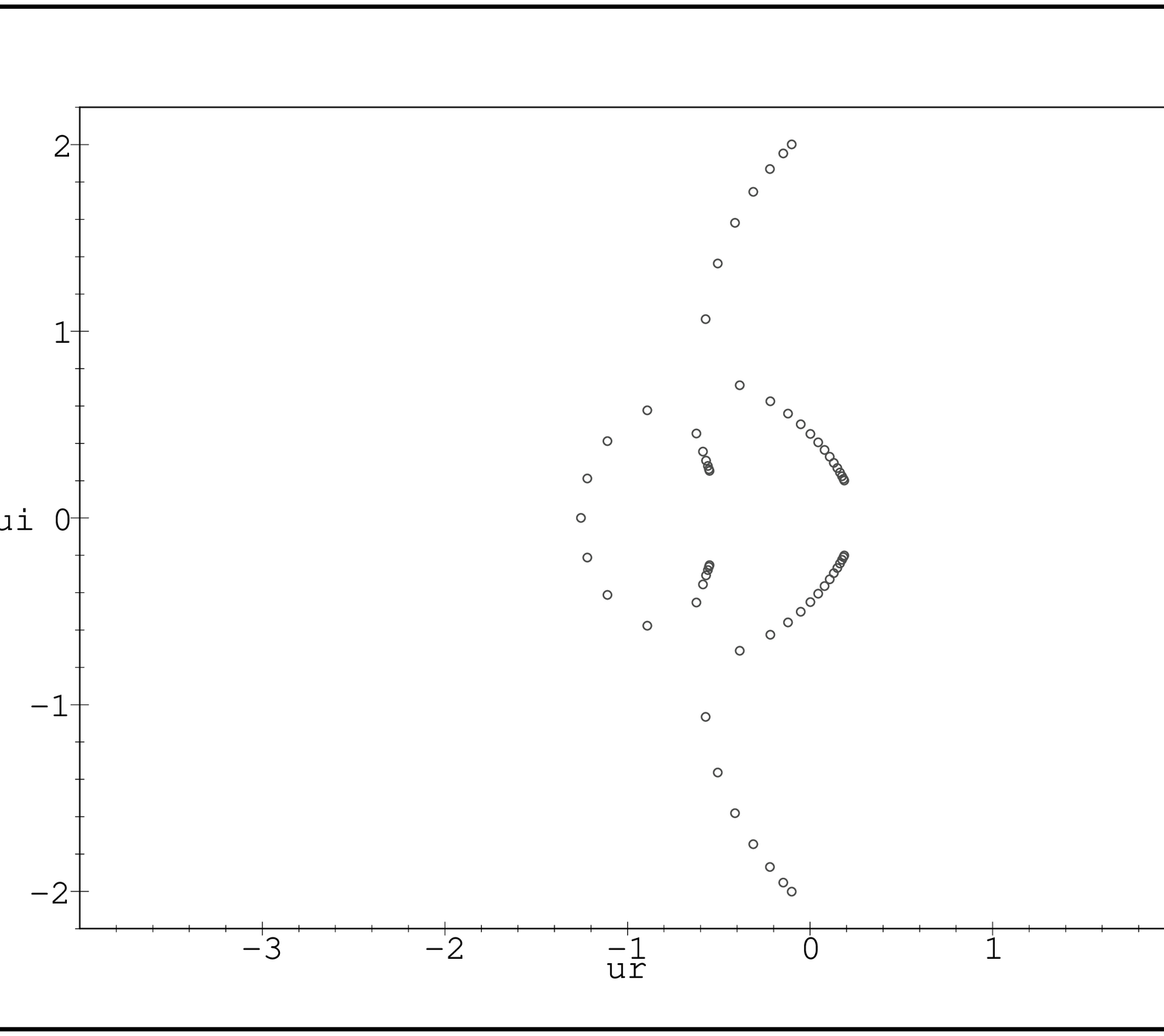}
\end{center}
\vspace{-2cm}
\caption{\footnotesize{Same as Fig. \ref{sqffuq10} for $q=3$.}}
\label{sqffuq3}
\end{figure}

\begin{figure}
\vspace{-4cm}
\centering
\leavevmode
\epsfxsize=4.0in
\begin{center}
\leavevmode
\epsffile{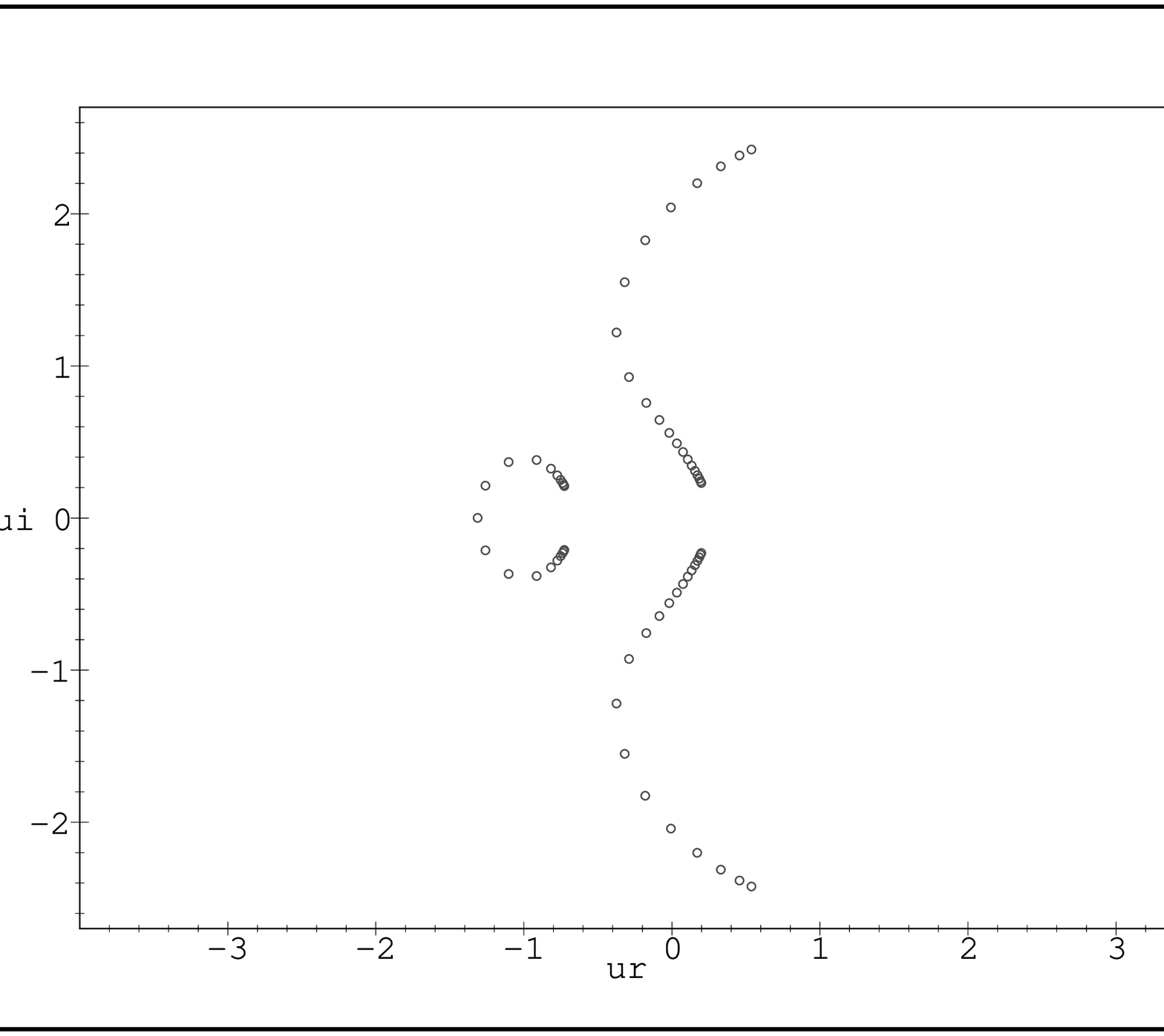}
\end{center}
\vspace{-2cm}
\caption{\footnotesize{Same as Fig. \ref{sqffuq10} for $q=2.5$.}}
\label{sqffuq2p5}
\end{figure}

As $q$ decreases, the endpoints of the self-conjugate arc retract
toward the real axis, it curls over to be more concave to the right,
and the point at which it crosses the real axis moves to the left.
For example, for $q=4$, the self-conjugate arc crosses the real axis
at $u=-1$ and has its endpoints at $q \simeq -0.4341 \pm 1.3178i$, and
the c.c arcs extend between endpoints at $q \simeq -0.3697 \pm
0.2394i$ and $q \simeq 0.1711 \pm 0.1593i$.  For $q=3$ (see
Fig. \ref{sqffuq3}), the self-conjugate arc crosses the real axis at
$u \simeq -1.2767$ and has endpoints at $q \simeq -0.5498 \pm
0.2489i$, and the c.c. arcs extend between $q \simeq -0.0839 \pm
2.0177i$ and $0.1892 \pm 0.1974i$, passing through the points $u, u^*
= e^{\pm 2\pi i/3}$.  These results for $q=3$ and 4 have interesting
implications that we shall discuss further below.  The changes in
${\cal B}_u$ as $q$ decreases further toward $q=2$ are illustrated in
Fig. \ref{sqffuq2p5} where we show the $q=2.5$ case. The self-conjugate
arc crosses the real axis at $u \simeq -1.323$ and has endpoints at $u
\simeq -0.7249 \pm 0.2083i$ while the c.c pairs of arcs have endpoints
at $q \simeq 0.2006 \pm 0.2272i$ and $q \simeq 0.56505 \pm 2.4351i$.

\begin{figure}
\vspace{-4cm}
\centering
\leavevmode
\epsfxsize=4.0in
\begin{center}
\leavevmode
\epsffile{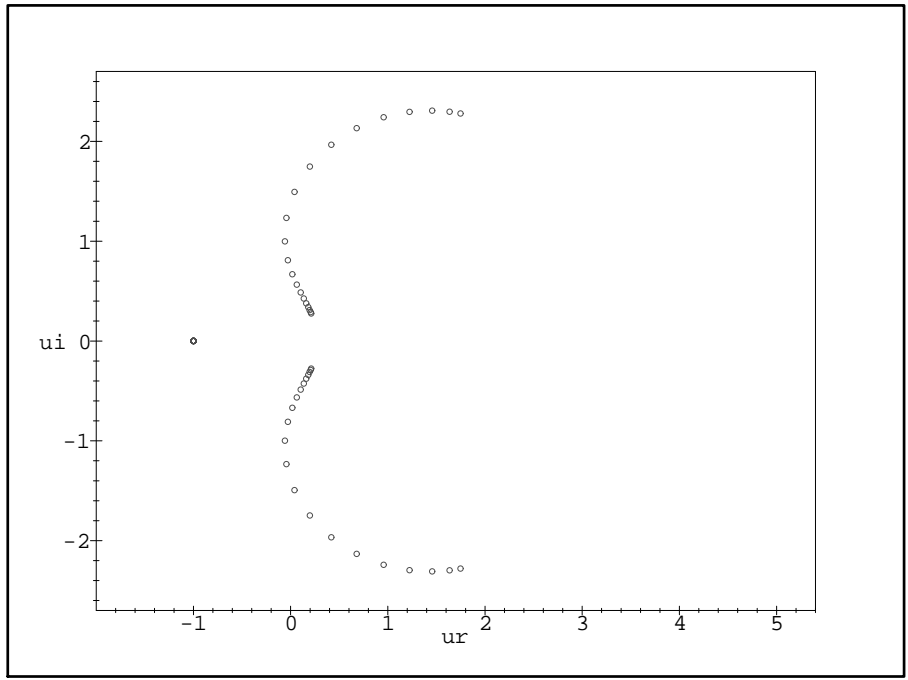}
\end{center}
\vspace{-2cm}
\caption{\footnotesize{Same as Fig. \ref{sqffuq10} for $q=2$.}}
\label{sqffuq2}
\end{figure}

For $q=2$ (Fig. \ref{sqffuq2}), the quartic polynomial in the square root of
eq. (\ref{lams}) factorizes into a quadratic polynomial times $(v+2)^2$,
yielding the result (\ref{lamsq12}).  Correspondingly, the self-conjugate arc
disappears, and the locus ${\cal B}$ consists of two complex-conjugate arcs
located in the half-plane $Re(q) \ge 0$, and touching the imaginary axis at
$u=\pm i$.  This locus is invariant under the inversion map $u \to 1/u$.  The
upper arc extends from the left endpoint at $u_{e1} \simeq 0.2138 + 0.2720i$
through $u=i$ to a right endpoint at $u_{e2} = 1/u_{e1}^* \simeq
1.78615-2.2720i$, and so forth for the c.c. arc.  As we have discussed before
in the context of ${\cal B}_q$ \cite{strip} (see also \cite{m1,wood}), these
endpoints are the zeros of the square root in (\ref{lamsq12}) where there are
finite branch point singularities in $\lambda_{S,1}$.  There is also a discrete
zero of $Z$ at the point $u=-1$ with multiplicity scaling proportional to the
lattice size.  As we proved in a previous theorem (Theorem 6 of
Ref. \cite{cmo}), this zero arises for the Ising model on a lattice with odd
coordination number; in the present case, all of the vertices of the strip $S$
except those on the four end-corners have $\Delta=3$.

Because of the unphysical nature of the Potts/random cluster model for 
$q < 2$, we shall not discuss this range except to mention another example
of the noncommutativity (\ref{fnoncomm}) at $q=0$. If one first sets 
$q=0$ and calculates $Z$, then, since $Z=0$ identically, the set of zeros of 
$Z$ is vacuous. However, if one takes the limit $n \to \infty$,
calculates the accumulation set ${\cal B}_u$, and then takes the limit $q \to
0$, one finds that $\lim_{q \to 1}{\cal B}_u$ is not the empty set.  This is 
clear from the fact that in the limit $q \to 0$ 
\beq
\lambda_{S,(1,2);q=0}= \frac{v^2}{2}\Bigl [ v+4 \pm (v^2+4v+12)^{1/2} \Bigr ]
\label{lamsq0}
\eeq
so that the degeneracy equation $|\lambda_{S,1;q=0}|=|\lambda_{S,2;q=0}|$ 
has a nontrivial solution, namely the section of the the circular arc in the 
$Re(u) < 0$ half-plane 
\beq
{\cal B}_u: \quad u = \frac{1}{3}e^{i(\pi \pm \theta)} \ , \quad 0 \le \theta
\le arctan(2\sqrt{2}) 
\label{arcq0}
\eeq
which crosses the real axis at $u=-1/3$ (i.e., $v=-4$) and has
endpoints at $u=(-1 \pm 2\sqrt{2}i)/9$. 

\pagebreak

\vspace{8mm}

\vspace{8mm}

\vspace{8mm}

\section{Cyclic and M\"obius Ladder Graphs}

\subsection{Results for $Z$}

By either using an iterative application of the deletion-contraction theorem 
for Tutte polynomials and converting the result to $Z$, or by using a 
transfer matrix method (in which one starts with a $q^2 \times q^2$ 
transfer matrix and generalizes to arbitrary $q$), one can calculate the 
partition function for the cyclic and M\"obius ladder graphs of arbitrary 
length, $Z(G,q,v)$, $G=L_m, \ ML_m$.  We have used both methods as checks on
the calculation.  Our results 
have the general form (\ref{zgsum}) with $N_\lambda=6$ and are
\beq
Z(L_m,q,v) = \sum_{j=1}^6 c_{L,j} (\lambda_{L,j})^m
\label{zlad}
\eeq
and
\beq
Z(ML_m,q,v) = \sum_{j=1}^6 c_{ML,j} (\lambda_{ML,j})^m
\label{zmblad}
\eeq
where
\beq
\lambda_{ML,j}=\lambda_{L,j} \ , \quad j=1,...,6
\label{lammb}
\eeq
\beq
\lambda_{L,1} = v^2
\label{lam1}
\eeq
\beq
\lambda_{L,2} = v(v+q)
\label{lam2}
\eeq
\beq
\lambda_{L,(3,4)} = \frac{v}{2}\Bigl [ q+v(v+4) \pm 
(v^4+4v^3+12v^2-2qv^2+4qv+q^2)^{1/2} \Bigr ] 
\label{lam34ex}
\eeq
and
\beq
\lambda_{L,5}=\lambda_{S,1} \ , \quad \lambda_{L,6}=\lambda_{S,2}
\label{lam56ex}
\eeq
where $\lambda_{S,j}$, $j=1,2$ for the open ladder were given above in
eq. (\ref{lams}).  We note that $\lambda_{L,3}\lambda_{L,4}=(1+v)(q+v)v^3$ and 
$\lambda_{L,5}\lambda_{L,6}=(1+v)(q+v)^2v^2$. Chromatic and Tutte polynomials
for recursive families of graphs obey certain recursion relations 
\cite{bds,hs}.  In terms of the equivalent Tutte polynomial, given in 
the appendix, the results (\ref{zlad}) and (\ref{zmblad}) agree with 
a recursion relation given in Ref. \cite{bds} (see also \cite{sands}).  

The coefficient functions for the cyclic and M\"obius ladders are 
\beq
c_{L,1} = q^2-3q+1
\label{c1lad}
\eeq
\beq
c_{L,2} = c_{L,3} = c_{L,4} = c_{ML,3} = c_{ML,4} = q-1
\label{c234lad}
\eeq
\beq
c_{L,5} = c_{L,6} = c_{ML,5} = c_{ML,6} = 1
\label{c56lad}
\eeq
\beq
c_{ML,1}=-1
\label{c1mob}
\eeq
\beq
c_{ML,2}=1-q \ . 
\label{c2mob}
\eeq
Because of the equalities $c_{G,3}=c_{G,4}$ and $c_{G,5}=c_{G,6}$ for $G=L$
and for $G=ML$, we can again apply the theorem on symmetric
polynomial functions of roots of algebraic equations \cite{uspensky} to
confirm that, although the $\lambda_{G,j}$'s for nonpolynomial algebraic
functions of $q$ and $v$ for $j=3,4,5,6$, $Z(G_m,q,v)$ is a polynomial function
of these variables $q$ and $v$, as it must be by (\ref{cluster}).

\subsection{Special values and expansions of $\lambda$'s}

We discuss some special cases. First, for the zero-temperature Potts
antiferromagnet, i.e. the case $a=0$ ($v=-1$), the partition functions
$Z(L_m,q,v)$ and $Z(ML_m,q,v)$ reduce, in accordance with the general result
(\ref{zp}), to the respective chromatic polynomials $P(L_m,q)$ and $P(ML_m,q)$
calculated in \cite{bds}.  In this special case, we have $\lambda_{L,1}=1$,
$\lambda_{L,2}=1-q$, and (for an appropriate choice of sign of terms of the
form $\sqrt{(q-3)^2}$ and $\sqrt{(D_4)^2}$ \ ) $\lambda_{L,3}=3-q$,
$\lambda_{L,4}=0$, $\lambda_{L,5}=D_4=q^2-3q+3$, and $\lambda_{L,6}=0$.  For
the infinite-temperature value $a=1$, we have $\lambda_{L,j}=0$ for
$j=1,2,3,4,6$, while $\lambda_{L,5}=q^2$, so that $Z(G,q,a=1)= q^{2m} = q^n$
for $G=L_m, \ ML_m$, in accord with the general result (\ref{za1}).

At $q=0$, besides the $q$-independent $\lambda_{L,1}$, we find 
\beq
\lambda_{L,2}=(a-1)^2
\label{lam2q0}
\eeq
\beq
\lambda_{L,3}=\lambda_{L,5} = 
\frac{1}{2}(a-1)^2\biggl [ a+3 + (a^2+2a+9)^{1/2} \biggr ]
\label{lam35q0}
\eeq
\beq
\lambda_{L,4}=\lambda_{L,6}=\frac{1}{2}(a-1)^2\biggl [ a+3 - 
(a^2+2a+9)^{1/2} \biggr ] \ . 
\label{lam46q0}
\eeq
Since $\lambda_{L,3}$ and $\lambda_{L,5}$ are leading and are 
degenerate at this point, it follows that
\beq
q=0 \quad {\rm is \ on} \quad {\cal B}_q(\{L\}) \quad \forall \ a \ . 
\label{q0onb}
\eeq
At $q=1$, $c_{L,j}=0$ for $j=2,3,4$ so that the corresponding $\lambda_{L,j}$,
$j=2,3,4$, do not contribute to $Z$.  Further, $c_{L,1}=-1=-c_{L,j}$, $j=5,6$
and $\lambda_{L,1}=\lambda_{L,6}=(a-1)^2$ so that the contributions of these 
terms cancel in $Z$, leaving only the contribution of $\lambda_{L,5}$:
$Z(L_m,q=1,a)=(\lambda_{L,5})^m=a^{3m}$, in agreement with the general 
formula (\ref{zq1}). 

\vspace{4mm}

In order to study the zero-temperature critical point in the ferromagnetic 
case and also the properties of the complex-temperature phase diagram, we 
calculate the $\lambda_{G,j,u}$'s corresponding to the $\lambda_{G,j}$'s, 
using eq. (\ref{lamu}).  This gives $\lambda_{L,1,u}=u(1-u)^2$, 
$\lambda_{L,2,u}=u(1-u)[1+(q-1)u)]$, and so forth for the others. 
In the vicinity of the point $u=0$ one has 
\beq
\lambda_{L,1,u}=u-2u^2+u^3
\label{lam1rtaylor}
\eeq
\beq
\lambda_{L,2,u}=u+(q-2)u^2+(1-q)u^3
\label{lam2rtaylor}
\eeq
and the Taylor series expansions
\beq
\lambda_{L,3,u}=1-u^2+2(q-2)u^3+O(u^4) 
\label{lam3rtaylor}
\eeq
\beq
\lambda_{L,4,u}=u+(q-4)u^2+(7-3q)u^3+O(u^4)
\label{lam4rtaylor}
\eeq
\beq
\lambda_{L,5,u}=1+(q-1)u^2\Bigl [ 1 + 4u + O(u^2) \Bigr ]
\label{lam5rtaylor}
\eeq
\beq
\lambda_{L,6,u}=u+2(q-2)u^2+(q^2-7q+7)u^3 + O(u^4) \ . 
\label{lam6rtaylor}
\eeq
Hence, at $u=0$, $\lambda_{L,3,u}$ and $\lambda_{L,5,u}$ are dominant and 
$|\lambda_{L,3,u}|=|\lambda_{L,5,u}|$, so that the point $u=0$ is on 
${\cal B}_u$ for any $q \ne 0,1$, where the noncommutativity (\ref{fnoncomm})
occurs.  For $q > 0$, $\lambda_{L,5,u}$ is dominant on the real $u$ axis in 
the vicinity of $u=0$ and hence in the PM and O phases that can be 
reached by analytic continuation therefrom, while the term $\lambda_{L,3,r}$ 
is dominant on the imaginary $u$ axis in the neighborhood of the origin, and 
hence in the O phases that can be reached by analytic continuation from this
neighborhood. 

To determine the angles at which the branches of ${\cal B}_u$ cross
each other at $u=0$, we write $u$ in polar coordinates as $u=re^{i\theta}$,
expand the degeneracy equation $|\lambda_{L,3,u}|=|\lambda_{L,5,u}|$, for small
$r$, and obtain $qr^2\cos(2\theta)=0$, which implies that (for $q \ne 0,1$) in
the limit as $r=|u| \to 0$, 
\beq 
\theta = \frac{(2j+1)\pi}{4} \ , \quad j=0,1,2,3 
\label{thetau}
\eeq
or equivalently, $\theta=\pm \pi/4$ and $\theta=\pm 3\pi/4$.  
Hence there are four branches of ${\cal B}_u$ intersecting at $u=0$ and these
branches cross at right angles.  The point $u=0$ is thus a multiple point on 
the algebraic curve ${\cal B}_u$, in the technical terminology of algebraic 
geometry (i.e., a point where several branches of an algebraic curve cross 
\cite{alg}).  

In order to investigate how these crossings depend on $L_y$, we have calculated
$Z$ for the cyclic strip graph of the square lattice with the next larger
width, $L_y=3$. Since the $T=0$ critical point for the Potts ferromagnet is
present for each $q \ne 0,1$, it suffices to do this calculation for the simple
$q=2$ Ising case (bearing in mind the noncommutativity that applies at special
values $q_s$ as discussed above).  We find that there are two $\lambda_j$'s
that are dominant near $u=0$, and the small--$u$ expansion of the degeneracy
equation yields the condition $r^3\cos(3\theta)=0$, so that there are six
curves on ${\cal B}_u$ crossing $u=0$, at the angles $\theta=(2j+1)\pi/6$,
$j=0,1,...,5$.  This leads to the generalization that for the cyclic strip
graph of the square lattice with width $L_y$, there are $2L_y$ curves on ${\cal
B}_u$ that cross each other at $u=0$, at the angles $\theta=(2j+1)\pi/(2L_y)$,
$j=0,1,...,2L_y-1$.  This inference implies, in turn, that in the limit $L_y
\to \infty$, an infinite number of curves on ${\cal B}_u$ intersect at $u=0$,
and the complex-temperature (Fisher) zeros become dense in the neighborhood of
this point. Since the origin of this phenomenon is not dependent in detail on
the lattice type, one would also infer that it occurs for infinite-length
width $L_y$ cyclic strips of other lattices.

For $q=0,1$ the (\ref{bnoncomm}) occurs, with  
$({\cal B}_u)_{nq}=\emptyset$ by eqs. (\ref{bnq0}) and (\ref{bnq1}), but 
$({\cal B}_u)_{qn} \ne \emptyset$.  
While ${\cal B}_u$ is compact for $q \ne 2$, it is noncompact for $q = 2$, 
where the symmetry (\ref{bq2sym}) holds.

Our exact calculations yield the following general result
\beq
{\cal B}(\{L\}) = {\cal B}(\{ML\}) \ . 
\label{bcycmob}
\eeq
This is in accord with the conclusion that the singular locus is the same for
an infinite-length finite-width strip graph for given transverse boundary
conditions, independent of the longitudinal boundary condition.  
This generalizes our previous finding that ${\cal B}_q$ was independent of the
longitudinal boundary conditions for the case $a=0$ \cite{wcy,pm,bcc}.  In the
present case, the result (\ref{bcycmob}) follows immediately because 
$Z(L_m,q,v)$ and $Z(ML_m,q,v)$ involve the same $\lambda_j$'s.  We note that
this is a sufficient, but not necessary condition for the loci to be the same
for a given family of graphs when one changes the longitudinal boundary
conditions; it may be recalled that for the $T=0$ Potts antiferromagnet on the
width $L_y=3$ strip of the square \cite{tk} with 
periodic transverse boundary conditions, when one changed from periodic to 
twisted periodic longitudinal boundary conditions, i.e. toroidal to Klein 
bottle topology, three of the $N_\lambda=8$ terms were absent.  However, since
none of these was a dominant term anywhere, the locus ${\cal B}_q$ was the 
same for either toroidal or Klein bottle boundary conditions.  From our
calculation of the chromatic polynomial for the width $L_y=3$ strip of the
triangular lattice with both free and periodic transverse boundary
conditions and periodic and twisted periodic longitudinal boundary conditions 
\cite{t} we found that a similar situation occured for the toroidal versus
Klein bottle boundary conditions: six of the $N_\lambda=11$ terms in the 
toroidal case were absent in the Klein bottle case, but again none of these was
dominant anywhere. 
Owing to the equality (\ref{bcycmob}), we shall henceforth, for brevity of
notation, refer to both ${\cal B}(\{L\})$ and ${\cal B}(\{ML\})$ as 
${\cal B}(\{L\})$ and similarly for specific points on ${\cal B}$, such as 
$q_c(\{L\})=q_c(\{ML\})$, etc.  

\subsection{${\cal B}_q(\{L\})$ for fixed $a$}

We find that ${\cal B}_q(\{L\})$ crosses the real $q$ axis at
\beq
q_c(\{L\}) = (1-a)(a+2) \ . 
\label{qclad}
\eeq
This is the solution to the degeneracy equation of leading terms
$|\lambda_{L,5}|=|\lambda_{L,3}|=|\lambda_{L,2}|$.  As $a$ increases from 0 
to 1, $q_c(\{L\})$ decreases monotonically from 2 to 0.  From eq. 
(\ref{qclad}) it
follows that there are, in general, two values of $a$ that correspond to this 
value of $q$ on ${\cal B}(\{L\})$, viz.,  
\beq
a_{c,\pm}(\{L\}) = \frac{1}{2}[ -1 \pm \sqrt{9-4q} \ ] \ ,  \quad i.e, \quad 
u_{c,\pm}(\{L\}) = \frac{-1\pm \sqrt{9-4q} \ }{2(q-2)} \ . 
\label{auclad}
\eeq

\subsection{Antiferromagnetic Case, $T=0$}

We start with the $T=0$ antiferromagnet, i.e. the case $a=0$.  After initial
studies in Refs. \cite{bds,readcarib,read91}, the locus ${\cal B}$ was
determined in Ref. \cite{w}.  As is shown in Fig. 3 of Ref. \cite{w}, ${\cal
B}_q$ separates the $q$ plane into four regions, and $q_c(\{L\})=2.$  
The outermost region is $R_1$, and includes the segments $2 \le q$ and
$q < 0$ on the real $q$ axis; in this region $\lambda_{L,5}$ is dominant.
The innermost region, denoted $R_3$, includes the segment $0 \le q \le 2$ on
the real axis; in this region, the term $\lambda_{L,3}$ is dominant.  In
addition, there are two other complex-conjugate regions, $R_2$ and $R_2^*$,
which touch the real axis at $q=q_c(\{L\})=2$ 
and stretch outward to triple points at
\beq
q_{L,trip.}, \ q_{L,trip.}^* = 2 \pm \sqrt{2} \ i \ . 
\label{qladtrip}
\eeq
The part of ${\cal B}$ separating region $R_3$ from regions $R_2$, $R_2^*$ is
the line segment $Re(q)=2$, $-\sqrt{2} \le Im(q) \le \sqrt{2}$.  In regions
$R_2$, $R_2^*$, $\lambda_2$ is dominant.  At $q=2$ all four terms are
degenerate (recall that for $a=0$, $\lambda_{L,4}=\lambda_{L,6}=0$).  At the 
triple points $q_{L,trip.}$, there are three degenerate leading terms, 
with $|\lambda_{L,5}|=|\lambda_{L,3}|=|\lambda_{L,2}|$.
All four regions are contiguous at $q_c(\{L\})$.

\subsection{Antiferromagnet Case for $T > 0$}

We proceed to consider the regions in the $q$ plane for the Potts
antiferromagnet at arbitrary nonzero temperature, i.e. the range $0 < a \le 1$.
The zeros of $Z$ in the $q$ plane are shown for several values of $a$ in the
figures.  In this range $0 < a < 1$ we find a number of general features.  As
was true at $T=0$, ${\cal B}_q$ continues to separate the $q$ plane into
different regions and, as indicated in eq. (\ref{q0onb}) and (\ref{qclad}),
this locus crosses the real axis at $q=0$ and $q_c(\{L\})$.  ${\cal B}_q$
consists of a single connected component made up of several curves.  Commenting
on the regions in the $q$ plane, starting for $a$ near 0, we note that again
the region $R_1$ is the outermost, and includes the semi-infinite line segment
on the real axis $q > q_c(\{L\})$ and $q < 0$; region $R_3$ is the innermost
region, and includes the line segment $0 \le q \le q_c(\{L\})$. The
complex-conjugate regions $R_2$ and $R_2^*$ extend upward and downward from
$q_c(\{L\})$ to triple points.  As $a$ increases, the complex-conjugate regions
$R_2$ and $R_2^*$ are reduced in size.  As is 
evident in the figures, as $a$ increases from 0 to 1, the locus ${\cal B}_q$
contracts toward the origin, $q=0$ and in the limit as $a \to 1$,
it degenerates to a point at $q=0$.  This also describes the general behavior
of the partition function zeros themselves.  That is, for finite graphs, there
are no isolated partition function zeros whose moduli remains large as $a \to
1$. This is clear from continuity arguments in this limit, given
eq. (\ref{za1}).

\vspace{8mm}

\begin{figure}
\vspace{-4cm}
\centering
\leavevmode
\epsfxsize=4.0in
\begin{center}
\leavevmode
\epsffile{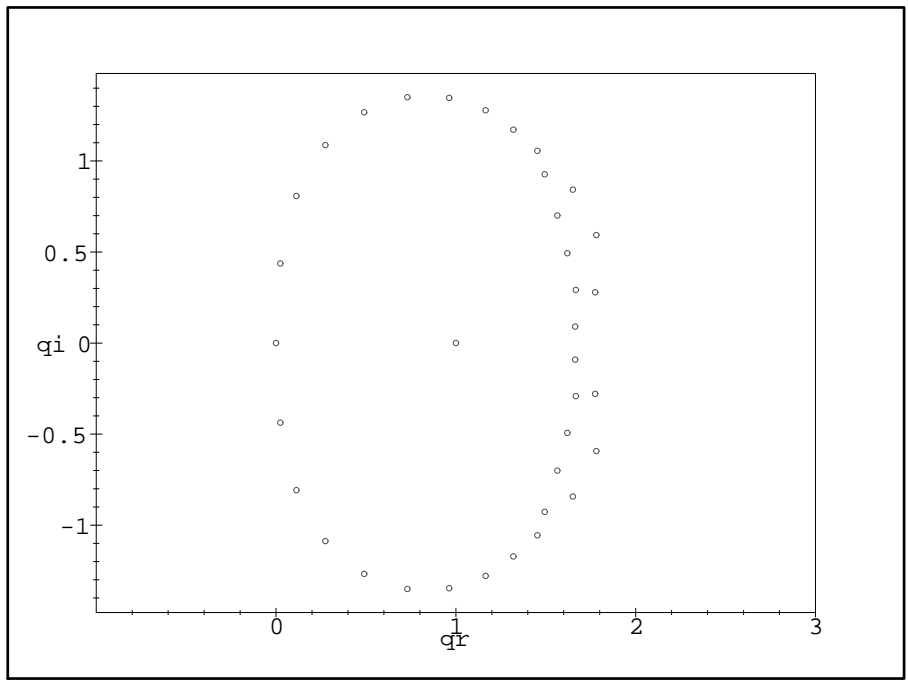}
\end{center}
\vspace{-2cm}
\caption{\footnotesize{Zeros of $Z(L_m,q,a)$ in the $q$ plane for $a=0.25$ and
$m=18$ ($n=36$).}}
\label{ladqa0p25}
\end{figure}

\vspace{8mm}

\begin{figure}
\vspace{-4cm}
\centering
\leavevmode
\epsfxsize=4.0in
\begin{center}
\leavevmode
\epsffile{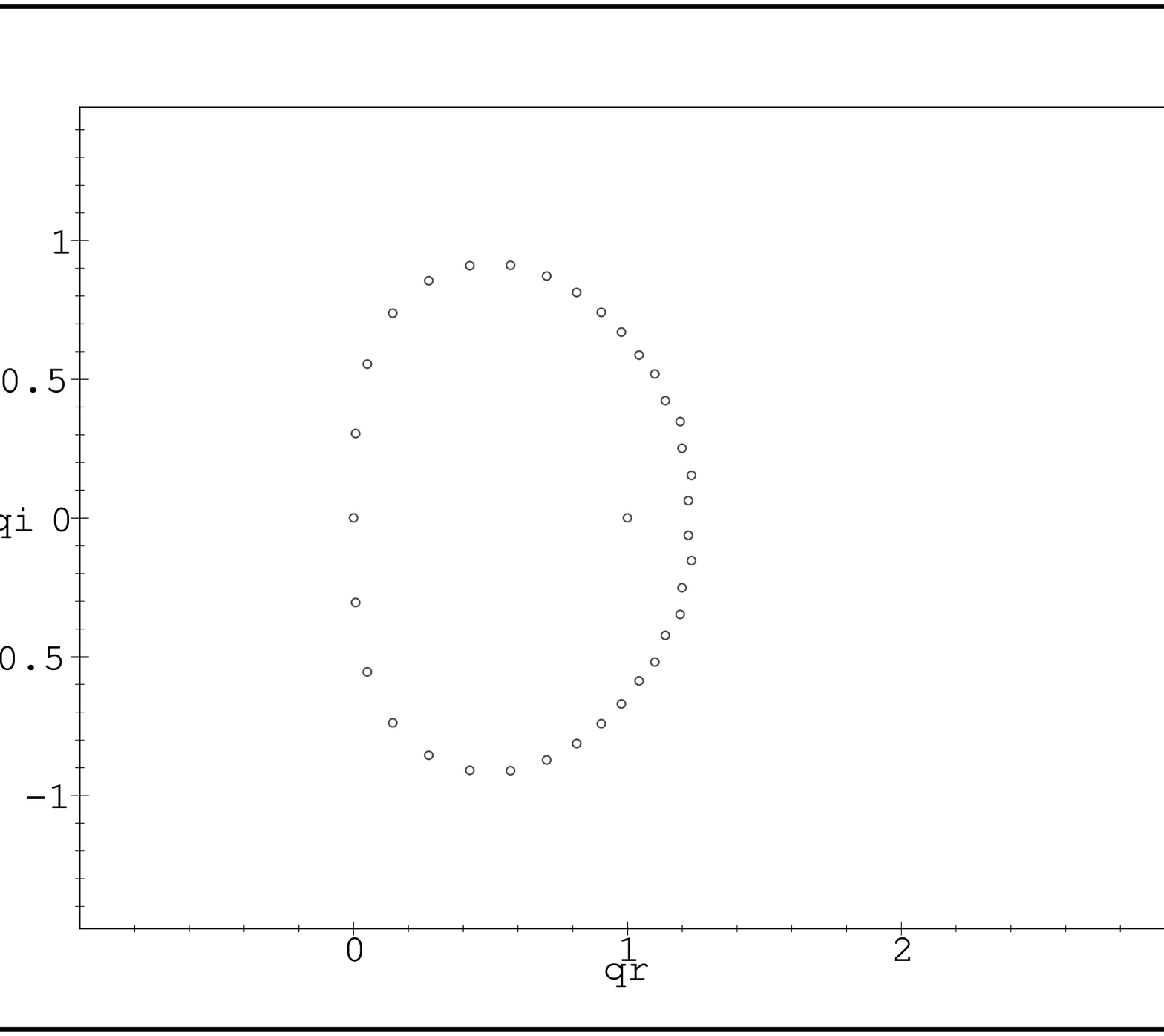}
\end{center}
\vspace{-2cm}
\caption{\footnotesize{Same as Fig. \ref{ladqa0p25} for $a=0.5$.}}
\label{ladqa0p5}
\end{figure}

\vspace{8mm}

\begin{figure}
\vspace{-4cm}
\centering
\leavevmode
\epsfxsize=4.0in
\begin{center}
\leavevmode
\epsffile{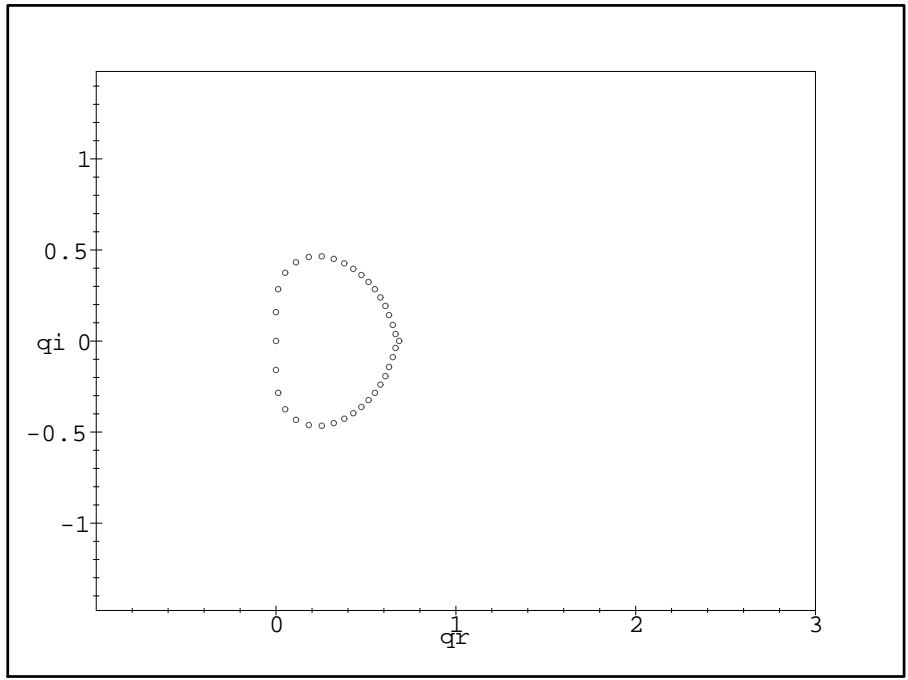}
\end{center}
\vspace{-2cm}
\caption{\footnotesize{Same as Fig. \ref{ladqa0p25} for $a=0.75$.}}
\label{ladqa0p75}
\end{figure}

\subsection{Ferromagnetic Case}

\begin{figure}
\vspace{-4cm}
\centering
\leavevmode
\epsfxsize=4.0in
\begin{center}
\leavevmode
\epsffile{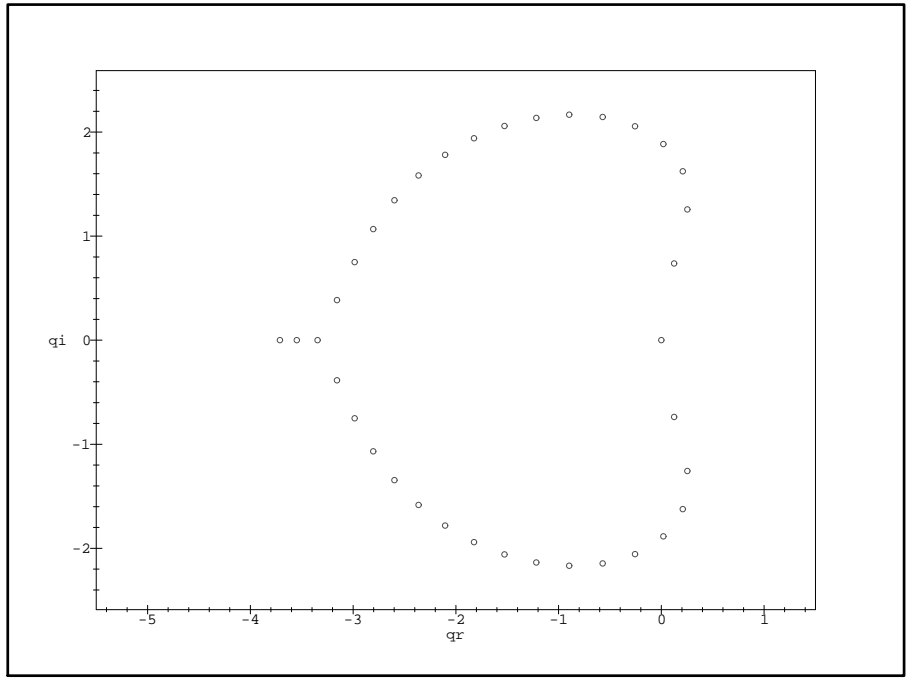}
\end{center}
\vspace{-2cm}
\caption{\footnotesize{Zeros of $Z(L_m,q,a)$ in $q$ for $a=2$ and $m=18$
($n=36$).}}
\label{ladqa2}
\end{figure}

In Fig. \ref{ladqa2} we show the zeros of $Z$ for a typical ferromagnetic
values, $a=2$.  We find the following general features of ${\cal B}_q$ for the
ferromagnetic Potts model for the full range of temperature.  The locus ${\cal
B}_q$ contains a heart-shaped figure and a finite line segment on the negative
real $q$ axis.  The line segment occurs because the expression in the square
roots in $\lambda_{L,5}$ and $\lambda_{L,6}$, given as $R_{S12}$ in
eq. (\ref{rs12}), is negative in an interval of the negative real axis,
yielding a pure imaginary square root so that, given that $T_{S12}$ is real,
$|\lambda_{L,5}|=|\lambda_{L,6}|$.  For example, for the case shown in
Fig. \ref{ladqa2}, $R_{S12} < 0$ for $-3.73 < q < -2.10$; within this interval,
$|\lambda_{L,5}|=|\lambda_{L,6}|$ are leading for $-3.73 < q < 3.35$, thereby
producing the line segment.  (For the remaining part of the interval, $-3.35 <
q < -2.10$, these eigenvalues have smaller magnitudes than $|\lambda_{L,3}|$
and hence do not determine the locus ${\cal B}_q$.)  The size of the
heart-shaped boundary increases as $a$ increases.  Since $q_c(\{L\})$, given in
eq. (\ref{qclad}), is negative, ${\cal B}_q$ does not intersect the positive
real $q$ axis. As was true for the Potts AF, in the region exterior to ${\cal
B}_q$ in the $q$ plane, the dominant $\lambda_j$ is $\lambda_{L,5}$, so that
the (reduced) free energy is 
\beq
f=\frac{1}{2}\ln \lambda_{L,5} = \frac{1}{2}\ln \lambda_{S,1} 
\label{fstrip}
\eeq
where $\lambda_{S,1}$ was given in eq. (\ref{lams}). 
For the range $q > 2$ where this system has acceptable physical behavior, the
above expression for the free energy holds for all (physical) temperatures. 
We shall discuss the thermodynamics further below. 
In the region interior to ${\cal B}_q$, $\lambda_{L,3}$
is dominant, so $|e^f|=|\lambda_{L,3}|^{1/2}$.

\vspace{8mm}

\vspace{8mm}

\subsection{${\cal B}_q$ for $a < 0$}

\begin{figure}
\vspace{-4cm}
\centering
\leavevmode
\epsfxsize=4.0in
\begin{center}
\leavevmode
\epsfxsize=4.0in
\epsffile{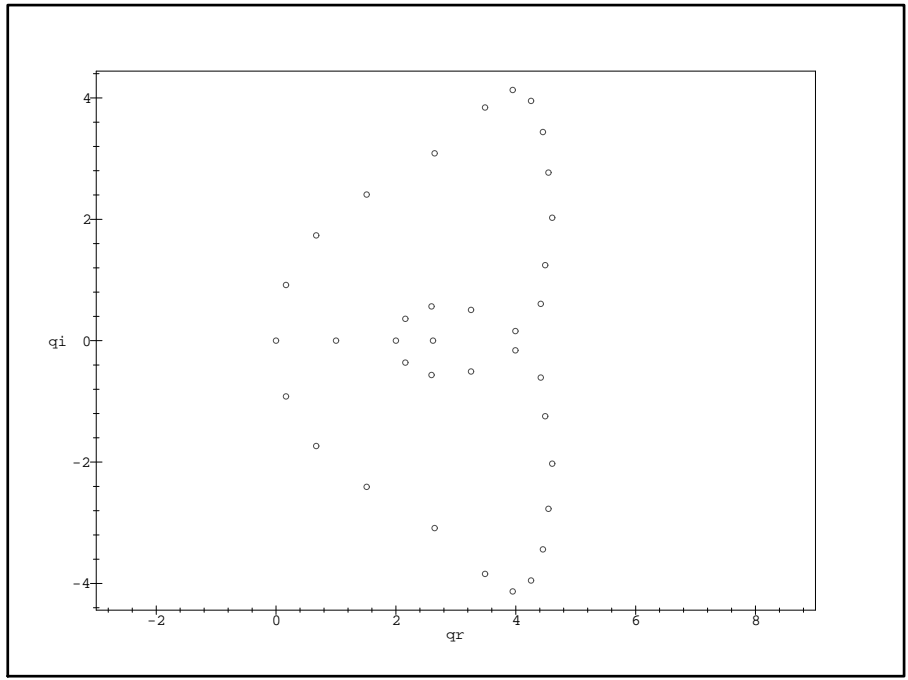}
\end{center}
\vspace{-2cm}
\caption{\footnotesize{Zeros of $Z(L_m,q,a)$ in $q$ for $a=-1$ and $m=18$.}}
\label{ladqam1}
\end{figure}

We briefly comment on ${\cal B}_q$ for negative real values of $a$, which 
correspond to complex temperature (as well as complex values of $a$ not 
considered here). For the interval
\beq
-\frac{1}{2} < a < 0
\label{int}
\eeq
the regions $R_2$ and $R_2^*$, which were separate, although contiguous at 
$q=q_c(\{L\})$, for $a$ in the interval $0 \le a < 1$, now merge to 
form one region, which we shall call $R_{22*}$ to indicate
this merger.  In this region, $\lambda_{L,2}$ is dominant.  The $R_{22*}-R_1$
boundary is determined by the degeneracy equation $|\lambda_{L,2}|=
|\lambda_{L,5}|$ and crosses the real axis at $q_c(\{L\})$.  The point
at which the $R_3-R_{22*}$ boundary crosses the real axis is 
determined by the relevant root of the threefold degeneracy equation 
$|\lambda_{L,3}|=|\lambda_{L,2}|=|\lambda_{L,5}|$, and is 
\beq
q_{b\ell}=2(1-a) \quad {\rm for} -\frac{1}{2} < a < 0 \ . 
\label{qbleft}
\eeq
The width of the merged $R_{22^*}$ region on the real axis is thus 
$(-a)(1-a)$ for this range of $a$. As $a$ decreases
in the interval (\ref{int}), $q_{b\ell}$ and $q_c$ both increase above 2.  When
$a$ decreases through the value $a=-1/2$, at which point $q_{b\ell}=9/4$, the
square root in $\lambda_{L,3}$ becomes complex. Viewed the other way,
solving eq. (\ref{qbleft}) for $a$ gives
\beq
a_{b\pm}=\frac{1}{2}\Bigl [ -1 \pm \sqrt{9-4q} \ \Bigr ] \ . 
\label{ableft}
\eeq
We are interested in the larger solution, $a_{b+}$.  When $q$ increases
through 9/4, corresponding to $a_{b+}=-1/2$, the square root becomes complex,
and there is no longer a real solution for $a_{b\pm}$.  The region diagram
changes qualitatively for $a < -1/2$.  The illustrative case $a=-1$ is shown
in Fig. \ref{ladqam1}.  Note that $Z(L_m,q,a=-1)$ has an overall factor 
$q(q-2)$. 
Here ${\cal B}_q$ crosses the real $q$ axis at $q=2$ and $q=q_c=4$ as 
well as at $q=0$.  The crossing at $q=4$ is a multiple point on the 
algebraic curve. 
Other interesting changes occur for larger negative values of
$a$, but we shall forgo discussing them to proceed to the physical range of
real $a \ge 1$.

\vspace{8mm}

\vspace{8mm}

\subsection{Thermodynamics of the Potts Model on the $L_y=2$ Strip} 

\subsubsection{$q \ge 2$} 

\begin{figure}
\vspace{-4cm}
\centering
\leavevmode
\epsfxsize=4.0in
\begin{center}
\leavevmode
\epsffile{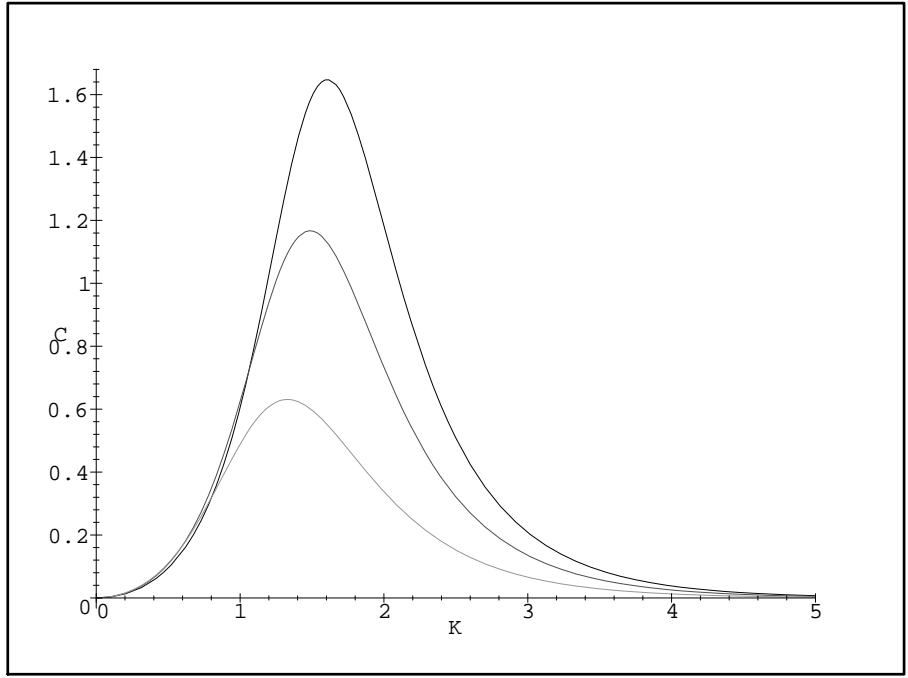}
\end{center}
\vspace{-2cm}
\caption{\footnotesize{Specific heat for the Potts ferromagnet on the
infinite-length, width $L_y=2$ strip (ladder) as a function
of $K=J/(k_BT)$.  Going from bottom to top in order of the heights of the
maxima, the curves are for $q=2,3,4$.}}
\label{cfmladder}
\end{figure}

\begin{figure}
\vspace{-4cm}
\centering
\leavevmode
\epsfxsize=4.0in
\begin{center}
\leavevmode
\epsffile{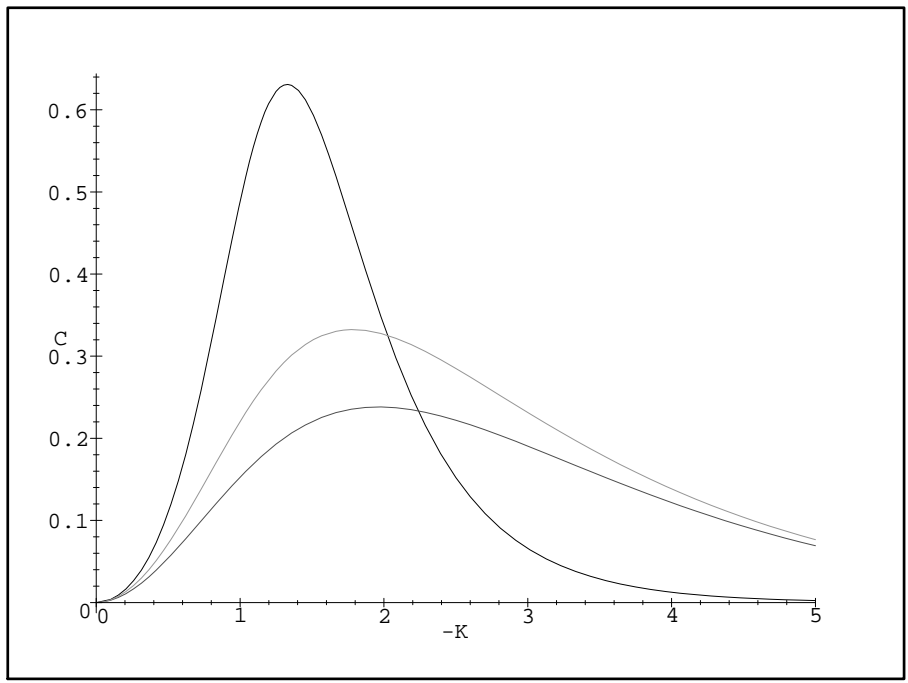}
\end{center}
\vspace{-2cm}
\caption{\footnotesize{Specific heat for the Potts antiferromagnet on the 
infinite-length, width $L_y=2$ strip (ladder) as a function of 
$-K = -J/(k_BT)$.  Going downward in order of the heights of the 
maxima, the curves are for $q=2,3,4$.}}
\label{cafmladder}
\end{figure}

In this section we first restrict to the range $q \ge 2$ where the Potts/random
cluster model has physical behavior for both the ferromagnetic and
antiferromagnetic cases, and then consider the behavior for $0 < q < 2$.  
For $q \ge 2$, the free energy is given for all temperatures by
(\ref{fstrip}).  It is straightforward to obtain the internal energy $U$ and
specific heat $C$ from this free energy; since the expressions are somewhat
complicated we do not list them here but instead concentrate on their high- and
low-temperature expansions and general features, as compared with those for the
$L_y=1$ case.  The high-temperature expansion of $U$ is
\beq
U=-\frac{3J}{2q}\biggl [ 1+\frac{(q-1)}{q}K + O(K^2) \biggr ] \ . 
\label{ustriphigh}
\eeq
The expression in brackets is the same as that for the $L_y=1$ strip up to and
including the $K^2$ term.  For the specific heat we have
\beq
C=\frac{3k_B(q-1)K^2}{2q^2}\biggl [ 1 + \frac{(q-2)}{q}K + O(K^2) \biggr ]
\label{cstriphigh}
\eeq
(here the order $K^2$ term differs from that for $L_y=1$.) 
The low-temperature expansions for the ferromagnet ($K \to \infty$) and
antiferromagnet ($K \to -\infty$) are 
\beq
U = J\Biggl [-\frac{3}{2}+(q-1)e^{-2K}\biggl [ 1 + 6e^{-K} + 7(q-1)e^{-2K} 
+ O(e^{-3K}) \biggr ] \Biggr ] \quad {\rm as} \quad K \to \infty
\label{ustriplowfm}
\eeq
and
\beq
U = \frac{(-J)e^K}{2(D_4)^2}\biggl [ t_1 + t_2 e^K + O(e^{2K}) \biggr ] 
\quad {\rm as} \quad K \to -\infty
\label{ustriplowafm}
\eeq
where $D_4=q^2-3q+3$ was given in (\ref{d4}) and 
\beq
t_1=(q-2)(3q^2-9q+8)
\label{t1}
\eeq
\beq
t_2=-\frac{(3q^6-42q^5+211q^4-532q^3+734q^2-534q+162)}{(D_4)^2} 
\label{t2}
\eeq
\beq
C = 2k_BK^2(q-1)e^{-2K}\biggl [1+9e^{-K} + 14(q-1)e^{-2K} + O(e^{-3K}) 
\biggr ] \quad {\rm as} \quad K \to \infty
\label{cstriplowfm}
\eeq
and
\beqs 
C = \frac{k_BK^2e^K}{2(D_4)^2}\biggl [ t_1 + 2t_2 e^K + O(e^{2K}) \biggr ] \ . 
\quad {\rm as} \quad K \to -\infty
\label{cstriplowafm}
\eeqs
Again, we observe that for the Ising case $q=2$, these expansions satisfy the
symmetry relations (\ref{ising_urel}) and (\ref{ising_crel}).  (In passing, we
mention the generalization of the first term in eq. (\ref{ustriplowfm}) 
to arbitrary $L_y$: in the $T=0$ limit of the Potts ferromagnet, 
\beq
U=-\frac{\Delta_{ave}J}{2} = -2\Bigl [ 1-\frac{1}{2L_y} \Bigr ]J \ . 
\label{ut0}
\eeq
We show plots of $C$ (with $k_B=1$) for the
ferromagnetic and antiferromagnetic Potts model on the $L_y=2$ strip (in the
$L_x \to \infty$ limit) in Figs. \ref{cfmladder} and \ref{cafmladder}.  As was
true for $L_y=1$, in the antiferromagnetic case, $C$ is a decreasing function
of $q$ for all finite temperature, while in the ferromagnetic case, $C$
increases (decreases) with $q$ at low (high) temperatures and has a maximum
that increases with $q$.  For a fixed $q$, by comparing the previous
plots of the specific heat on the line ($L_y=1$ case) with the corresponding 
plots for the $L_y=2$ strip, for the ferromagnet, and for the antiferromagnet,
one can see quantitatively how the behavior of this function changes as $L_y$
increases.  

\vspace{8mm}

\vspace{8mm}

\vspace{8mm}

For both the $L_y=1$ and $L_y=2$ strips, we observe that the exponential zero
in the specific heat as $T \to 0$ for both the ferromagnetic and
antiferromagnetic cases is $C \sim (q-1)K^2 e^{-L_y|K|}$.  For comparative
purposes we have also calculated the partition function, free energy, and these
thermodynamic functions for the Ising model on the strip with the next larger
width, $L_y=3$.  We find that the above dependence on $L_y$ is again exhibited,
namely $C \sim K^2e^{-3|K|}$.  

In view of the fact that the Potts ferromagnet has a zero-temperature critical
point for the infinite-length, finite-width strip graphs of the square lattice
(as does the Potts antiferromagnet in the $q=2$ case where it is equivalent to
the ferromagnet on these graphs), it is of interest to investigate the
dependence of the singularities in thermodynamic functions
on the strip width $L_y$.  As is typical for systems at their lower critical
dimensionality, these are essential singularities.  We have done this
comparative study above for the specific heat.  We next consider the 
(divergent) exponential singularities in the correlation length and 
the (uniform, zero-field) susceptibility (again, in the $q=2$ case, under the
replacement $J \to -J$ and uniform $\to$ staggered, this subsumes the
antiferromagnetic case).  Define
the ratio of the subleading eigenvalue divided by the leading eigenvalue of 
the transfer matrix for the strip graphs with periodic longitudinal 
boundary conditions considered here as $\rho_{L_y}$.  
Thus
\beq
\rho_1 = \frac{\lambda_{C,2}}{\lambda_{C,1}} = \frac{v}{q+v}
\label{rho1}
\eeq
and
\beq
\rho_2 = \frac{\lambda_{L,3}}{\lambda_{L,5}} \ . 
\label{rho2}
\eeq
We have also calculated $\rho_3$ for the Ising case, but since it is a rather
messy expression involving cube roots, we do not display it here. These ratios 
$\rho_{L_y}$ 
control the asymptotic decay of the spin-spin correlation function.  For
example, in the $n \to \infty$ limit, the spin-spin correlation function in 
the 1D case is given by $\langle \sigma_r \sigma_{r^\prime}\rangle \propto 
(\rho_1)^{|r-r^\prime|}$.  The correlation length can be written as 
\beq
\xi = - \frac{1}{\ln \rho_{L_y}} \ . 
\label{xiinv}
\eeq
For the 1D case, one knows that the correlation length has an 
exponential divergence as $T \to 0$: $\xi \sim q^{-1}e^K +O(1)$.  For 
the $L_y=2$ strips we find
\beq
\xi \sim q^{-1}e^{2K} + O(e^{K}) 
 \ , \quad {\rm as} \quad T \to 0 \quad {\rm for} \quad \{G\}=\{L\} \ .
\label{xily2}
\eeq
and for the Ising model on the width $L_y=3$ strip, we obtain 
$\xi \sim (1/2)e^{3K} + O(e^{2K})$.  These results show that the exponential
divergence in the correlation length is more rapid for larger width $L_y$  
and are consistent with an inference that 
$\xi \sim q^{-1}e^{L_yK} + O(e^{(L_y-1)K})$ as $T \to 0$.  The fact that the
correlation length diverges more rapidly as $L_y$ increases is easily explained
since this is due to the spin-spin interactions and the average effect of these
interactions, as determined by the average coordination number, 
$\Delta_{ave}$ in eq. (\ref{delta}), increases as $L_y$ increases. 

The zero-field susceptibility (per site) is well known for the 1D case: 
$\chi=\beta(1+\rho_1)/(1-\rho_1)$, which diverges as a function of $K$ like 
$\chi \sim Ke^K$ as $K \to \infty$.  Our results for $L_y=2,3$ support the
inference that $\chi \sim K e^{L_yK}$ as $K \to \infty$.  The more rapid
divergence in $\chi$ as $L_y$ increases can be explained in the same way as was
done for the correlation length.  

The inferred $L_y$ dependence of the divergences in the correlation length
$\xi$ and susceptibility $\chi$ at the zero-temperature critical point of the
Potts ferromagnet dramatically illustrate the fact that the thermodynamic
behavior of the model on this sequence of infinite-length, width $L_y$ strips
of the square lattice is quite different, even in the limit $L_y \to \infty$,
from the behavior of the model on the square lattice.  In the latter case, the
thermodynamic limit is $L_x \to \infty$, $L_y \to \infty$, with $\lim_{L_x \to
\infty} L_y/L_x$ equal to a finite nonzero constant.  For the strips, for any
$L_y$ no matter how large, the ferromagnet is critical only at $T=0$, and as $T
\to 0$ and $\xi \to \infty$, the strip acts as a one-dimensional system, since
$lim_{L_x \to \infty} L_y/L_x=0$.  In contrast, for the Potts model on the
square lattice, the phase transition occurs at finite temperature, at the known
value $K_c = \ln(1+\sqrt{q} \ )$.  These studies of the thermodynamic behavior
of the Potts model for general $q$ on $L_y \times \infty$ strips thus
complement studies such as those on the approach to the thermodynamic limit of
the Ising model on $L_x \times L_y$ rectangular regions, in which $L_x$ and
$L_y$ both get large with a fixed finite ratio $L_y/L_x$ \cite{ff}, and
finite-size scaling analyses \cite{bf}.  These differences are also evident in
the behavior of ${\cal B}_u$; we have inferred above that as $L_y \to \infty$,
there are an infinite number of curves on ${\cal B}_u$ that cross each other at
the ferromagnetic zero-temperature critical point, $u=0$, so that the Fisher
zeros become dense in the neighborhood of this point.  This is quite different
from the accumulation set of the Fisher zeros for the square lattice; although
this is known exactly only for the Ising case, the existence of low-temperature
expansions with a finite radius of convergence for the $q$-state Potts model is
equivalent to the statement that the singular locus ${\cal B}_u$ does not pass
through $u=0$.

In the case of the antiferromagnet,
as we have shown \cite{w2d}, for $q$ values that are only moderately above the
value of $q=3$ where the Potts antiferromagnet is critical on the square
lattice, the ground state entropy of infinite-length, finite-width strips
rapidly approaches its value for the square lattice.  For the ($L_x \to \infty$
limit of the) $L_y=2$ strip, this is given by $S_0 = (1/2)k_B\ln(q^2-3q+3)$,
which is nonzero for $q > 2$.  The analytic expressions for the $L_y=3,4$ cases
are given in \cite{w2d}.  This can be understood because the ground state
entropy is a disorder quantity and, for $q > 3$ is not associated with any
large correlation length.

\subsubsection{$0 < q < 2$: Phase Transition for Antiferromagnet}

For the range $0 < q < 2$, our result for $a_{c,+}$ in eq. (\ref{auclad}) 
shows that ${\cal B}$ crosses the positive real $a$ axis in the interval $0 < a
< 1$, so that the Potts/random cluster antiferromagnet has a 
finite-temperature phase transition, at the temperature
\beq
T_{L,p} = \frac{J}{k_B\ln \Bigl [ \frac{1}{2}\{-1 + \sqrt{9-4q} \ \} \Bigr ]} 
\ , \quad 0 < q < 2
\label{tlp}
\eeq
(where both $J$ and the log are negative, yielding a positive $T_{L,p}$).  For
$q=1$, it is understood that one takes $n \to \infty$ first and then $q \to 1$,
i.e., that one uses the free energy $f_{qn}$. As
$q$ decreases from 2 to 0, the phase transition temperature $T_{L,p}$ increases
from 0 to infinity.  In the high- and low-temperature phases, the free energy 
is given by eq. (\ref{fstrip}) and by $f=(1/2)\ln \lambda_{L,3}$, 
respectively.  These results may be compared with the temperature $T_p$ in
eq. (\ref{tp}) for the circuit graph.  The same comments that we made in that
case apply here; this result does not contradict the usual theorem that 1D (and
quasi-1D) spin systems with short-range interactions do not have any
finite-temperature phase transition because the phase transition here 
is intrinsically connected with
unphysical behavior of the model in the low-temperature phase, including
negative specific heat, negative partition function, and non-existence of
an $n \to \infty$ limit for thermodynamic functions that is independent of 
boundary conditions. Indeed, the last pathology is obvious from the fact that
for the $n \to \infty$ limit of the ladder graph with open longitudinal
boundary conditions, the free energy is given by eq. (\ref{fs}) for all
temperatures, the singular locus ${\cal B}$ does not cross the positive $a$
axis, and there is no such phase transition at finite temperature.  

Evidently, the temperature value at which the phase transition takes place in
the Potts/random cluster antiferromagnet on the infinite-length limits of both
the circuit graph and the cyclic and M\"obius $L_y=2$ strip graphs is
determined by the respective formulas relating $q_c$ to $a$, eqs. (\ref{qcc})
and (\ref{qclad}).  From the point of view of ${\cal B}_q$ in the $q$ plane, as
we have discussed, we find, as a general feature, that in the 
antiferromagnetic case, as one increases $T$ from 0 to infinity, the value
of $q_c(\{G\})$ for a given family $\{G\}$ decreases from its $T=0$ value to
the origin, $q=0$.  Correspondingly, for the
$n \to \infty$ limit of a given family $\{G\}$ with periodic (or twisted
periodic) longitudinal boundary conditions, the antiferromagnet will exhibit a
finite-temperature phase transition at a temperature $T_{\{G\},p}$ for the 
range $0 < q < q_c(\{G\})$:
\beq
\exists \ T_{\{G\},p} > 0 \quad {\rm for} \quad 0 < q < q_c(\{G\}) \ . 
\label{tpgeneral}
\eeq

Thus, for example, for the $L_y=3$ square strip of the square lattice with
cyclic or M\"obius boundary conditions, for which we determined ${\cal B}_q$
for the $T=0$ antiferromagnet \cite{wcy} and, in particular,
$q_c(sq,L_y=3,cyc.) \simeq 2.33654$, it follows that the random cluster
antiferromagnet has a finite-temperature phase transition for $0 < q <
q_c(sq,L_y=3,cyc.)$.  Just as we have discussed above, at special integer
values $q_s$ in the range $0 < q < q_c(\{G\})$, it is understood that one takes
the limit $n \to \infty$ first, and then $q \to q_s$ in calculating $f=f_{qn}$
and ${\cal B}_u = ({\cal B}_u)_{qn}$.  Similarly, on the
$L_y=2$ cyclic and M\"obius triangular lattice strips, where we found that
$q_c(tri,L_y,cyc.)=3$ for $L_y=2$ \cite{wcy} and $L_y=3$ \cite{t}, it follows
that the Potts/random cluster antiferromagnet has a finite-temperature phase
transition for $0 < q < 3$.  In all cases, however, this transition involves
unphysical aspects, among which is the non-existence of a unique $n \to \infty$
limit that is independent of boundary conditions.

\subsection{${\cal B}_u(\{L\})$ for $q > 4$}

We next proceed to the slices of ${\cal B}$ in the plane defined by the 
temperature Boltzmann variable $u$, for given values of $q$, starting with 
large $q$. In the limit $q \to \infty$, the locus ${\cal B}_u$ is
reduced to $\emptyset$.  This follows because for large $q$, there is only a
single dominant $\lambda_j$, namely 
\beq 
\lambda_{L,5} \sim q^2 + 3qv + O(1) \quad {\rm as } \quad q \to \infty \ . 
\label{lambdar1asymp}
\eeq
Note that in this case, one gets the 
same result whether one takes $q \to \infty$ first and then 
$n=2m \to \infty$, or $n \to \infty$ and then $q \to \infty$, so that these
limits commute as regards the determination of ${\cal B}_u$. 

\vspace{8mm}

\begin{figure}
\vspace{-4cm}
\centering
\leavevmode
\epsfxsize=4.0in
\begin{center}
\leavevmode
\epsffile{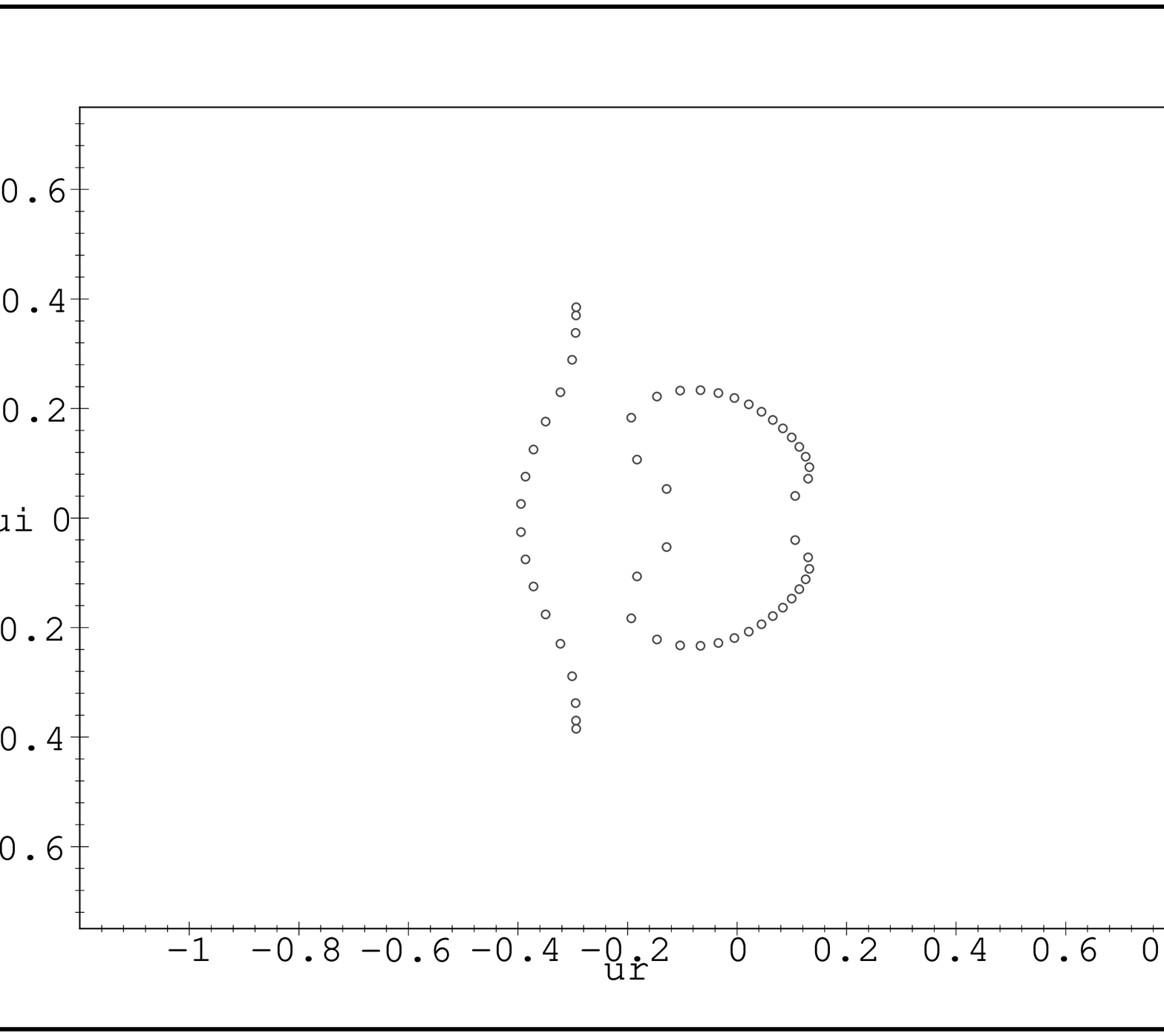}
\end{center}
\vspace{-2cm}
\caption{\footnotesize{Zeros of $Z(L_m,q,a)$ in the $u=1/a$ plane for
$q=10$ and $m=18$ ($n=36$).}}
\label{laduq10}
\end{figure}

\vspace{8mm}

\begin{figure}
\vspace{-4cm}
\centering
\leavevmode
\epsfxsize=4.0in
\begin{center}
\leavevmode
\epsffile{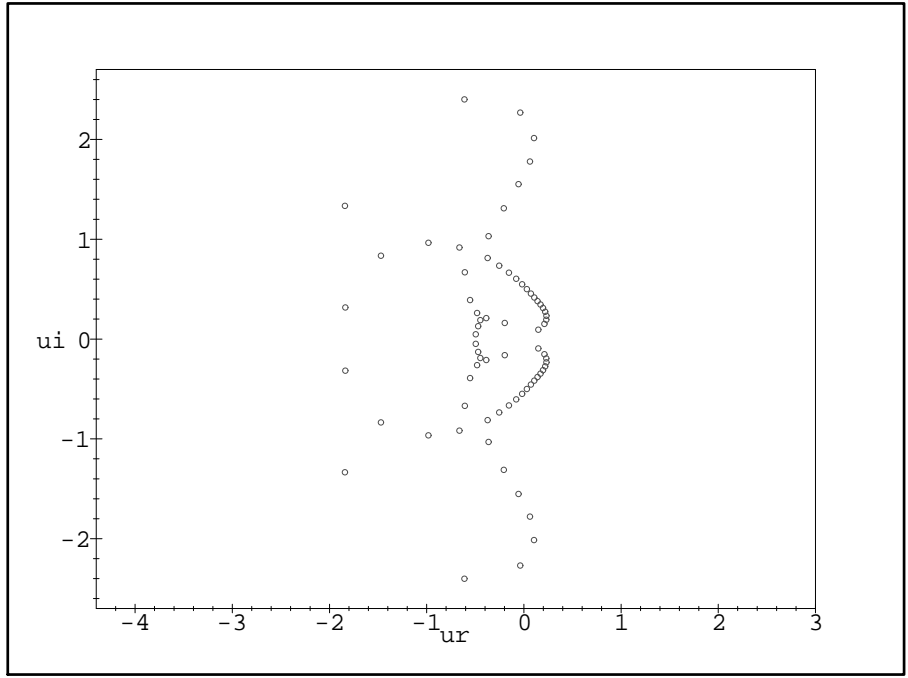}
\end{center}
\vspace{-2cm}
\caption{\footnotesize{Zeros of $Z(L_m,q,a)$ in the $u=1/a$ plane for 
$q=3$ and $m=24$ ($n=48$).}}
\label{laduq3}
\end{figure}

\vspace{8mm}

In the large--$q$ region we find that the locus ${\cal B}_u$ consists of two
complex-conjugate curves that pass through $u=0$ at the angles (\ref{thetau}),
hence intersecting at right angles and forming a distorted figure-8, together
with a separate self-conjugate arc.  Thus, ${\cal B}_u$ is comprised of two
disconnected parts.  For real $q$, the degeneracies in magnitude among leading
terms are $|\lambda_{L,3,u}|=|\lambda_{L,5,u}|$ at $u=0$ and
$|\lambda_{L,5,u}|=|\lambda_{L,6,u}|$ at the point on the negative real axis
where $T_{S12}=0$.  As a typical illustration of ${\cal B}_u$ for the large-$q$
region, we show the complex-temperature zeros for $q=10$ calculated for $m=18$,
i.e., $n=2m=36$, in Fig. \ref{laduq10}.  It is instructive to compare this plot
with the analogous plot for the open strip with $q=10$ given above in
Fig. \ref{sqffuq10}.  The self-conjugate arc is the same in the two plots,
crossing the real $u$ axis at $u \simeq -0.3954$, where $T_{S12}=0$ and having
endpoints at two of the zeros of the expression in the square root in
$\lambda_{L,j}$, $j=5,6$, which are identical to $\lambda_{S,j}$, $j=1,2$ in
eq. (\ref{lams}).  A notable
difference between the locus ${\cal B}_u$ for the cyclic or M\"obius ladder and
the analogous locus for the open square strip is that while the latter does not
separate the $u$ plane into different regions, the former does.  Specifically,
there are three regions: the physical PM phase occupying the interval $0 \le u
\le \infty$ and its maximal analytic continuation, together with
an $O_1$ phase in the interior of the upper curve and its complex-conjugate
phase $O_1^*$.  As is evident in the figure, the density of zeros on the curves
decreases strongly as $u$ approaches the multiple point at $u=0$. In the PM 
phase $\lambda_{L,5,u}$ is dominant, and so the reduced free energy is given by
\beq 
f = \frac{1}{2}\ln \lambda_{L,5} \quad {\rm for} \quad u \in PM \ .
\label{fpmlargeq}
\eeq
In the other phases only the magnitude $|e^f|$ can be determined unambiguously,
and, with appropriate choices of branch cuts, we have
\beq
|e^f| = |\lambda_{L,3}|^{1/2} \quad {\rm for} \quad u \in O_1, \ O_1^* \ . 
\label{fo1largeq}
\eeq
As $q$ increases, ${\cal B}_u$ contracts toward $u=0$, just as was true for the
open strip.  As $q$ decreases, the point at which the self-conjugate arc
crosses the negative real $u$ axis (i.e. where $T_{S12}=0$) moves toward the
left, and the  elongate toward the left.

\subsection{${\cal B}_u$ for $3$}

We next discuss the complex-temperature phase diagram for the case $q=3$.  An
important conclusion that we shall draw from our studies of ${\cal B}_u$ for
$q=3$ and $q=4$ (as well as the $q=2$ case), building on our earlier
comparative studies of complex-temperature phase diagrams for the 1D and 2D
Ising model with both spin 1/2 and higher spin $s$ \cite{is1d,hsi}, is that
although 1D and quasi-1D systems with short-range spin-spin interactions
(infinite-length circuit or cyclic or M\"obius strips) have qualitatively
different physical thermodynamic properties than the same systems in higher
dimensions, the complex-temperature phase diagrams of these 1D and quasi-1D
systems can give insight into the corresponding phase diagrams of the model on
lattices of dimensionality $d=2$.  Since no exact solution has been obtained
for the Potts model in $d \ge 2$ (except for the $d=2$, $q=2$ case), whereas we
have exact solutions on infinite-length, finite-width strips, this means that
one can use these as a tool to suggest properties of the complex-temperature
properties of the Potts model in 2D (and perhaps in $d > 2$).

The complex-temperature zeros of $Z$ in the variable $u$ are shown for 
$q=3$ in Fig. \ref{laduq3}.  In addition to the (CTE)PM phase,
which includes the intervals $u \ge 0$ and $u \le -2$ on the real $u$ axis and
the intervals $0.52 \le |Im(u)| \le 1.7$ and $|Im(u)| \ge 2.2$ on the imaginary
axis, and extends outward to the circle at infinity, one also has several O
phases.  Among these are an $O_1$ phase that contains the interval $0 \le
Im(u) \le 0.52$ and an $O_2$ phase which includes the interval $1.7 \le Im(u)
\le 2.2$, together with the complex conjugates of these phases, which are
denoted $O_1^*$ and $O_2^*$.  The dominant terms in these phases are:
$\lambda_{L,5}$ in PM; $\lambda_{L,3}$ in $O_1, \ O_1^*$; and
$\lambda_{L,2}$ in $O_2, \ O_2^*$.  Two phases that are self-conjugate and
include intervals of the real axis are the $O_3$ phase, containing the real
interval $-1/2 \le u \le 0$, in which $\lambda_{L,5}$ is dominant; and the
$O_4$ phase, containing the real interval $-2 \le u \le -1/2$, in which
$\lambda_{L,1}$ is dominant.  The point $u=-2$ here is the same as the point 
$u_c(q)$ in eq. (\ref{ucc}) for the infinite-length limit of the circuit
graph.  There are also phases that have no support on the
real axis.  The locus ${\cal B}_u$ has several multiple
points (in the technical terminology of algebraic geometry, meaning
points where several branches of an algebraic curve intersect).  Anticipating
our results for other values of $q$, we find that the point on the negative 
real $u$ axis where the PM phase, on the left, is contiguous with the $O_4$ 
phase, on the right, is given by the same $u_c$ as for the circuit graph, i.e.,
\beq
u_{PM-O_4}(\{L\}) = u_c(\{C\}) = -\frac{2}{q-2} \ . 
\label{upm04}
\eeq In a similar manner, we label the point on the negative real $u$ axis
where the $O_4$ phase, on the left, is contiguous with the $O_3$ phase, on the
right, as $u_{O_4-O_3}$; as indicated, this has the value $-1/2$ for $q=3$.
The degeneracies in magnitude between leading terms $\lambda_{L,j,u}$ at these
multiple points are as follows (with appropriate conventions for branch cuts in
square roots): 
\beq 
|\lambda_{L,3,u}|=|\lambda_{L,5,u}| \quad {\rm at} \quad u=0
\label{u0deg}
\eeq
\beq
|\lambda_{L,1,u}|=|\lambda_{L,5,u}| \quad {\rm at} \quad u=u_{O_4-O_3}
\label{urdeg}
\eeq
\beq
|\lambda_{L,1,u}|=|\lambda_{L,2,u}|=|\lambda_{L,5,u}| \quad {\rm at} \quad 
u=u_{PM-O_4}
\label{uelldeg}
\eeq
\beq
|\lambda_{L,1,u}|=|\lambda_{L,3,u}|=|\lambda_{L,5,u}| \quad {\rm at} \quad 
u \simeq -0.44 \pm 0.22i
\label{ucompint}
\eeq
\beq
|\lambda_{L,j,u}| \quad {\rm all \ equal} \quad {\rm at} \quad 
u=e^{\pm 2i\pi/3}
\ . 
\label{ue}
\eeq
Note that these points $u_e$ are the same as for the open square strip 
discussed above. 

Motivated by our previous work \cite{is1d,hsi}, we next explore relations
between the exactly determined complex-temperature phase diagram for the Potts
model on these strip graphs and on the square lattice.  For $q=3$ and higher
integers, where the 2D $q$-state Potts model is not exactly solved, the 
complex-temperature phase diagram and associated Fisher zeros for various 
lattices have been studied in a number of works (e.g. \cite{mr}-\cite{kch},
\cite{ssbounds}).  One exact result concerning the complex-temperature phase 
boundary ${\cal B}_a$ for the $q=3$ Potts model on the square lattice is that,
as a result of the duality property (\ref{zdual}) and the fact that the 
square lattice is self-dual, ${\cal B}_a$ maps to itself when one replaces 
$a$ with $a_d$ given by (\ref{adual}); hence the fact that the $q=3$ Potts 
antiferromagnet on the square lattice has a zero-temperature critical point
\cite{lieb}, so that $a=0$ is on
${\cal B}_a$, means that the dual of this point, namely $a=-2$, is also on
${\cal B}_a$, or equivalently, $u=-1/2$ is on ${\cal B}_u$  \cite{pfef}. 
Our exact results for the 1D case with periodic boundary conditions,
eq. (\ref{ucircle}) and for the infinite-length cyclic or M\"obius ladder 
have the same feature, viz., that ${\cal B}_u$ contains the point $u=-1/2$:
\beq 
{\cal B}_u \ni u=-1/2 \quad {\rm for} \quad \{C\}, \ \{L\}  \quad 
{\rm and} \quad \Lambda_{sq} \quad {\rm with} \quad q=3
\label{buqmp5}
\eeq
where $\Lambda_{sq}$ denotes the square lattice. 

This interesting similarity of a feature of the complex-temperature phase
boundary ${\cal B}_u$ leads us to investigate whether there are also other such
connections.  Our results suggest that the point $u=-2$ where for $q=3$ the
general formula (\ref{upm04}) shows that ${\cal B}_u(\{G\})$ crosses the 
negative real axis for $\{G\}=\{C\}$ and $\{L\}$, and the points 
$u=e^{\pm 2 i \pi/3}$ in eq. (\ref{ue}) where degeneracies in $|\lambda_j|$'s
and associated multiple points on ${\cal B}(\{L\})$ occur, have analogues for
${\cal B}_u$ in the $q=3$ Potts model on the square lattice.  
These conjectures could, in principle, be tested by calculations of Fisher
zeros for $q=3$ on finite square lattices, and these have been done; however, 
the considerable scatter of the zeros in the $Re(a) < 0$ region
\cite{mr,mm,mbook,chw,pfef} renders it difficult to test the conjectures 
at present.  One could also calculate the Potts model free energy and the
boundary ${\cal B}_u$ on infinite-length strips of greater width, $L_y \ge 3$
and check to see if for $q=3$, the points $u=-2$ and $u=e^{\pm 2 i \pi/3}$ are
on the resultant locus ${\cal B}_u$.  For the analogous multiple points at 
$u=\pm i$ on ${\cal B}_u$ for the $q=2$ Ising model we have done this (see
below) and have found that these points are on this locus not only for $L_y=1$
and 2 but also for $L_y=3$, just as they are for the full 2D square lattice.

Let us comment further on the correspondences of special complex-temperature
points for the present strip and for the square lattice.  There are close
relations with the proposed formulas $(a-1)^2 = q$ for $q \ge 2$ and $(a+1)^2 =
4-q$ for $2 \le q \le 3$ \cite{baxter82} for the Potts ferromagnet and
antiferromagnet, respectively.  The second formula, $(a+1)^2=4-q$ 
(generalized to consider complex as well as physical temperatures) has the
solutions $a=-1 \pm \sqrt{4-q}$ (which are duals of each other under the map $a
\to a_d = 1 + q/(a-1)$.  For $q=3$, these are 0 and $-2$.  For $q=4$, the two
solutions coincide at $a=u=-1$.  Both of these complex-temperature points,
$u=a^{-1}=-2$ for $q=3$ and $u=-1$ for $q=4$, agree with eq. (\ref{upm04}) for
the infinite cyclic or M\"obius square strip.

One of the earliest conjectures for the complex-temperature phase boundary
${\cal B}_a$ of the $q=3$ Potts model on the square lattice was that it
consisted of the square-lattice Potts model consists of the two circles
$|a-1|=\sqrt{q}$ and $|a+1|=\sqrt{4-q}$ \cite{mr}.  From later studies of
Fisher zeros, it was concluded that the ${\cal B}_a$ was not this simple 
\cite{mm,mbook,chw,pfef}. This was further established combining the
calculations of Fisher zeros with the analysis of 
low-temperature series expansions \cite{pfef,p}, which showed the existence of
complex-temperature singularities not on this locus, which were associated with
prongs or cusps formed by the zeros.  Nevertheless, if, indeed, the points
$a,a^* = e^{\pm 2\pi i/3}$ are on ${\cal B}_a$, (equivalently, ${\cal B}_u$
since the set of these points is the same under inversion) for the $q=3$
square-lattice Potts model, as might be inferred from our exact results on the
strips considered here, then this makes a very interesting connection with the
old conjecture of Ref. \cite{mr}, since the two circles are $|a-1|=\sqrt{3}$
and $|a+1|=1$ for $q=3$, and these intersect precisely in the two points $a,a^*
= e^{\pm 2\pi i/3}$. We recall that if one uses self-dual boundary conditions,
then one finds that the Fisher zeros lie nicely on the circle $|\zeta|=1$,
where $\zeta=(a-1)/\sqrt{q}$, at least for $Re(\zeta) > 0$ \cite{chw} and,
moreover, in the $q \to \infty$ limit, the complex-temperature phase boundary
is $|\zeta|=1$, where $\zeta=(a-1)/\sqrt{q}$ \cite{wuetal}.

\subsection{${\cal B}_u$ for $3 < q \le 4$}

As $q$ increases in the real interval $3 \le q \le 4$, the $O_4$ phase
contracts, as can be seen from the fact that the point $u_{PM-O_4}$ moves to
the right, from $-2$ to $-1$, while the right-hand
boundary at the point $u_{O_4-O_3}$ moves to the left, from $-1/2$ to $-1$; 
thus, at $q=4$, these coincide:
\beq
u_{PM-O_4}=u_{O_4-O_3}=-1 \quad {\rm for} \quad q=4 \ . 
\label{ucoincide}
\eeq 
In this interval $3 \le q < 4$, the degeneracies in magnitude of leading
terms $|\lambda_{L,j}|$ at $u_{PM-O_4}$ in eq. (\ref{uelldeg}) and at
$u_{O_4-O_3}$ in eq. (\ref{urdeg}) continue to hold.  For $q=4$, all
$|\lambda_{L,j}|$ are equal at $u=-1$.  The confluence of $u_{PM-O_4}$ and
$u_{O_4-O_3}$ at $-1$ and the equality of all $|\lambda_{L,j}|$'s at this point
reflect a special role for the value $q=4$ for the complex-temperature phase
diagram.  (However, for the physical thermodynamics of these 1D and quasi-1D
systems, there is no qualitative change in the nature of the singularity at the
zero-temperature critical point of the Potts ferromagnet at this value $q=4$.)
It may be recalled that the value $q=4$ is also special, albeit in a different
way, for the 2D Potts model in that for physical values of $q$ below 4 the 
Potts ferromagnet has a second-order phase transition while for $q \ge 5$ this 
transition is first order.

\subsection{ ${\cal B}_u$ for $2 < q < 3$} 

We next discuss the complex-temperature phase diagram as $q$ decreases through
real values from 3 to 2.  From the point of view of this phase diagram, the
limit $q \to 2$ is singular, since ${\cal B}_u$ is compact if $q \ne 2$
but is noncompact for $q=2$, passing through $1/u=0$.   As $q$
decreases through this range, the point $u_{PM-O_4}$ moves leftward,
approaching $-\infty$ as $q \to 2^+$.  The point $u_{O_4-O_3}$ 
at which ${\cal B}_u$ crosses the
negative real $u$ axis separating the $O_4$ phase on the left from the $O_3$
phase on the right moves toward the right, from $-1/2$ to $-0.453398..$ 
(a root of the cubic equation $u^3+2u+1=0$) as $q$ decreases from 3 to 2.  
In Fig. \ref{laduq2} we show a plot of complex-temperature zeros for $q=2.5$.
One can observe how the intersection points which occurred at 
$u=e^{\pm 2 i \pi/3}$ at $q=3$ have shifted outward from the real axis and 
toward the right.  When $q$ decreases all the way to 2, these intersection 
points reach $\pm i$ (see below).  For $q=2.5$, the point 
$u_{PM-O_4}=-4$, while the crossing at $u=u_{O_4-O_3}$ is clearly visible, near
to its $q=2$ limiting location at $u \simeq -0.4534$. 

\vspace{8mm}

\begin{figure}
\vspace{-4cm}
\centering
\leavevmode
\epsfxsize=4.0in
\begin{center}
\leavevmode
\epsffile{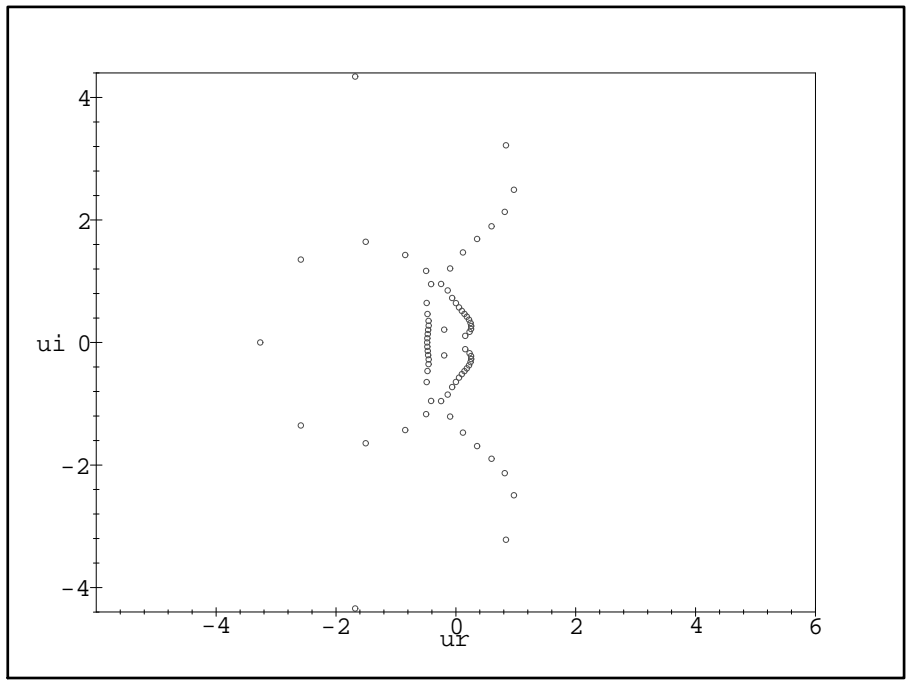}
\end{center}
\vspace{-2cm}
\caption{\footnotesize{Zeros of $Z(L_m,q,a)$ in the $u=1/a$ plane for
$q=2.5$ and $m=24$ (i.e., $n=48$).}}
\label{laduq2p5}
\end{figure}

\vspace{8mm}

\vspace{8mm}
\begin{figure}
\vspace{-4cm}
\centering
\leavevmode
\epsfxsize=4.0in
\begin{center}
\leavevmode
\epsffile{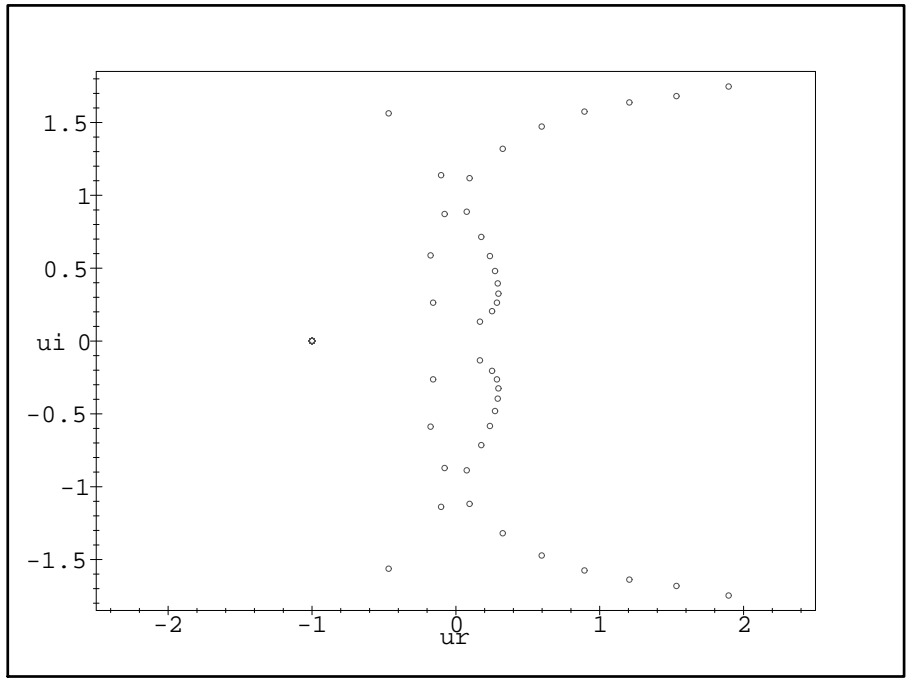}
\end{center}
\vspace{-2cm}
\caption{\footnotesize{Same as Fig. \ref{laduq2p5} for $q=2$.}}
\label{laduq2}
\end{figure}

\vspace{8mm}

\subsection{ ${\cal B}_u$ for $q=2$}

Here we again encounter noncommutativity in the definition of the free energy
and the locus ${\cal B}_u$.  We first discuss $f_{nq}$ and $({\cal B}_u)_{nq}$,
obtained by setting $q=2$ and then taking $n \to \infty$.  For $q=2$, besides 
the $q$-independent $\lambda_{L,1}=(a-1)^2$, we have 
\beq
\lambda_{L,2}=(a-1)(a+1)
\label{lam2q2}
\eeq
\beq
\lambda_{L,3}=a(a-1)(a+1)
\label{lam3q2}
\eeq
\beq
\lambda_{L,4}=(a-1)^2
\label{lam4q2}
\eeq
\beq
\lambda_{L,(5,6)}=\frac{1}{2}(a+1)\biggl [a^2+1 \pm \Bigl ( 
a^4-4a^3+10a^2-4a+1 \Bigr )^{1/2} \ \biggr ]
\label{lam56q2}
\eeq
so that for this value of $q$, $\lambda_{L,1}=\lambda_{L,4}$.  Further, 
$c_{L,1}=-c_{L,4}=-1$ so that the $(\lambda_{L,1})^m$ and 
$(\lambda_{L,4})^m$ terms cancel each other in $Z$, which reduces to 
\beq
Z(L_m,q=2,a)=\sum_{j=2,3,5,6} (\lambda_{L,j})^m \ . 
\label{zladq2}
\eeq
Since each of the $\lambda_j$'s contributing to $Z(L_m,q=2,a)$ has an 
$(a+1)$ factor, it follows that 
\beq
Z(L_m,q=2,a)=2(a+1)^m \times polyn.
\label{zladq2fac}
\eeq
The result (\ref{zladq2fac}) implies that $Z(L,q=2,a=-1,m)=0$, which is implied
more generally by our earlier Theorem 6 of Ref. \cite{cmo}.  This theorem
states that for lattices with odd coordination number, the (zero-field)
partition function of the Ising model vanishes at $a=-1$.  This point may or
may not be on $({\cal B}_u)_{nq}$; for example, for the honeycomb lattice, it 
is, while for the Archimedean 3-12 lattice it is not \cite{cmo}. 
In the present case we will find that it is not on $({\cal B}_u)_{nq}$. 

In Fig. \ref{laduq2} we show a plot of Fisher zeros for the $q=2$ case.
Because of the general fact that $({\cal B}_u)_{nq}$ passes through $u=0$ and 
the special inversion symmetry (\ref{zbipq2}), (\ref{bq2sym})
that holds for $q=2$, it follows that $({\cal B}_u)_{nq}$
passes through $a=0$ also.  The complex-temperature phase diagram consists of
six phases, bounded by the locus ${\cal B}_u$ which is the continuous
accumulation set of the Fisher zeros.  The phase diagram in the $u$ plane
consists of (i) the complex-temperature extension of the ${\mathbb
Z}_2$-symmetric, paramagnetic phase, PM, including the real axis $u > 0$,
together with five O phases: (ii) $O_1$, including the interval on the
imaginary $u$ axis from $u=0$ to $u=i$, and its complex-conjugate, (iii)
$O_1^*$; (iv) $O_2$, including the interval on the imaginary axis from $u=i$ to
$u=i\infty$, and (v) its complex conjugate, $O_2^*$; and (vi) $O_3$, including
the negative real axis, $u \le 0$.  The PM and $O_3$ phases map to
themselves under the inversion $u\to 1/u=a$, while $O_1 \leftrightarrow O_2^*$
and $O_2 \leftrightarrow O_1^*$ under this inversion.   As in our previous work
\cite{chisq,cmo} we shall henceforth suppress the qualifier (CTE) and refer 
to the PM phase simply as the PM phase.  In this phase, 
$\lambda_{L,5}$ is the dominant term in
(\ref{zlad}) so that reduced free energy is given by 
$f = (1/2)\ln \lambda_{L,5}$ in the PM phase, as in (\ref{fpmlargeq}). 
In other regions that are not analytically connected with the PM region, 
only $|e^f|$ can be determined unambiguously:
\beq
|e^f| = |\lambda_{L,5}|^{1/2} \quad {\rm for} \quad u \in O_3, O_3^*
\label{flado3}
\eeq
\beq
|e^f| = |\lambda_{L,3}|^{1/2} \quad {\rm for} \quad u \in O_1, O_1^*
\label{flado1}
\eeq
and
\beq
|e^f| = |\lambda_{L,2}|^{1/2} \quad {\rm for} \quad u \in O_2, O_2^* \ . 
\label{flado2}
\eeq
where here $f=f_{nq}$. 
The locus ${\cal B}_u$ has multiple points (in the algebraic geometry sense) 
at $u=0$ and $u=\pm i$ and at complex infinity, i.e., at $a=1/u=0$
corresponding to the zero-temperature ferromagnetic and antiferromagnetic Ising
critical points.  We have also calculated $Z(G,q,v)$ exactly for the Ising
model on the cyclic $L_y=3$ strip and have found that ${\cal B}_u$ again has
multiple points at $u=\pm i$ (as well as at $u=0$ and $1/u=0$).  A possible 
inference would be that this is true for cyclic or M\"obius 
strips of the square graph for all finite values of $L_y \ge 2$.  Given our
exact results for $L_y=2,3$, we see once again a very interesting connection
with the complex-temperature phase diagram of the same model -- in this case,
the Ising model -- on the two-dimensional square lattice, for which ${\cal
B}_a$ consists of the Fisher circles $|u \pm 1|=\sqrt{2}$, which intersect
precisely in the points $u=\pm i$.  
 
If one takes $n \to \infty$ first and then $q \to 2$, the resulting locus
$({\cal B}_u)_{qn}$ differs from $({\cal B}_u)_{nq}$ in several respects.
First, $({\cal B}_u)_{qn}$ does not satisfy the inversion symmetry
(\ref{bq2sym}).  Second, while $\lambda_{L,5,u}$ is dominant on the negative
$u$ axis in the vicinity of the origin $u=0$, $\lambda_{L,1,u}$ (equal, in the
$q \to 2$ limit, to $\lambda_{L,4,u}$) becomes dominant for $u < -0.454$ and
similarly for radial paths emanating outward from the origin in the upper and
lower $Re(u) < 0$ half-plane.  This gives rise to another region boundary.
(Recall that the contributions of these $\lambda$'s cancelled if one took $q=2$
first, so that they did not affect $Z$ or the locus $({\cal B})_{nq}$ in the $n
\to \infty$ limit.)  The absence of the inversion symmetry in $({\cal
B}_u)_{qn}$ is clear since a region boundary on this locus passes through $u
\simeq -0.454$ but not the inverse of this point.  This is, then, an example of
the noncommutativity $({\cal B}_u)_{qn} \ne ({\cal B}_u)_{nq}$ for a case where
both of these loci are nontrivial.  Finally, one can also discuss ${\cal B}_u$
for negative real $q$ and for complex $q$, but we shall forgo this.

\section{Summary and Conclusions}

In summary, we have calculated exact closed-form expressions for the Potts
model/random cluster partition function for general $q$ and temperature $T$, or
equivalently, the Whitney/Tutte polynomial for the open, cyclic, and M\"obius
square strips (ladder graphs) of width $L_y=2$ and arbitrary length
$L_x$. Taking the limit $L_x \to \infty$, we have determined the free energy
$f$ (and $|e^f|$ in unphysical phases) and the continuous locus ${\cal B}$
where the free energy is singular, which arises as the continuous accumulation
set of the partition function zeros in the ${\mathbb C}^2$ space of the
variables $q$ and $u$.  The divergences in the correlation length and
susceptibility of the Potts ferromagnet at its zero-temperature critical point
were shown to be more rapidly approached as the strip width increases, and the
physical reason for this was given.  Our comparison of different strip widths
suggests the inference that for infinite-length, width $L_y$ cyclic (or
M\"obius) strip graphs, as $L_y \to \infty$, an infinite number of curves on
${\cal B}_u$ pass through the point $u=0$ and the Fisher zeros become dense in
the neighborhood of this point.  It was shown that the Potts/random cluster
antiferromagnet on both the infinite-length circuit graph and ladder graph with
cyclic or M\"obius boundary conditions exhibits a phase transition at finite
temperature if $0 < q < 2$, but with unphysical properties.  We discussed a
subtlety in the definition of the free energy of the random cluster model due
to the noncommutativity at certain special values $q_s$: $\lim_{n \to \infty}
\lim_{q \to q_s} Z^{1/n} \ne \lim_{q \to q_s} \lim_{n \to \infty} Z^{1/n}$.
Several generalizations of results for the $T=0$ limit of the Potts
antiferromagnet (chromatic polynomials) were presented.  Among these is the
general form (\ref{zgsum}) for the partition function of recursive strip graphs
(and its further generalization to the case of nonzero external field,
(\ref{zghsum})).  The analysis of \cite{bkw} for the singular locus ${\cal
B}_q$ in the case of chromatic polynomials was generalized to the present case
of the full temperature-dependent Potts partition function.  The dependence of
the locus ${\cal B}$ as a function of the longitudinal boundary conditions was
studied, and it was shown that this locus is the same for the strips considered
here with cyclic and twisted cyclic (M\"obius) longitudinal boundary
conditions.  For the Potts antiferromagnet, it was found that as the
temperature increases from 0 to infinity, the singular locus ${\cal B}_q$
contracts in to the origin, $q=0$.  For the Potts ferromagnet, this locus was
found not to cross the positive $q$ axis, in contrast to the antiferromagnetic
case, where, for the cyclic strip, it does. In both the antiferromagnet and
ferromagnet cases, for the strips studied here, ${\cal B}_q$ passes (does not
pass) through $q=0$ if one uses periodic (free) longitudinal boundary
conditions.  Generalizing our previous result for chromatic polynomials, we
found that for the strips with periodic longitudinal boundary conditions,
${\cal B}_q$ encloses regions in the $q$ plane for values of $a$ where it is
nontrivial (i.e., $a \ne 1$).  Several advantages of periodic, as opposed to
free, longitudinal boundary conditions were noted, including the fact that with
such periodic longitudinal boundary conditions, the locus ${\cal B}_u$ passes
through $u=0$, in 1-1 correspondence with the zero-temperature critical point
of the Potts ferromagnet.  Finally, certain properties of the
complex-temperature phase diagrams and loci ${\cal B}_u$ for these
infinite-length, finite-width strips were shown to be the same as known
properties of the model on the square lattice, including the multiple points at
$u=\pm i$ for $q=2$, which also occur in the exactly solved square-lattice
Ising model and the point $u=-1/2$ for $q=3$, which is known to lie on ${\cal
B}_u$ for the square lattice since it is dual to the zero-temperature critical
point at $a=0$.  This shows that exact solutions on infinite-length strips
could provide a way of generating plausible conjectures for complex-temperature
properties of the Potts model on two-dimensional lattices and some conjectures
were made.

\vspace{12mm}

I would like to thank Prof. N. L. Biggs for kindly sending me a copy of
\cite{sands} and Prof.  F. Y. Wu for discussions, particularly on spanning
trees, and for hospitality during a visit to the National Center for
Theoretical Science (NCTS) and Academia Sinica, Taiwan, when some of this
research was performed.  I have also benefited from recent collaborations with
H. Kluepfel and S.-C. Chang \cite{ks,t} and thank A. Sokal and J. Salas for
informing me about their work.  The present research was supported in part at
Stony Brook by the NSF grant PHY-97-22101 and at Brookhaven by the U.S. DOE
contract DE-AC02-98CH10886.\footnote{\footnotesize{Accordingly, the
U.S. government retains a non-exclusive royalty-free license to publish or
reproduce the published form of this contribution or to allow others to do so
for U.S. government purposes.}}

\vspace{6mm}

\section{Appendix}

\subsection{Connection Between Potts Model Partition Function and Tutte 
Polynomial}

The Potts model partition function $Z(G,q,v)$ is related to the Tutte
polynomial $T(G,x,y)$ as follows.  
The graph $G$ has vertex set $V$ and edge set $E$,
denoted $G=(V,E)$.  A spanning subgraph $G^\prime$ is defined as a subgraph
that has the same vertex set and a subset of the edge set:
$G^\prime=(V,E^\prime)$ with $E^\prime \subseteq E$.  The Tutte polynomial
of $G$, $T(G,x,y)$, is then given by \cite{tutte1}-\cite{tutte3} 
\beq
T(G,x,y)=\sum_{G^\prime \subseteq G} (x-1)^{k(G^\prime)-k(G)}
(y-1)^{c(G^\prime)}
\label{tuttepol}
\eeq
where $k(G^\prime)$, $e(G^\prime)$, and $n(G^\prime)=n(G)$ denote the number 
of components, edges, and vertices of $G^\prime$, and
\beq
c(G^\prime) = e(G^\prime)+k(G^\prime)-n(G^\prime)
\label{ceq}
\eeq
is the number of independent circuits in $G^\prime$ (sometimes called the
co-rank of $G^\prime$).   Note that the first factor can also be written as 
$(x-1)^{r(G)-r(G^\prime)}$, where
\beq
r(G) = n(G)-k(G)
\label{rank}
\eeq
is called the rank of $G$.  The graphs $G$ that we consider here are 
connected, so that $k(G)=1$.  Now let 
\beq
x=1+\frac{q}{v}
\label{xdef}
\eeq
and
\beq
y=a=v+1
\label{ydef}
\eeq
so that $q=(x-1)(y-1)=(x-1)v$. Then 
\bigskip
\beq
Z(G,q,v)=(x-1)^{k(G)}(y-1)^{n(G)}T(G,x,y) \ . 
\label{ztutte}
\eeq
There is also a connection with the Whitney rank polynomial, 
$R(G,\xi,\eta)$, defined as \cite{whit,bbook}
\beq
R(G,\xi,\eta)=\sum_{G^\prime \subseteq G}\xi^{r(G^\prime)}\eta^{c(G^\prime)}
\label{whitney}
\eeq
where the sum is again over spanning subgraphs $G^\prime$ of $G$. Then 
\beq
T(G,x,y)=(x-1)^{r(G)}R(G,\xi=(x-1)^{-1},\eta=y-1)
\label{tr}
\eeq
and 
\beq
Z(G,q,v)=q^{n(G)} R(G,\xi=\frac{v}{q},\eta=v) \ . 
\label{zwhit}
\eeq
Note that the chromatic polynomial is a special case of the Tutte polynomial:
\beq
P(G,q)=q^{k(G)}(-1)^{k(G)+n(G)}T(G,x=1-q,y=0)
\label{tprel}
\eeq
(recall eq. (\ref{zp})). 

 From the representation (\ref{ztutte}) and the duality property of the Tutte
 polynomial 
\beq
T(G,x,y) = T(G^*,y,x)
\label{tuttedual}
\eeq
where $G^*$ is the dual graph corresponding to $G$, it follows that 
\beq
Z(G,q,v) = v^{e(G)}q^{-c(G)} Z(G^*,q,v_d)
\label{zdual}
\eeq
where $G^*$ denotes the graph that is dual to $G$ and $v_d$ is the dual image
of $v$:
\beq
v_d = \frac{q}{v} 
\label{vdual}
\eeq
or equivalently, in terms of the variable $a$, 
\beq
a_d = \frac{a-1+q}{a-1} \ . 
\label{adual}
\eeq

Corresponding to the form (\ref{zgsum}) we find that the Tutte polynomial for
recursively defined graphs comprised of $m$ repetitions of some subgraph has
the form 
\beq
T(G_m,x,y) = \sum_{j=1}^{N_\lambda} c_{T,G,j}(\lambda_{T,G,j})^m
\label{tgsum}
\eeq

\subsection{ Square Strip with Free Longitudinal Boundary Conditions}

The generating function representation for the Tutte polynomial for the open 
square strip $S_m$ is
\beq
\Gamma_T(S_m,x,y;z) = \sum_{m=0}^\infty T(S_m,x,y)z^m \ . 
\label{gammatfbc}
\eeq
We have 
\beq
\Gamma_T(S,x,y;z) = \frac{{\cal N}_T(S,x,y;z)}{{\cal D}_T(S,x,y;z)}
\label{gammas}
\eeq
where
\beq
{\cal N}_T(S,x,y;z)=A_{T,S,0}+A_{T,S,1}z = (y+x+x^2+x^3)-yx^3z
\label{numts}
\eeq
and
\beqs
{\cal D}_T(S,x,y,z) & = & 1-(y+1+x+x^2)z+yx^2z^2 \cr\cr
                    & = & \prod_{j=1}^2 (1-\lambda_{T,S,j}z)
\label{dents}
\eeqs
with
\beq
\lambda_{T,S,(1,2)} = \frac{1}{2}\biggl [ (1+y+x+x^2) \pm \Bigl ( 
 y^2 +2y(1+x-x^2) + (x^2+x+1)^2 \Bigr )^{1/2} \biggr ] \ . 
\label{lams12}
\eeq
The corresponding closed-form expression is given by the general formula from
\cite{hs}, as applied to Tutte, rather than chromatic, polynomials, namely
\beq
T(S_m,x,y)=\biggl [ 
\frac{A_{T,S,0}\lambda_{T,S,1}+A_{T,S,1}}{\lambda_{T,S,1}-\lambda_{T,S,2}} 
\biggr ] (\lambda_{T,S,1})^m + \biggl [ 
\frac{A_{T,S,0}\lambda_{T,S,2}+A_{T,S,1}}
{\lambda_{T,S,2}-\lambda_{T,S,1}} \biggr ] (\lambda_{T,S,2})^m \ . 
\label{tssumform}
\eeq

An alternative expression for $T$ that explicitly shows that it is a symmetric
function of the $\lambda_{S,j}$, $j=1,2$, is
\beqs
& & T(S_m,x,y) = \frac{1}{2}(y+x+x^2+x^3)\Bigl [ (\lambda_{T,S,1})^m+
(\lambda_{T,S,2})^m \Bigr ] + \cr\cr & & 
\frac{1}{2}\Bigl [y^2+y+2yx+2yx^2+x+2x^2+3x^3+2x^4-yx^3+x^5 \Bigr ] 
\Biggl [ \frac{(\lambda_{T,S,1})^m-(\lambda_{T,S,2})^m}{\lambda_{T,S,1} -
\lambda_{T,S,2}} \Biggr ] \ . 
\label{tssymform}
\eeqs

\subsection{Cyclic and M\"obius Square Strips}

We write the Tutte polynomials for the cyclic and M\"obius square strips $L_m$
and $ML_m$ as 
\beq
T(L_m,x,y) = \sum_{j=1}^6 c_{T,L,j}(\lambda_{T,L,j})^m
\label{tlxy}
\eeq
and 
\beq
T(ML_m,x,y) = \sum_{j=1}^6 c_{T,ML,j}(\lambda_{T,ML,j})^m
\label{tmbxy}
\eeq
where it is convenient to extract a common factor from the coefficients:
\beq
c_{T,G,j} \equiv \frac{\bar c_{T,G,j}}{x-1} \ , \quad G = L, ML \ . 
\label{cbar}
\eeq
Of course, although the individual terms contributing
to the Tutte polynomial are thus rational functions of $x$ rather than
polynomials in $x$, the full Tutte polynomial is a polynomial
in both $x$ and $y$.  We have 
\beq
\lambda_{T,ML,j}=\lambda_{T,L,j} \ , \quad j=1,...,6
\label{lamtutlmb}
\eeq
\beq
\lambda_{T,L,1} = 1 
\label{lamtutt1}
\eeq
\beq
\lambda_{T,L,2} = x
\label{lamtutt2}
\eeq
\beq
\lambda_{T,L,(3,4)} = \frac{1}{2}\biggl [x+y+2 \pm \Bigl ( (x-y)^2 + 
4(x+y+1) \Bigr )^{1/2} \ \biggr ]
\label{lamtutt34}
\eeq
and
\beq
\lambda_{T,L,5} = \lambda_{T,S,1} \ , \quad \lambda_{T,L,6} = \lambda_{T,S,2} 
\ . 
\label{lamtutt56}
\eeq
Our result for $T(G,x,y)$, $G=L,ML$ agrees with a recursion relation given 
in \cite{bds} (see also \cite{sands}).  
\beq
\bar c_{T,L,1} = [(x-1)(y-1)]^2-3(x-1)(y-1)+1
\label{c1tutt}
\eeq
\beq
\bar c_{T,L,2} = \bar c_{T,L,3} = \bar c_{T,L,4} = xy-x-y
\label{c234tutt}
\eeq
\beq
\bar c_{T,L,5} = \bar c_{T,L,6} = 1
\label{c56tutt}
\eeq
\beq
\bar c_{T,ML,1}=-1
\label{c1mbtutt}
\eeq
\beq
\bar c_{T,ML,2}=-xy+x+y
\label{c2mbtutt}
\eeq
\beq
\bar c_{T,ML,3}= \bar c_{ML,4}=xy-x-y
\label{c34mbtutt}
\eeq
\beq
\bar c_{T,ML,5}= \bar c_{T,ML,6}=1 \ . 
\label{c56mbtutt}
\eeq
We note that $\lambda_{T,L,3}\lambda_{T,L,4}=xy$ and 
$\lambda_{T,L,5}\lambda_{T,L,6}=x^2y$. 

\subsection{Special Values of Tutte Polynomials for Square Strips}

For a given graph $G=(V,E)$, at certain special values of the arguments $x$ and
$y$, the Tutte polynomial $T(G,x,y)$ yields quantities of basic graph-theoretic
interest \cite{tutte5}-\cite{boll}, cite{wu77}-\cite{tzengwu}.  We
recall some definitions: a spanning subgraph $G^\prime=(V,E^\prime)$ of $G$
is a graph with the same vertex set $V$ and a subset of the edge set,
$E^\prime \subseteq E$.  Furthermore, a tree is a graph with no cycles, and a 
forest is a graph containing one or more trees.  Then the number of spanning
trees of $G$, $N_{ST}(G)$, is
\beq
N_{ST}(G)=T(G,1,1) \ , 
\label{t11}
\eeq
the number of spanning forests of $G$, $N_{SF}(G)$, is 
\beq
N_{SF}(G)=T(G,2,1) \ , 
\label{t21}
\eeq
the number of connected spanning subgraphs of $G$, $N_{CSSG}(G)$, is 
\beq
N_{CSSG}(G)=T(G,1,2) \ , 
\label{t12}
\eeq
and the number of spanning subgraphs of $G$, $N_{SSG}(G)$, is 
\beq
N_{SSG}(G)=T(G,2,2) \ . 
\label{t22}
\eeq
Clearly, 
$N_{SSG}(G)-N_{CSSG}(G)$ is the number of disconnected spanning subgraphs of
$G$ and $N_{CSSG}(G)-N_{ST}(G)$ is the number of connected spanning subgraphs
of $G$ that contain one or more cycles.  One thus has the inequality
\beq
N_{SSG}(G) \ge N_{CSSG}(G) \ge N_{ST}(G) \quad i.e., \quad 
T(G,2,2) \ge T(G,1,2) \ge T(G,1,1) \ . 
\label{tineq1}
\eeq
Also, clearly 
\beq
N_{SSG}(G) \ge N_{SF}(G) \quad i.e., \quad T(G,2,2) \ge T(G,2,1)
\label{tineq2}
\eeq
and 
\beq
N_{SF}(G) \ge N_{ST}(G) \quad i.e., \quad T(G,2,1) \ge T(G,1,1) \ . 
\label{tineq3}
\eeq
The set of spanning forests differs from the set of connected spanning
subgraphs by the removal of the condition that the subgraph is connected but
the imposition of the condition that the subgraph have no cycles, and hence 
there is no general inequality between $N_{SF}(G)$ and $N_{CSSG}(G)$. 

We recall the results for tree graphs $T_n$ and
circuit graphs $C_n$: $T(T_n,1,1)=T(T_n,1,2)=1$,
$T(T_n,2,1)=T(T_n,2,2)=2^{n-1}$, $T(C_n,1,1)=n$, $T(C_n,2,1)=2^n-1$, 
$T(C_n,1,2)=n+1$, and $T(C_n,2,2)=2^n$.  

For the open square strip $S_m$ we find 
\beq
N_{ST}(S_m)=
2\Bigl [ (2+\sqrt{3} \ )^m + (2-\sqrt{3} \ )^m \Bigr ] + 
\frac{7}{2\sqrt{3}}\Bigl [ (2+\sqrt{3} \ )^m - (2-\sqrt{3} \ )^m \Bigr ]
\label{ts11}
\eeq
\beq
N_{SF}(S_m)=\frac{15}{2}\biggl [ (2(2+\sqrt{3}))^m + (2(2-\sqrt{3}))^m 
\biggr ] + \frac{13}{\sqrt{3}}\biggl [ (2(2+\sqrt{3}))^m - 
 (2(2-\sqrt{3}))^m \biggr ] 
\label{ts21}
\eeq
\beq
N_{CSSG}(S_m)=\frac{5}{2}\biggl [ \biggl (\frac{5+\sqrt{17}}{2} \ 
\biggr )^m
+ \biggl (\frac{5-\sqrt{17}}{2} \ \biggr )^m \biggr ]
+\frac{21}{2\sqrt{17}}\biggl [ \biggl (\frac{5+\sqrt{17}}{2} \ \biggr )^m - 
\biggl (\frac{5-\sqrt{17}}{2} \ \biggr )^m \biggr ]
\label{ts12}
\eeq
and 
\beq
N_{SSG}(S_m)=2^{3m+4} \ . 
\label{ts22}
\eeq
That eqs. (\ref{ts11})-(\ref{ts12}) yield integers follows from the theorem 
on symmetric polynomial functions of roots of an algebraic equation, as 
discussed in \cite{pm}. 

With the definition 
\beq
\eta_G = \cases{+1 & if $G=L$ \cr
              -1 & if $G=ML$ \cr} 
\label{eta}
\eeq
our calculations of the Tutte polynomials for the cyclic strip 
$L_m$ and the M\"obius strip $ML_m$ yield 
\beq
N_{ST}(G_m) = m \biggl \{ -\eta_G + \frac{1}{2}\biggl [ (2+\sqrt{3})^m + 
(2-\sqrt{3})^m \biggr ] \biggr \} \ , \quad G_m=L_m, \ ML_m
\label{nstlm}
\eeq
\beqs
N_{SF}(G_m) & = & \eta_G(1-2^m) - \biggl [ 
\biggl ( \frac{5+\sqrt{17}}{2} \ \biggr )^m 
+ \biggl ( \frac{5-\sqrt{17}}{2} \ \biggr )^m \biggr ] \cr\cr
& + & \Bigl (2(2+\sqrt{3})\Bigr )^m
+ \Bigl (2(2-\sqrt{3})\Bigr )^m \ , \quad G_m=L_m, \ ML_m
\label{nsfl}
\eeqs
\beqs
N_{CSSG}(L_m) & = & N_{CSSG}(ML_m)-2m-1 = -(m+2) + \frac{(m+2)}{2}\biggl [  
\biggl ( \frac{5+\sqrt{17}}{2} \ \biggr )^m 
+ \biggl ( \frac{5-\sqrt{17}}{2} \ \biggr )^m \biggr ] \cr\cr
& & 
-\frac{m}{2\sqrt{17}}\biggl [ \biggl ( \frac{5+\sqrt{17}}{2} \ \biggr )^m
- \biggl ( \frac{5-\sqrt{17}}{2} \ \biggr )^m \biggr ] 
\label{ncssgl}
\eeqs
and 
\beq
N_{SSG}(L_m) = N_{SSG}(ML_m) = 2^{3m} \ . 
\label{nssgl}
\eeq
The results for the spanning trees for the cyclic and M\"obius strips are known
\cite{fhspan} (for higher-dimensions, see \cite{tzengwu}); we are not aware
of the other quantities having been published. 

Several comments are in order. 
Since $T(G_m,x,y)$ grows exponentially as $m
\to \infty$ for the families $G_m=S_m$, $L_m$, and $ML_m$ for $(x,y)=(1,1)$, 
(2,1), (1,2), and (2,2) (as well as for $G_m=C_m$ and $T_m$ for $(x,y)=(2,1)$
and (2,2)), it is natural to define corresponding constants
\beq
z_{set}(\{G\}) = \lim_{n(g) \to \infty} n(G)^{-1} \ln N_x(G) \ , \quad 
set = ST, \ SF, \ CSSG, \ SSG
\label{zset}
\eeq
where, as above, the symbol $\{G\}$ denotes the limit of the graph family $G$
as $n(G) \to \infty$ (and the $z$ here should not be confused with the 
auxiliary expansion 
variable in the generating function (\ref{gammatfbc}) or the Potts partition 
function $Z(G,q,v)$.)  The general inequalities (\ref{tineq1}), 
(\ref{tineq2}), and (\ref{tineq3}) imply that, for a given $\{G\}$, 
\beq
z_{SSG} \ge z_{CSSG} \ge z_{ST}
\label{zineq1}
\eeq
\beq
z_{SSG} \ge z_{SF}
\label{zineq2}
\eeq
and
\beq
z_{SF} \ge z_{ST} \ . 
\label{aineq3}
\eeq

We find that for both the line $(L_y=1$) and the $L_y=2$ square strip, the 
quantity $z_{set}(\{G\})$ is independent of whether the longitudinal boundary 
conditions are free, periodic, or M\"obius:
\beq
z_{ST}(\{G\}) = \frac{1}{2}\ln(2+\sqrt{3}) \simeq 0.658479 \quad {\rm for} 
\quad G=S,L,ML
\label{zst}
\eeq
\beq
z_{SF}(\{G\}) = \frac{1}{2}\ln[2(2+\sqrt{3} \ )] \simeq 1.00505 \quad 
{\rm for} \quad G=S,L,ML
\label{zsf}
\eeq
\beq
z_{CSSG}(\{G\}) = \frac{1}{2}\ln \Biggl ( \frac{5+\sqrt{17}}{2} \Biggr ) 
\simeq 0.758832 \quad {\rm for} \quad G=S,L,ML
\label{tl12asymp}
\eeq
and
\beq
z_{SSG}(\{G\}) = \frac{3}{2}\ln 2 \simeq 1.03972 \quad {\rm for} \quad G=S,L,ML
\label{tl22asymp}
\eeq
(where the result for $z_{ST}$ can be extracted from \cite{fhspan}).

\subsection{Tutte Polynomials for Dual Graphs}

Since the Tutte polynomial satisfies the duality relation (\ref{tuttedual}),
our calculations of the Tutte polynomials $T(G_m,x,y)$ for the open, cyclic,
and M\"obius square strips, $G_m=S_m$, $L_m$, and $ML_m$, also yield the
corresponding results for the duals of these graphs.  The dual of the square
$L_y=2$ strip with $m+1$ squares, i.e., length $L_x=m+1$ edges, $(S_m)^*$,
can be described as follows: for $m \ge 1$, consider a line of $m+1$ vertices,
with successive vertices $v_i$ and $v_{i+1}$ connected to each other by an 
edge $e_i$; the vertices
on this line are connected to a single external vertex by double edges, except
for the first and last vertices on the line, each of which is connected to the
external vertex via three edges.  This is $(S_m)^*$ for $m \ge 1$.  Note
that this is a multigraph, since a (proper) graph is normally defined not to
have loops or multiple edges.  For the case $m=0$, i.e., a single square, the
dual is $(S_1)^*=TL_4$, where in the mathematical literature, $TL_\ell$,
denoted ``thick link'', is the multigraph consisting of two vertices connected
by $\ell$ edges.  Our results for $T(S_m,x,y)$ in eq. (\ref{tssumform}) thus
give the Tutte polynomial for this dual graph as 
$T((S_m)^*,x,y) = T(S_m,y,x)$. 

For the dual of the cyclic strip graph, $(L_m)^*$, we recall a definition from
graph theory: given two graphs $G$ and $H$, the ``join'' $G+H$ is the graph
obtained by connected each vertex of $G$ to each vertex of $H$ with edges.  We
also recall the notation $\bar K_p$ for the complement of $K_p$, i.e. the 
graph consisting of $p$ vertices with no edges.  Then for $m \ge 3$,
\beq
(L_m)^* = \bar K_2 \ + \ C_m
\label{lmdual}
\eeq
where $C_m$ is the circuit graph with $m$ vertices.   Hence our results for 
$T(L_m,x,y)$ in eqs. (\ref{zlad}) with (\ref{lam1})-(\ref{c56lad}) also 
determine $T((L_m)^*,x,y) = T(L_m,y,x)$.

\vfill
\eject


\begin{thebibliography}{99}

\bibitem{potts}{R. B. Potts, Proc. Camb. Phil. Soc. {\bf 48} 106 (1952).}

\bibitem{wurev}{F. Y. Wu, Rev. Mod. Phys. {\bf 54} (1982) 235.} 

\bibitem{birk}{G. D. Birkhoff, Ann. of Math. {\bf 14} (1912) 42.}

\bibitem{whit}{H. Whitney, Ann. of Math. {\bf 33} (1932) 688.}

\bibitem{bl}{G. D. Birkhoff and D. Lewis, Trans. Amer. Math. Soc. {\bf 60} 
(1946) 355.} 

\bibitem{kf}{P. W. Kasteleyn and C. M. Fortuin, J. Phys. Soc. Jpn. {\bf 26}
(1969) (Suppl.) 11; C. M. Fortuin and P. W. Kasteleyn, Physica {\bf 57} (1972)
536.}

\bibitem{tutte1}{W. T. Tutte, Proc. Cam. Phil. Soc. {\bf 43} (1947) 26.}

\bibitem{tutte2}{W. T. Tutte, Can. J. Math. {\bf 6} (1954) 80.}

\bibitem{tutte3}{W. T. Tutte, J. Combin. Theory {\bf 2} (1967) 301.}

\bibitem{tutte4}{W. T. Tutte, ``Chromials'', in Lecture Notes in Math. v. 411
(1974) 243.}

\bibitem{tutte5}{W. T. Tutte, {\it Graph Theory}, vol. 21 of Encyclopedia of
Mathematics and Applications (Addison-Wesley, Menlo Park, 1984).} 

\bibitem{bbook}{N. L. Biggs, {\it Algebraic Graph Theory} (2nd ed., Cambridge 
Univ. Press, Cambridge, 1993).}

\bibitem{welsh}{D. J. A. Welsh, {\it Complexity: Knots, Colourings, and
Counting}, London Math. Soc. Lect. Note Ser. 186 (Cambridge University Press,
Cambridge, 1993).}

\bibitem{boll}{B. Bollob\'as, {\it Modern Graph Theory} (Springer, New
York, 1998).}

\bibitem{onsager}{L. Onsager, Phys. Rev. {\bf 65} (1944) 117.}

\bibitem{cft}{See, e.g., J. Cardy, in C. Domb and J. L. Lebowitz, eds., {\it
Phase Transitions and Critical Phenomena} (Academic Press, New York, 1987),
vol. 11, p. 55; C. Itzykson, H. Saleur, and J.-B. Zuber, {\it Conformal
Invariance and Applications to Statistical Mechanics} (World Scientific,
Singapore, 1988); P. Di Francesco, P. Mathieu, and D. S\'en\'echal, {\it
Conformal Field Theory} (Springer, New York, 1997), and references therein.}

\bibitem{lieb}{E. H. Lieb, Phys. Rev. {\bf 162}, 162 (1967).}

\bibitem{al}{M. Aizenman and E. H. Lieb, J. Stat. Phys. {\bf 24} (1981) 279;
Y. Chow and F. Y. Wu, Phys. Rev. {\bf B36} (1987) 285.} 

\bibitem{rrev}{R. C. Read, J. Combin. Theory {\bf 4} (1968) 52.}

\bibitem{rtrev}{R. C. Read and W. T. Tutte, ``Chromatic Polynomials'',
in {\it Selected Topics in Graph Theory, 3}, eds. L. W. Beineke and
R. J. Wilson (Academic Press, New York, 1988.).}

\bibitem{w}{R. Shrock and S.-H. Tsai, Phys. Rev. {\bf E55} (1997) 5165.}

\bibitem{bds}{N. L. Biggs, R. M. Damerell, and D. A. Sands, J. Combin. Theory 
B {\bf 12} (1972) 123.}

\bibitem{sands}{Sands, D. A., Ph.D. Thesis, Univ. of London, 1972
(unpublished).}

\bibitem{bm}{N. L. Biggs and G. H. Meredith, J. Combin. Theory B{\bf 20} 
(1976) 5; N. L. Biggs, Bull. London Math. Soc. {\bf 9} (1976) 54.}

\bibitem{bkw}{S. Beraha, J. Kahane, and N. Weiss, J. Combin. Theory B
{\bf 27} (1979) 1; {\it ibid.} {\bf 28} (1980) 52.}

\bibitem{readcarib}{R. C. Read, in Proc. 3rd Caribbean Conf. on Combin. and
Computing (1981); Proc. 5th Caribbean Conf. on Combin. and Computing (1988).}

\bibitem{baxter}{R. J. Baxter, J. Phys. A {\bf 20} (1987) 5241.}

\bibitem{read91}{R. C. Read and G. F. Royle, in {\it Graph Theory,
Combinatorics, and Applications} (Wiley, NY, 1991), vol. 2, p. 1009.}

\bibitem{rw}{R. C. Read and E. G. Whitehead, Discrete Math. {\bf 204} (1999)
337 and unpublished reports.}

\bibitem{wc}{R. Shrock and S.-H. Tsai, Phys. Rev. {\bf E56} (1997) 1342,
2733, 3935, 4111.}

\bibitem{w2d}{R. Shrock and S.-H. Tsai, Phys. Rev. {\bf E58} (1998) 4332; 
cond-mat/9808057.}

\bibitem{strip}{M. Ro\v{c}ek, R. Shrock, and S.-H. Tsai, Physica 
{\bf A252} (1998) 505.}

\bibitem{strip2}{M. Ro\v{c}ek, R. Shrock, and S.-H. Tsai, Physica {\bf A259}
(1998) 367.}

\bibitem{hs}{R. Shrock and S.-H. Tsai, Physica {\bf A259} (1998) 315.}

\bibitem{wa3}{R. Shrock and S.-H. Tsai, J. Phys. A {\bf 31} (1998) 9641;
Physica {\bf A265} (1999) 186.}

\bibitem{pg}{R. Shrock and S.-H. Tsai, J. Phys. A Lett. {\bf 32} (1999) L195.}

\bibitem{wcy}{R. Shrock and S.-H. Tsai, Phys. Rev. {\bf E60} (1999) 3512; 
Physica A {\bf 275} (1999) 429.}

\bibitem{sokal}{A. Sokal, Combin. Prob. Comput., in press (cond-mat/9904146); 
cond-mat/9910503.}

\bibitem{nec}{R. Shrock and S.-H. Tsai, J. Phys. {\bf 32} (1999) 5053.} 

\bibitem{matmeth}{N. L. Biggs, LSE report LSE-CDAM-99-03 (May 1999), to
appear.}

\bibitem{pm}{R. Shrock, Phys. Lett. {\bf A261} (1999) 57.}

\bibitem{tk}{N. L. Biggs and R. Shrock, J. Phys. A (Letts) {\bf 32}, L489 
(1999).} 

\bibitem{bcc}{R. Shrock, in the {\it Proceedings of the 1999 British 
Combinatorial Conference, BCC99} (July, 1999), Discrete Math., to appear.}

\bibitem{tw}{R. Shrock, in the Proceedings of the Taiwan Conference on
Equilibrium and Non-Equilibrium Phase Transitions (Academia Sinica, Taipei,
Aug. 1999).}

\bibitem{ks}{H. Kluepfel and R. Shrock, YITP-99-32,33; H. Kluepfel, Stony Brook
thesis (July, 1999).} 

\bibitem{t}{S.-C. Chang and R. Shrock, YITP-SB-99-50,58.} 

\bibitem{ss}{J. Salas and A. Sokal, work in progress.} 

\bibitem{zerofree}{e.g. G. Berman and W. T. Tutte, J. Combin. Theory {\bf 6}
(1969) 301; D. Woodall, Discrete Math. {\bf 101} (1992) 333; B. Jackson,
Combin. Prob. Comput. {\bf 2} (1993) 325; F. Brenti, G. Royle, and D. Wagner, 
Canad. J. Math. {\bf 46} (1994) 55; V. Thomassen,
Combin. Prob. Comput. {\bf 6} (1997) 497; J. Brown, J. Brown, J. Combin.
Theory {\bf 6} B (1998) 251.}

\bibitem{ipz}{C. Itzykson, R. B. Pearson, and J.-B. Zuber, Nucl. Phys. 
{\bf B220} (1983) 415.} 

\bibitem{ih}{V. Matveev and R. Shrock, J. Phys. A {\bf 28} (1995) 4859;
Phys. Rev. {\bf E53} (1996) 254; Phys. Lett. {\bf A215} (1996) 271.}

\bibitem{ly}{T. D. Lee and C. N. Yang, Phys. Rev. {\bf 87} (1952) 410; 
C. N. Yang and T. D. Lee, {\it ibid}, {\bf 87} (1952) 404.}

\bibitem{fisher}{M. E. Fisher, {\it Lectures in Theoretical Physics}
(Univ. of Colorado Press, Boulder, 1965), vol. 7C, p. 1.}

\bibitem{kc}{S.-Y. Kim, R. Creswick, C.-N. Chen, and C.-K. Hu, in the
Proceedings of the Taiwan Conference on Equilibrium and Non-Equilibrium Phase
Transitions (Academia Sinica, Taipei, Aug. 1999).}

\bibitem{chisq}{V. Matveev and R. Shrock, J. Phys. A {\bf 28} (1995) 1557.}

\bibitem{cmo}{V. Matveev and R. Shrock, J. Phys. A {\bf 28} (1995) 5235.} 

\bibitem{ssbounds}{J. Salas and A. Sokal, J. Stat. Phys. {\bf 86} (1997)
551.} 

\bibitem{ww}{Y. K. Wang and F. Y. Wu, J. Phys. A {\bf 9} (1976) 593.} 

\bibitem{is1d}{V. Matveev and R. Shrock, Phys. Lett. {\bf A204} (1995) 353.}

\bibitem{m1}{P. P. Martin, J. Phys. A {\bf 19}, 3267 (1986). See also
P. P. Martin, {\it ibid.}, {\bf 20}, L601 (1986).}

\bibitem{wood}{D. W. Wood, J. Phys. A {\bf 20}, 3471 (1987); 
D. W. Wood, R. W. Turnbull, and J. K. Ball, {\it ibid.} 3495 (1987).}

\bibitem{uspensky}{J. V. Uspensky, {\it Theory of Equations}
(McGraw-Hill, NY 1948), 264.}

\bibitem{alg}{R. Hartshorne, {\it Algebraic Geometry}
(Springer, New York, 1977).}

\bibitem{ff}{A. E. Ferdinand and M. E. Fisher, Phys. Rev. {\bf 185} (1969)
832.}

\bibitem{bf}{M. E. Fisher and M. N. Barber, Phys. Rev. Lett. {\bf 28} (1972)
1516; M. N. Barber, in C. Domb and J. Lebowitz, {\it Phase Transitions and
Critical Phenomena}, v. 8 (Wiley, New York, 1983).}

\bibitem{hsi}{V. Matveev and R. Shrock, J. Phys. A (Letts.) {\bf 28} (1995) 
L533.}

\bibitem{mr}{J. M. Maillard and R. Rammal, J. Phys. A {\bf 16} (1983) 353.} 

\bibitem{mm}{P. P. Martin and J. M. Maillard, J. Phys. A {\bf 19} (1986)
L547.}

\bibitem{mbook}{P. P. Martin, {\it Potts Models and Related Problems in
Statistical Mechanics} (World Scientific, Singapore, 1991).}

\bibitem{chw}{C. N. Chen, C. K. Hu, and F. Y. Wu, Phys. Rev. Lett. {\bf 76} 
(1996) 169.}

\bibitem{wuetal}{F. Y. Wu, G. Rollet, H. Y. Huang, J. M. Maillard, C. K. Hu,
and C. N. Chen, Phys. Rev. Lett. {\bf 76} (1996) 173.}

\bibitem{pfef}{V. Matveev and R. Shrock, Phys. Rev. {\bf E54} (1996) 6174.}

\bibitem{pafhc}{R. Shrock and S.-H. Tsai, J. Phys. A {\bf 30} (1997) 495.}

\bibitem{p}{H. Feldmann, R. Shrock, and S.-H. Tsai, J. Phys. A (Lett.) {\bf 30}
(1997) L663;  Phys. Rev. {\bf E57} (1998) 1335; H. Feldmann, A. J. Guttmann, 
I. Jensen, R. Shrock, and S.-H. Tsai), J. Phys. A {\bf 31} (1998) 2287.}

\bibitem{kch}{W.-Y. Kim and R. Creswick, Phys. Rev. {\bf E58} (1998) 7006.} 

\bibitem{baxter82}{R. J. Baxter, Proc. Roy. Soc. London, Ser. A
{\bf 383} (1982) 43.}

\bibitem{fhspan}{R. K. Guy and F. Harary, Univ. of Calgary Rept. 2, 1966;
J. Sedl\'acek, in Combinatorial Structures and Applications (Gordon and Breach,
New York, 1970), p. 387.} 

\bibitem{wu77}{F. Y. Wu, J. Phys. A {\bf 10} (1977) L113.} 

\bibitem{wt}{F. Y. Wu, C. King, and W. T. Lu, Ann. Inst. Fourier {\bf 49}
(1999) 101.}

\bibitem{tzengwu}{W.-J. Tzeng and F. Y. Wu, Northeastern-NCTS preprint; 
R. Shrock, F. Y. Wu, Northeastern-NCTS-Stony Brook preprint.}

\end{thebibliography}
\end{document}